\documentclass[11pt,a4paper]{article}
\usepackage[a4paper, total={6in, 8in}]{geometry}
\usepackage[utf8]{inputenc}
\usepackage[export]{adjustbox}
\usepackage{authblk}
\usepackage{hyperref}

\usepackage{amsmath}
\usepackage{amsthm}
\usepackage{amssymb}
\usepackage{lmodern}
\usepackage{multirow}
\usepackage{graphicx}
\usepackage{subcaption}
\usepackage{placeins}
\RequirePackage[dvipsnames]{xcolor}
\usepackage{float}
\usepackage{cleveref}
\usepackage{cite}
\usepackage{lineno}
\usepackage[normalem]{ulem}

\definecolor{mypink}{rgb}{0.858, 0.188, 0.478}

\begin{document}

\title{Deep Learning for the Classification of Quenched Jets}

\author[1,2]{L. Apolin\'{a}rio}
\author[1,3]{N. F. Castro}
\author[1]{M. Crispim Rom\~{a}o}
\author[1,2]{J. G. Milhano}
\author[1]{R. Pedro}
\author[1, 4, 5]{F. C. R. Peres}
\affil[1]{Laboratório de Instrumentação e Física Experimental de Partículas (LIP), Av. Professor Gama Pinto 2, 1649-003 Lisboa, Portugal}
\affil[2]{Instituto Superior T\'{e}cnico, Universidade de Lisboa, Av. Rovisco Pais 1, 1609-001, Lisboa, Portugal}
\affil[3]{Departamento de F\'{i}sica, Escola de Ci\^{e}ncias, Universidade do Minho, 4710-057 Braga, Portugal}
\affil[4]{International Iberian Nanotechnology Laboratory (INL), Av. Mestre José Veiga, 4715-330 Braga, Portugal}
\affil[5]{Departamento de F\'{i}sica e Astronomia, Faculdade de Ci\^{e}ncias, Universidade do Porto, R. do Campo Alegre, 4169-007 Porto, Portugal}

\maketitle

\begin{abstract}
    An important aspect of the study of Quark-Gluon Plasma (QGP) in ultra-relativistic collisions of heavy ions is the ability to identify, in experimental data, a subset of the jets that were strongly modified by the interaction with the QGP. In this work, we propose studying deep learning techniques for this purpose. Samples of $Z+$jet events were simulated in vacuum and medium and used to train deep neural networks with the objective of discriminating between \textit{medium}- and \textit{vacuum-like} jets. Dedicated Convolutional Neural Networks, Dense Neural Networks and Recurrent Neural Networks were developed and trained, and their performance was studied. Our results show the potential of these techniques for the identification of jet quenching effects induced by the presence of the QGP.
\end{abstract}

\clearpage
\tableofcontents


\section{Introduction}

The experimental observation \cite{Adler:2002xw,Adcox:2001jp} of suppression of high transverse momentum hadrons in ultra-relativistic collisions of heavy ions relative to appropriately scaled proton-proton collisions was a major step in establishing that a novel state of matter, a Quark-Gluon Plasma (QGP), is produced in heavy-ion (HI) collisions. 
The ability to systematically reconstruct full jets in the presence of the large and fluctuating underlying event of HI collisions, first at the LHC \cite{Aad:2010bu,Chatrchyan:2011sx,Abelev:2013kqa} and later at RHIC \cite{Adamczyk:2012eoa,Adare:2010mq}, significantly expanded the scope of the use of jets as tools to understand the inner workings of the QGP they develop within. 
Studies of QGP-induced jet modifications, commonly referred to as jet quenching, were initially based on global jet properties (e.g. jet total transverse momentum) but have since evolved into detailed studies of increasingly complex observables \cite{Apolinario:2017qay,Andrews:2018jcm} in the most part related to the internal structure of jets (their sub-structure).

While this program has significantly advanced the understanding of the dynamics of jet-QGP interaction, a fundamental difficulty underlies all but a few jet quenching studies. This difficulty can be illustrated by noting that in HI collisions a set of jets with total reconstructed transverse momentum $p_T$ within a given range includes jets that have experienced different levels of modification, that is jets that have been quenched to diverse extents. Together with the steeply falling spectrum for jet production, this makes any HI jet sample within any given $p_T$ range to be dominated by jets that experienced little or no modification. As such, quenching effects may present themselves as subtle modifications not because quenching is a small effect overall, but rather because jets that were significantly modified are diluted within a sample dominated by those with little modification.

The mitigation of this difficulty requires the ability to compare jets that were born alike rather than those that were detected with the same final reconstructed $p_T$ allowing for direct assessment of the modifications experienced by jets.

A path towards such mitigation involves the analysis of jets produced back-to-back with electroweak bosons ($\gamma$, Z or W) \cite{Sirunyan:2017jic,Sirunyan:2018qec,Aaboud:2018anc,Aaboud:2019oac}, as in this case the $p_T$ of the electroweak boson provides a close proxy to the $p_T$ of the parton from which the jet develops. However, these events have limited statistics.

Another possibility is to use data-driven procedures like the one proposed in \cite{Brewer:2018dfs}, which allows for the determination of the \textit{average} $p_T$ lost by jets being reconstructed in HI collisions with some final $p_T$, but not of the fluctuations of that energy loss.

More recently, a novel reclustering jet algorithm was proposed to study jet quenching effects in HI collisions~\cite{Apolinario:2020uvt}, allowing to identify jets that have different levels of QGP-induced modifications. This ability to identify within a jet sample a subset of the most modified jets is invaluable to augment the visibility of quenching effects and thus provide a cleaner slate to distinguish specific features of jet-QGP interaction.

In this work, we ask whether Deep Learning (DL) techniques can provide complementary criteria to distinguish strongly modified jets from essentially unmodified ones. A recent study \cite{Lai:2018ixk} showed that a convolutional neural network (CNN) trained on jet images for jets modified using the strong/weak Hybrid model \cite{Casalderrey-Solana:2014bpa} including effects of medium response \cite{Casalderrey-Solana:2020rsj} allowed for the extraction, on a jet-by-jet basis, of the $p_T$ the jet would have had if no QGP was present. These important results rely, at least in significant part, on the presence of a medium response component which is the leading feature identified by the CNN as signalling the strength of quenching effects. However, the medium response remains the most model-dependent component of state-of-the-art \cite{Zapp:2013vla,Casalderrey-Solana:2020rsj,Young:2011ug} jet quenching simulations and is not inconceivable that features highlighted by the CNN may be model-specific and absent in real data.

Dedicated HI jet observables have been proposed as being more resilient to such medium response component~\cite{Apolinario:2017qay}, while being, simultaneously, calculable in a pQCD prescription~\cite{Neill:2021std}. Nonetheless, some of the currently available jet substructure measurements \cite{Sirunyan:2017bsd,Acharya:2019djg} do not enjoy this feature, and part of the unrelated underlying event can lead to effects very similar to those that can be argued to arise from medium response contribution to jets~\cite{Milhano:2017nzm}. For these observables, the validation of a procedure that distinguishes quenched jets in HI collisions from those without any medium modification can only be made by comparing HI jets to vacuum jets embedded in the fully uncorrelated background (e.g. built with mixed event techniques \cite{Sirunyan:2021jty} or by embedding in real PbPb events without jets \cite{Acharya:2019djg}) with both samples undergoing the same analysis workflow including background subtraction.

In this work we ask a different, more fundamental, question: whether modifications imparted on jets by the QGP on a perturbative level, that is modifications of the parton shower, are sufficient for DL to attain a satisfactory discriminatory power.

The DL architectures considered in this work are trained without accounting for the medium response, thus maximizing the training on QCD in-medium emissions whose implementation in MC generators is solidly grounded in perturbative QCD. While some model dependence persists, since state-of-the-art jet quenching Monte Carlo event generators implement QGP-induced modifications in different ways, we believe it to be as small as presently possible.

The paper is organised as follows: in Section~\ref{sec:sec2}, we present our simulation setup and the procedure to prepare the data samples used by the different DL architectures. The DL models used throughout this work and their training results are presented in Section~\ref{sec:DL}. A careful analysis of the DL outputs and their possible interpretation to separate jet quenching effects is done in Section~\ref{sec:Results}. The final conclusions are presented in Section~\ref{sec:conclusions}.

\section{Simulated data}
\label{sec:sec2}

To understand if Deep Learning can be applied to identify jet quenching effects we will use JEWEL v2.2.0~\cite{Zapp:2013vla}, a Monte Carlo event generator that accounts for medium-induced effects during the QCD parton shower evolution. Since the main goal is to identify medium-induced modifications to the parton shower structure, medium recoils (the particular implementation of QGP response in JEWEL) are not considered in this work. 

We use the simple, parametrised medium described in detail in \cite{Zapp:2013vla} with  settings tuned to $T = 440~\rm{MeV}$ and $\tau_i = 0.4~\rm{fm}$, while the remaining parameters were set to the default values. These are known to reproduce current jet energy loss experimental observations even without medium recoil effects~\cite{KunnawalkamElayavalli:2017hxo}. From $10^6$ weighted Z+jet hadronic events at a $\sqrt{s_{NN}} = 5.02~\rm{TeV}$ whose particles are required to have a minimum transverse~\footnote{The transverse plane is defined with respect to the colliding beams axis.} momentum of $p_{T,part}^{min} = 500$~\rm{MeV}, we reconstruct the $Z$-boson from the pair of muons that result into a reconstructed object with a minimum transverse momentum of $p_{T,Z}^{min} = 90~\rm{GeV}$ and mass $m_Z \in [75; 105]~\rm{GeV}$. The anti-k$_{T}$~\cite{Cacciari:2008gp} reconstructed jet with $R = 0.5$ and minimum transverse momentum of $p_{T,jet}^{min}=30~\rm{GeV}$ is required to be within an azimuthal angle with respect to the reconstructed $Z$-boson of $|\Delta \phi| = |\phi_{Z} - \phi_{jet}| > 7\pi/8$ and to have an absolute pseudo-rapidity of $|\eta_{jet}| < 1.0$ to avoid projection effects in the resulting jet image. All jet reconstruction procedures were performed within the FastJet package~\cite{Cacciari:2011ma}.

\begin{figure}[]
    \centering
    \includegraphics[width=0.44\textwidth]{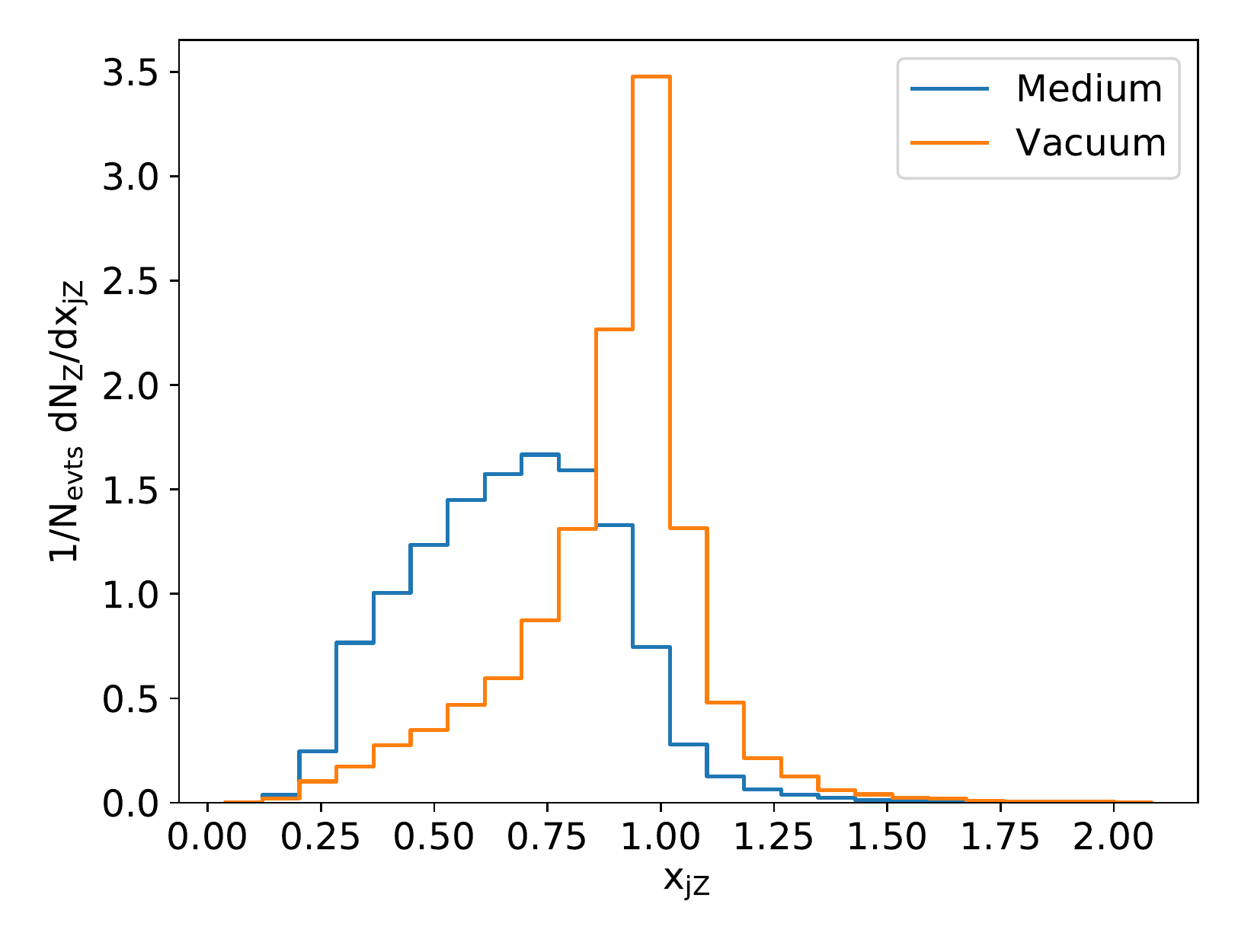} 
    \caption{\label{fig:xjz} Distribution of the transverse momentum ratio between the jet and $Z$-boson, $x_{jZ}$, as provided by PYTHIA+JEWEL for Vacuum and Medium.}
\end{figure}

\begin{figure}[]
    \centering
    \includegraphics[width=0.44\textwidth]{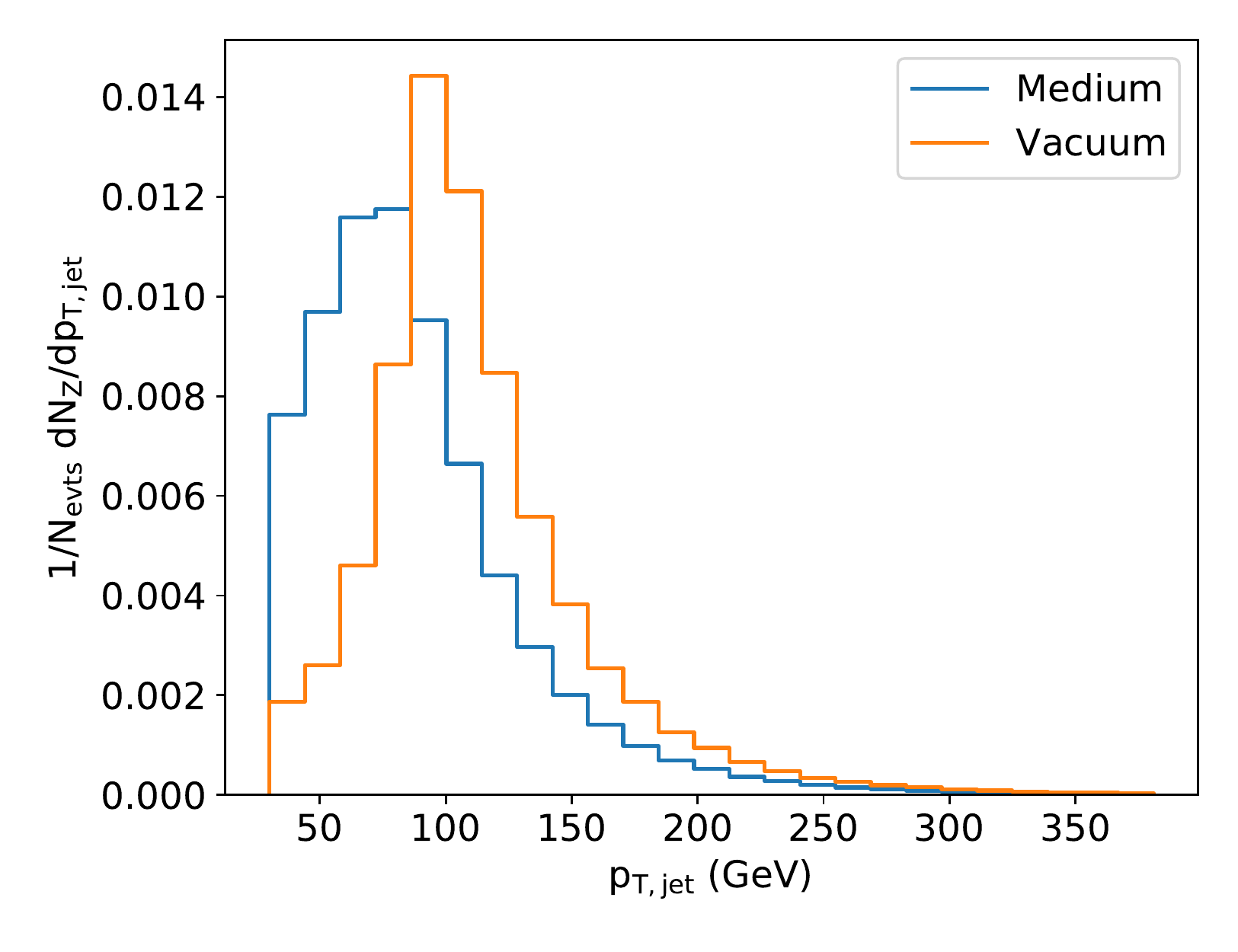}
    \includegraphics[width=0.44\textwidth]{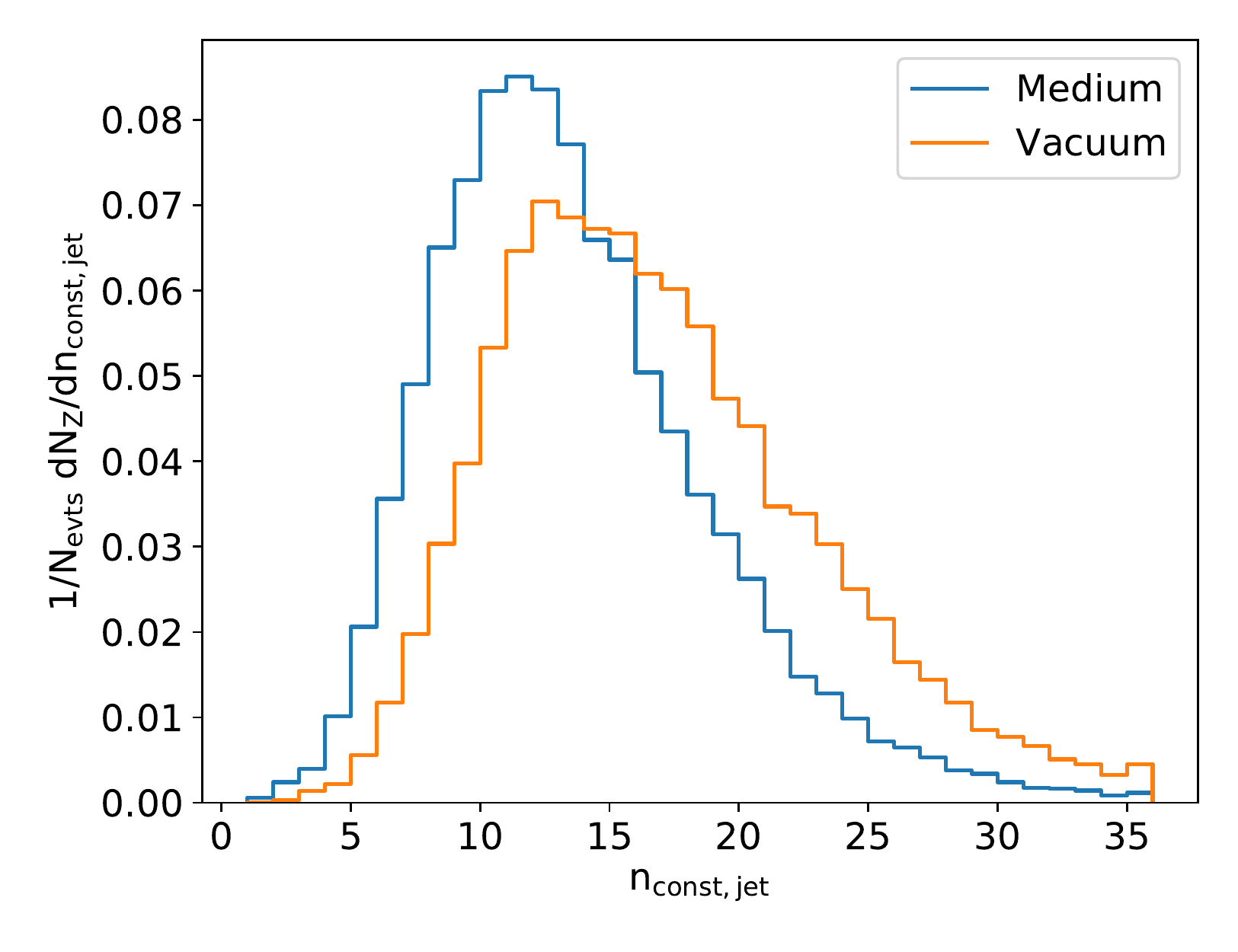}
    \caption{\label{fig:sample_distributions} Distribution of the (left) jet transverse momentum $p_{T,jet}$ and (right) number of jet constituents $n_{const}$ for the JEWEL+PYTHIA Vacuum and Medium samples.}
\end{figure}

\begin{figure}[]
    \centering
    \includegraphics[width=0.49\textwidth]{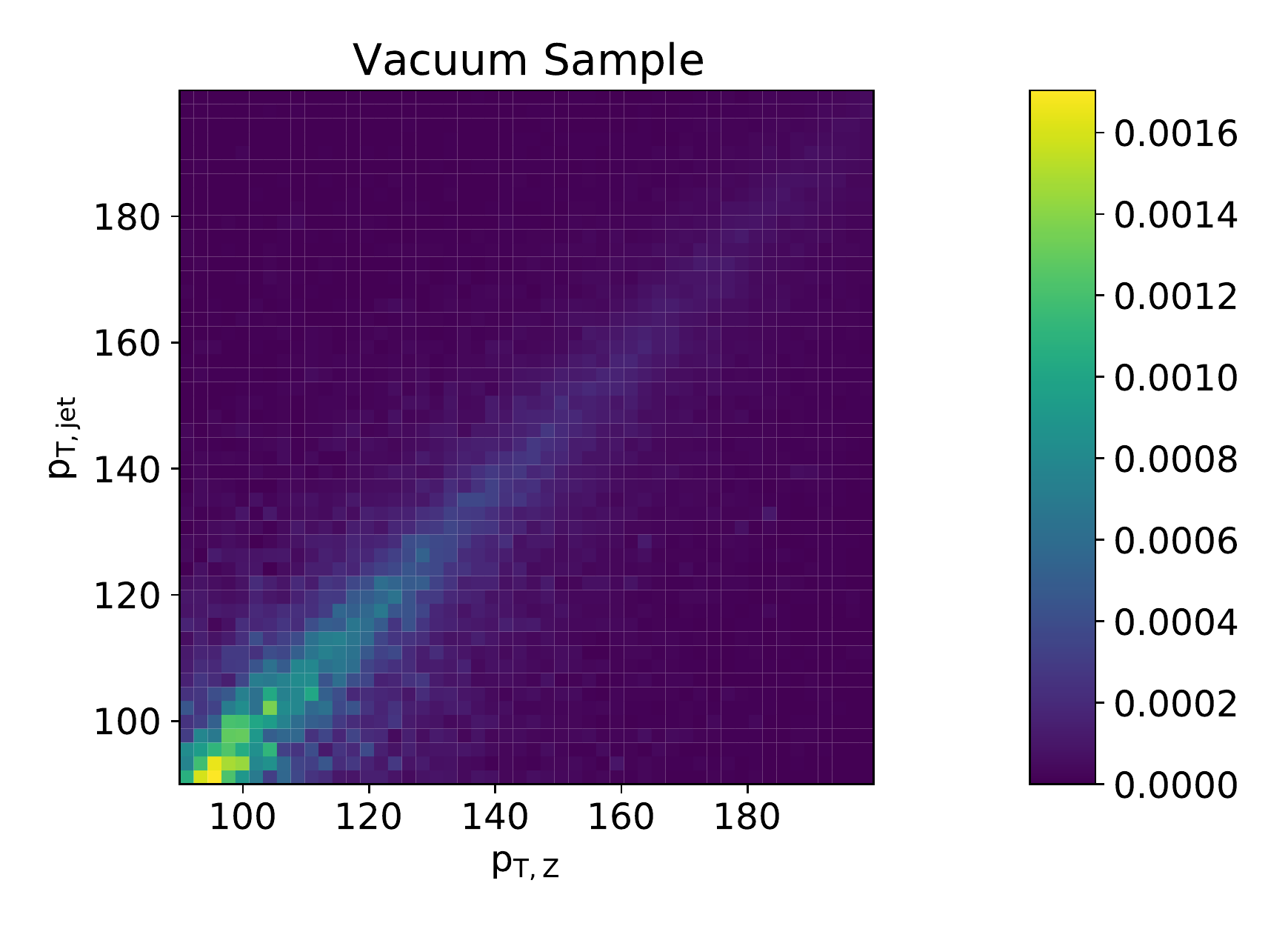}
    \includegraphics[width=0.49\textwidth]{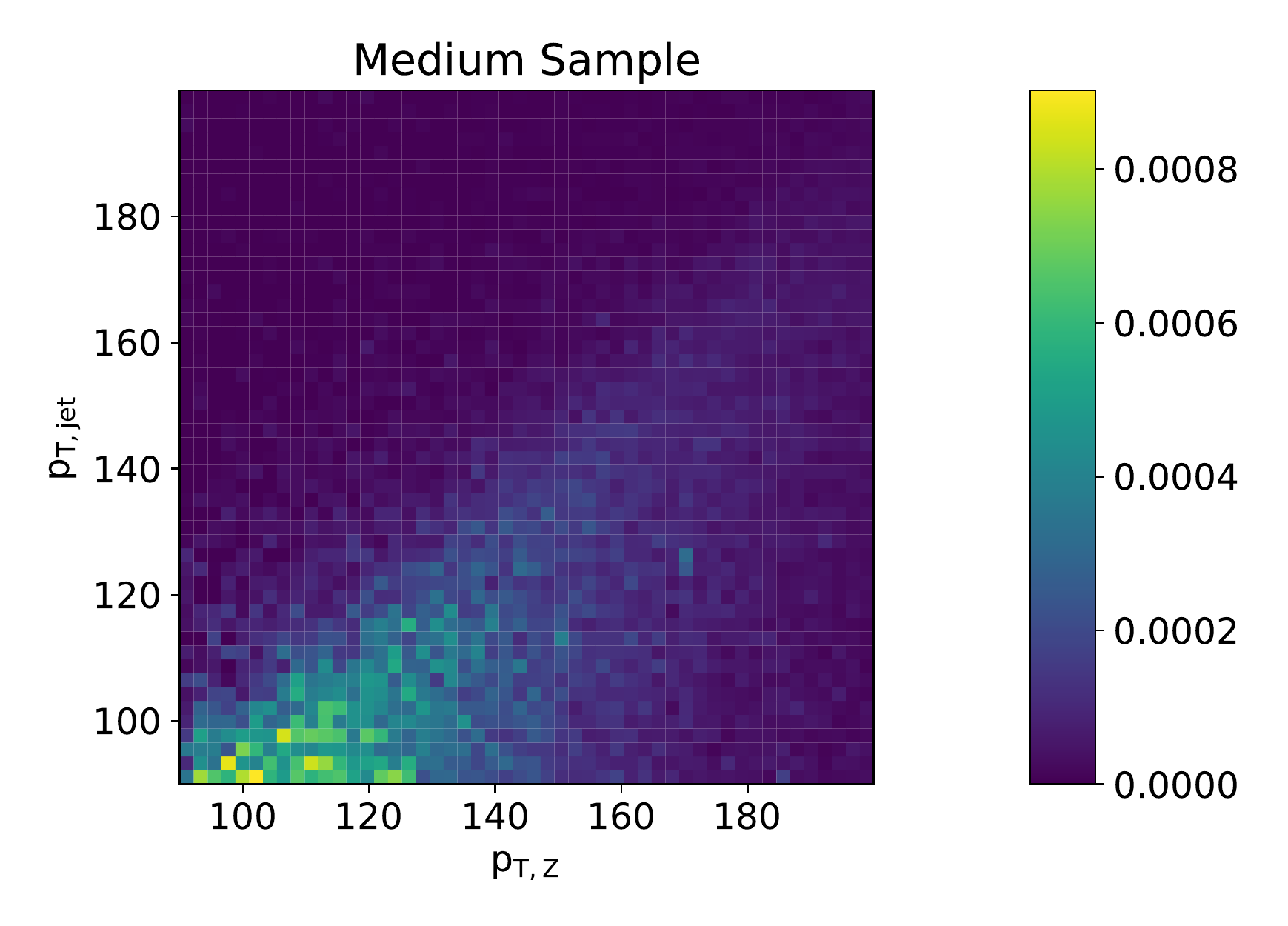} 
    \caption{\label{fig:ptcorrelation} Bi-dimensional distributions of the $Z$-boson and jet transverse momenta, $p_{T,Z}$ and $p_{T,jet}$, for the JEWEL+PYTHIA (left) Vacuum and (right) Medium samples.}
\end{figure}

The resulting transverse momentum ratio between the jet and $Z$-boson, $x_{jZ} = p_{T,jet}/p_{T,Z}$, $p_{T,jet}$ and jet multiplicity, $n_{const}$ are shown in \cref{fig:xjz} and \cref{fig:sample_distributions} for PYTHIA+JEWEL (Vacuum, without jet quenching effects) in orange and PYTHIA+JEWEL (Medium, with jet quenching effects) in blue. In proton-proton collisions, the transverse momentum ratio, $x_{jZ}$ is naturally peaked at $1$ with a spread towards small and larger values. The former is a consequence of not being able to fully recover the energy due to the finite radial extent of the jet and events where more than one jet is reconstructed~\footnote{JEWEL uses LO hard matrix elements and thus does not generate Z + 2 jets at hard matrix element level, the parton shower generates configurations where the initial parton radiates sufficiently hard and wide for the end result being 2 reconstructed jet.}, while the latter comes mainly from initial-state radiation contamination. Since the $Z$-boson and its decay products (muons) are colourless, they will not undergo any modification when medium effects are introduced. So, they provide a good proxy for the initial momentum of the jet-initiating parton. Nonetheless, the recoiling jet will experience several scattering processes, inducing extra radiation that is emitted at finite angles. While part of this radiation stays inside the jet under the form of softer fragments, collisional energy loss contributes further to the depletion of the reconstructed transverse momentum $p_{T,jet}$ (and $x_{jZ}$) and effectively reduces the number of particles $n_{const}$ since they are transported up to large radial distances in $(\eta, \phi)$~\cite{Sirunyan:2018jqr}. As such, while in vacuum there is a large correlation between $p_{T,jet}$ and $p_{T,Z}$, as shown in the left panel of \cref{fig:ptcorrelation}, the left shift on the $x_{jZ}$ medium distribution partially destroys the correlation between the boson and jet transverse momenta (right panel of \cref{fig:ptcorrelation}) 

In addition, the criterion on the minimum jet transverse momentum also induces a selection bias on the medium sample: pairs whose recoiling jet is below the cut-off will not be included. These configurations are usually dominated by jets with a larger number of constituents and a wider fragmentation pattern. As a result, the medium sample will be dominated by jets with a narrower fragmentation pattern, which in turn are naturally present in the vacuum sample.

The $p_{T,jet}$ and $n_{const}$ show significant differences between vacuum and medium samples (see \cref{fig:sample_distributions}). These will be used as input information to the training of some of the DL models used in this work. For such networks, we expect that the discriminating power will be significantly correlated to these variables. Nonetheless, the $x_{jZ}$ variable will not be included in any of the DL architectures. Since this is a powerful discriminant in itself, we preserve it as a physical benchmark against which to compare the different DL outputs.

\FloatBarrier
\subsection{Data representations}

The simulated data used in the present study was prepared in different formats, each representing the jets in a specific way, which encodes the information with different implicit biases. We explore three main jet representations: calorimeter images, Lund plane coordinates, and jet-wise $p_{T,jet}$ and $n_{const}$. Each representation of the jet carries different implicit features that are more suitable to study different substructure aspects of jet quenching.

\subsubsection*{Jet images}

The jet-image consists of displaying the transverse momentum and multiplicity of the jet constituents mimicking calorimeter towers. As such, the jet particles are drawn in a $(\Delta \eta, \Delta \phi)$ grid composed of $35\times35$ cells centred in the jet axis. Each cell will have two channels, where the first contains the accumulated transverse momentum of the particles contained in that cell while the second channel contains the particle multiplicity. When summing over of all cell's content we recover the jet $p_{T,jet}$ and $n_{const}$. This type of information contains, in principle, all possible angularity-type of variables \cite{Larkoski:2017jix}. The usage of calorimeter images with CNNs have been explored previously~\cite{de2016jet,Chien:2018dfn} in the context of the classification between jets initiated by quarks and jets initiated by gluons, both in proton-proton and heavy-ion collisions.

We work with two different types of jet images. In the first case, \emph{unnormalised}, we use the absolute values of the $p_T$ and multiplicity of each cell, while in the second approach, \emph{normalised}, each channel is normalised by the sum of its entries, \emph{i.e.}, the $p_{T,jet}$ and $n_{const}$. The purpose of this is to have a comparison in performance between a DL network that has access to the whole information, including the scale of $p_{T,jet}$ and $n_{const}$, and one that only has access to the relative fragmentation pattern in $(\Delta \phi, \Delta \eta)$.

\begin{figure}[t]
    \centering
    \includegraphics[width=0.9\textwidth]{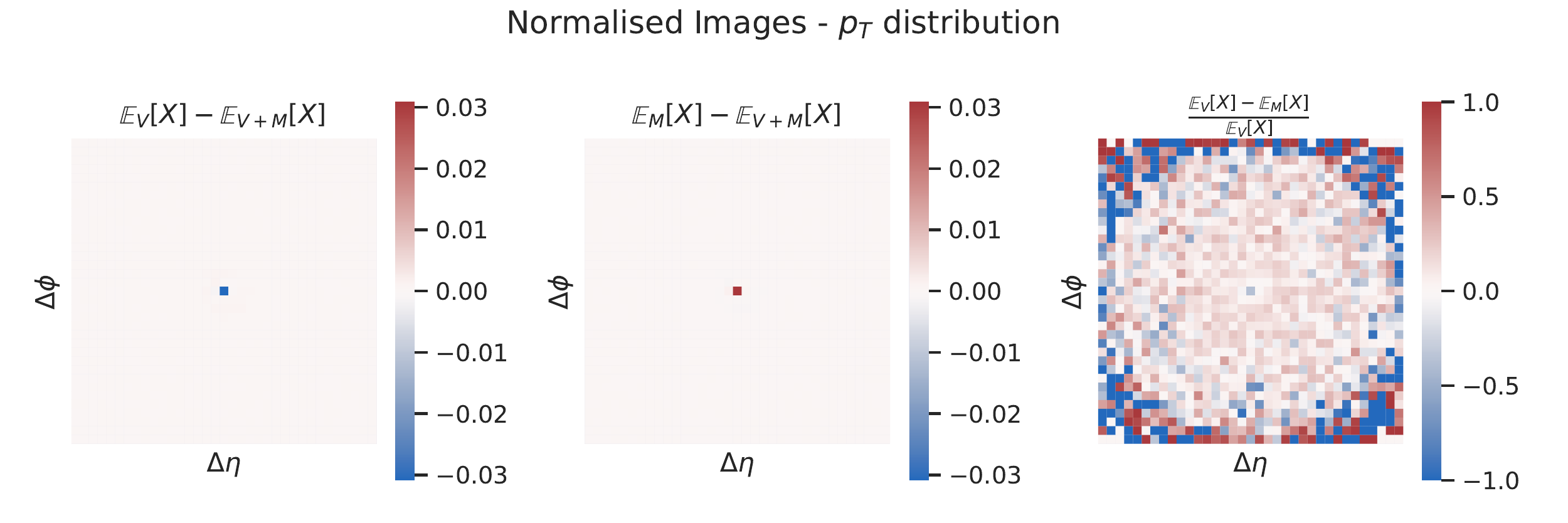}
    \includegraphics[width=0.9\textwidth]{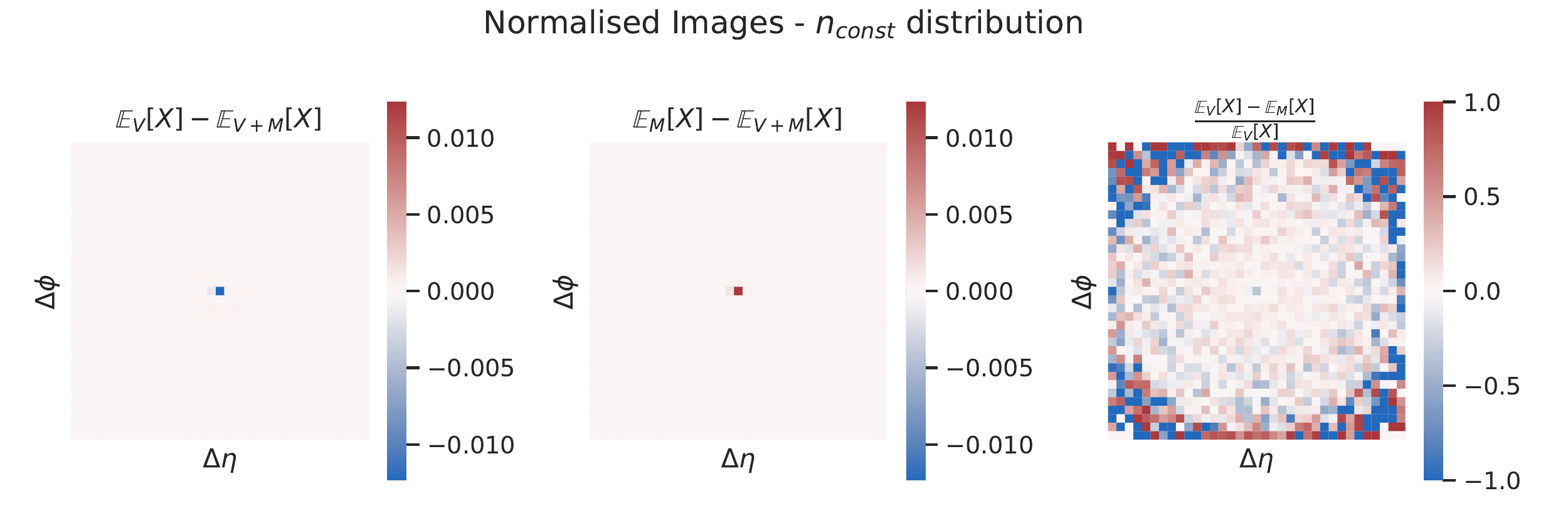}
    \caption{\label{fig:images_normalised} Representation of the deviation from the mean (top) jet transverse momentum and (bottom) number of jet constituents in the $(\Delta \eta, \Delta \phi)$-plane for the (left) Vacuum sample, (center) Medium sample and (right) the difference between the mean Vacuum and Medium images, relative to Vacuum. The images are individually normalised to the total jet transverse momentum and to the total number of jet constituents.}
\end{figure}

In~\cref{fig:images_normalised} we present both channels, the relative (normalized) $p_{T,jet}$ and $n_{const}$, of the mean image of each sample subtracted by the mean image of both samples, defined as
\begin{equation}
    \mathbb{E}_{V+M}[X] = \frac{1}{2} (\mathbb{E}_{V}[X]+\mathbb{E}_{M}[X]) \ ,
\end{equation}

\noindent where $X$ stands for the channel being shown, $\mathbb{E}$ is the expected value, and $V$ and $M$ representing the Vacuum and Medium samples, respectively. As we can see, the differences against the mean normalied image of both samples are very nuanced for both Vacuum and Medium samples. However, we do observe that the central pixel has, on average, a smaller value for the Vacuum sample than for the medium sample for both channels, signalling a narrower jet selection bias.

\begin{figure}[t]
    \centering
    \includegraphics[width=0.9\textwidth]{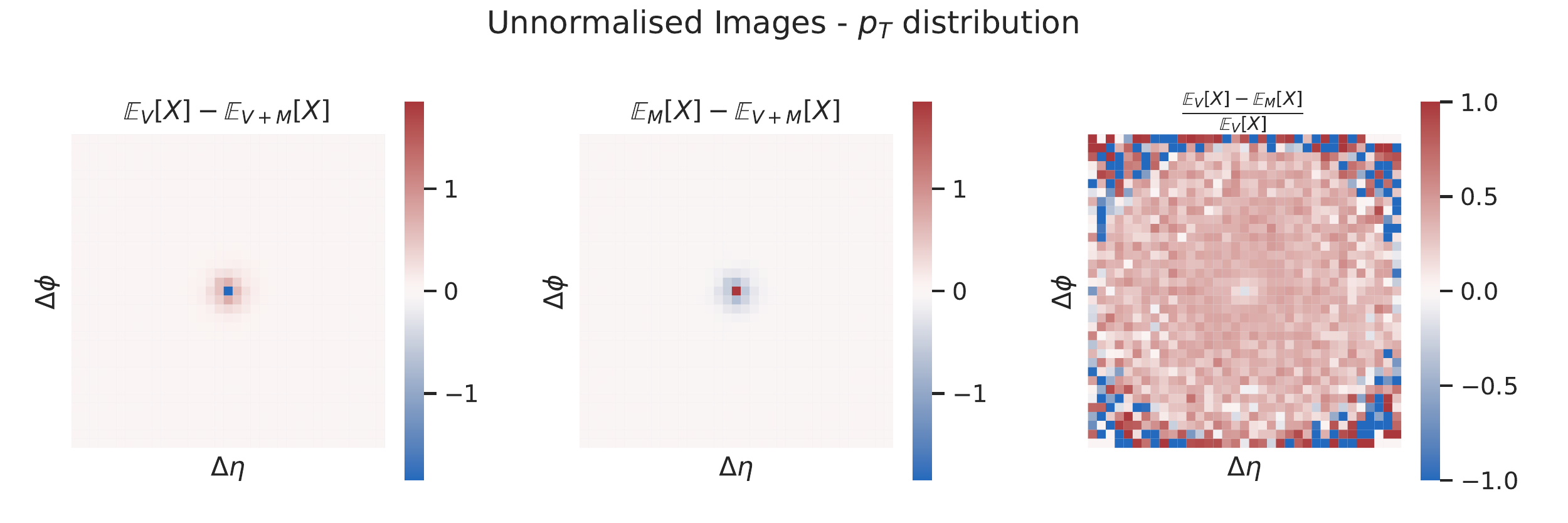}
    \includegraphics[width=0.9\textwidth]{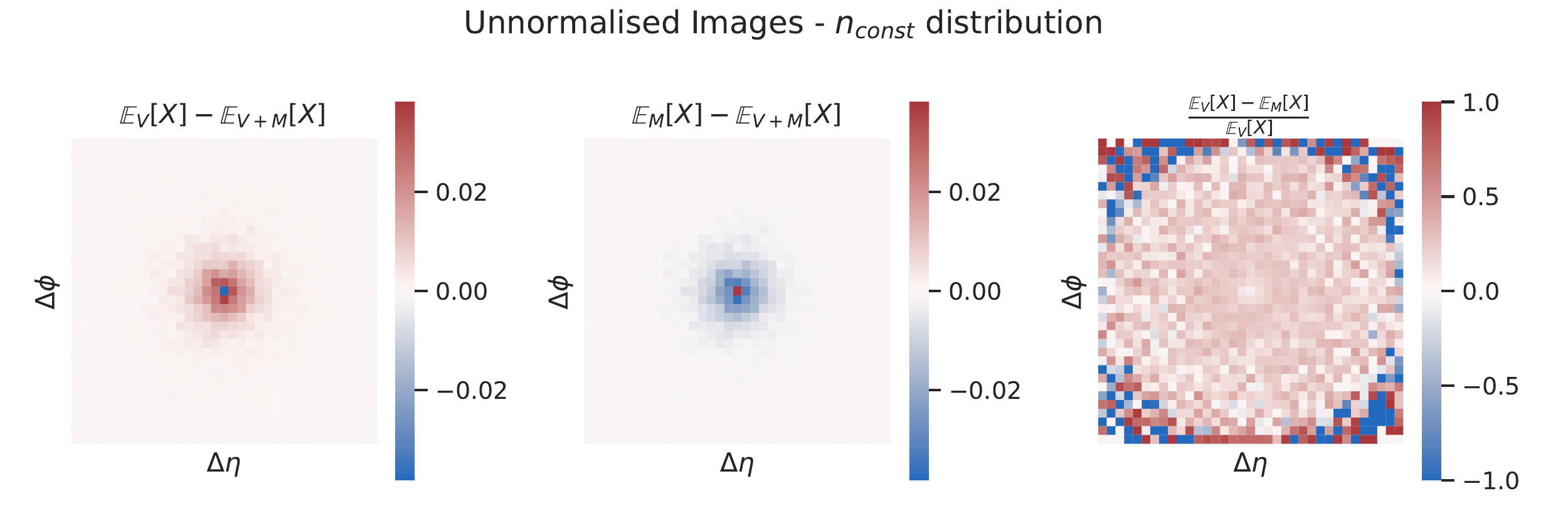}
    \caption{\label{fig:images_unnormalised} Representation of the deviation from the mean (top) jet transverse momentum and (bottom) number of jet constituents in the $(\Delta \eta, \Delta \phi)$-plane for the (left) Vacuum sample, (center) Medium sample and (right) the difference between the mean Vacuum and Medium images, relative to Vacuum. The images represent the total jet transverse momentum and the total number of jet constituents.}
\end{figure}

In~\cref{fig:images_unnormalised} we show the same image for the unnormalised case. Here, the differences between Vacuum and Medium are more noticeable, highlighting the expectation that providing the absolute scale of both $p_{T,jet}$ and $n_{const}$ will facilitate discrimination between both samples. We also see that the distribution of momentum and multiplicity inside of jets in the Medium sample is typically more suppressed with respect to the Vacuum sample. This observation is in agreement with the energy loss mechanism implemented within JEWEL, and the effect on the selection bias induced by the jet $p_{T}$ cut, discussed previously. Overall, jets with a narrower fragmentation pattern are more likely to survive in the presence of energy loss effects. 

\begin{figure}[]
    \centering
    \includegraphics[width=0.49\textwidth]{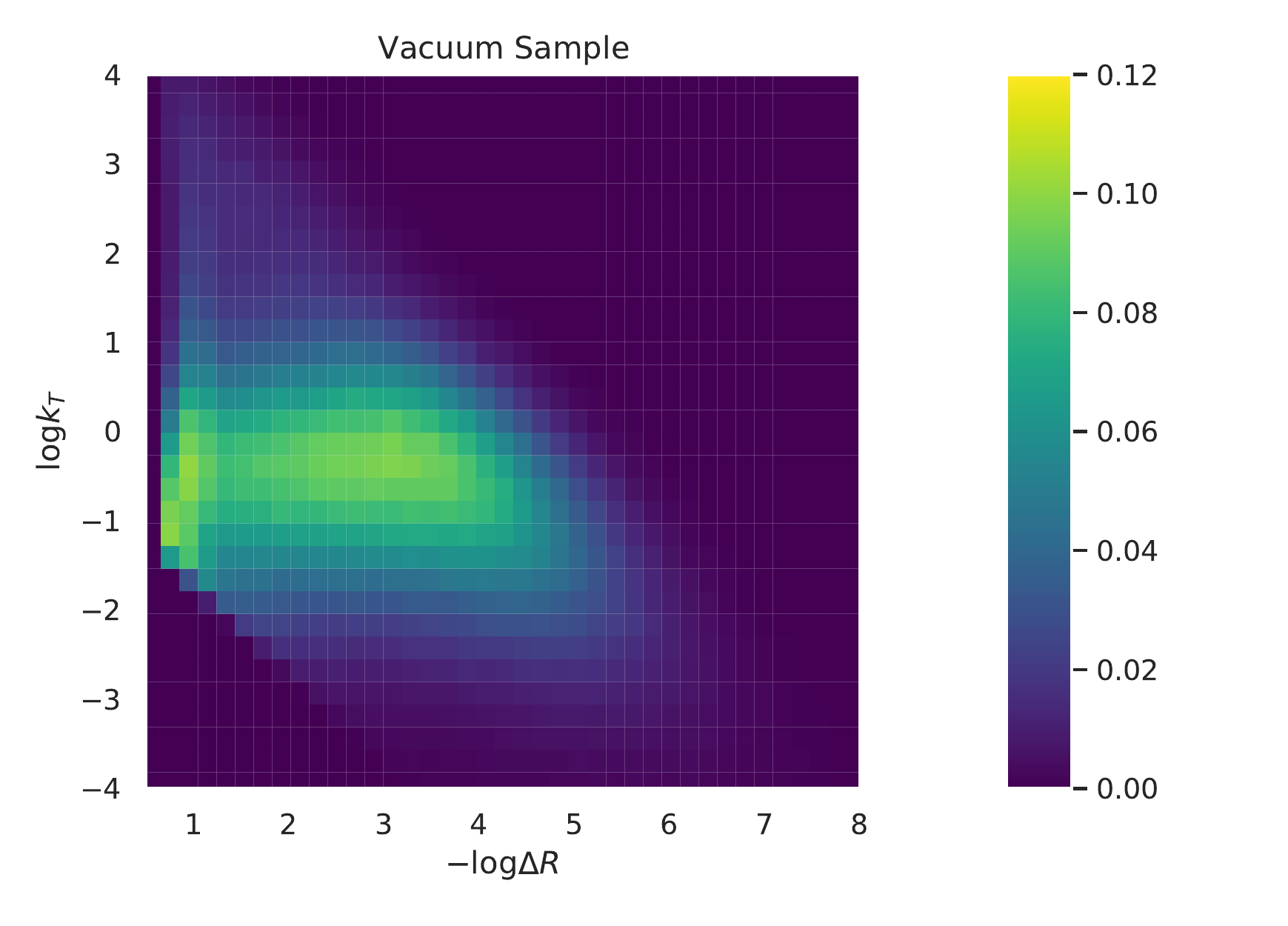}
    \includegraphics[width=0.49\textwidth]{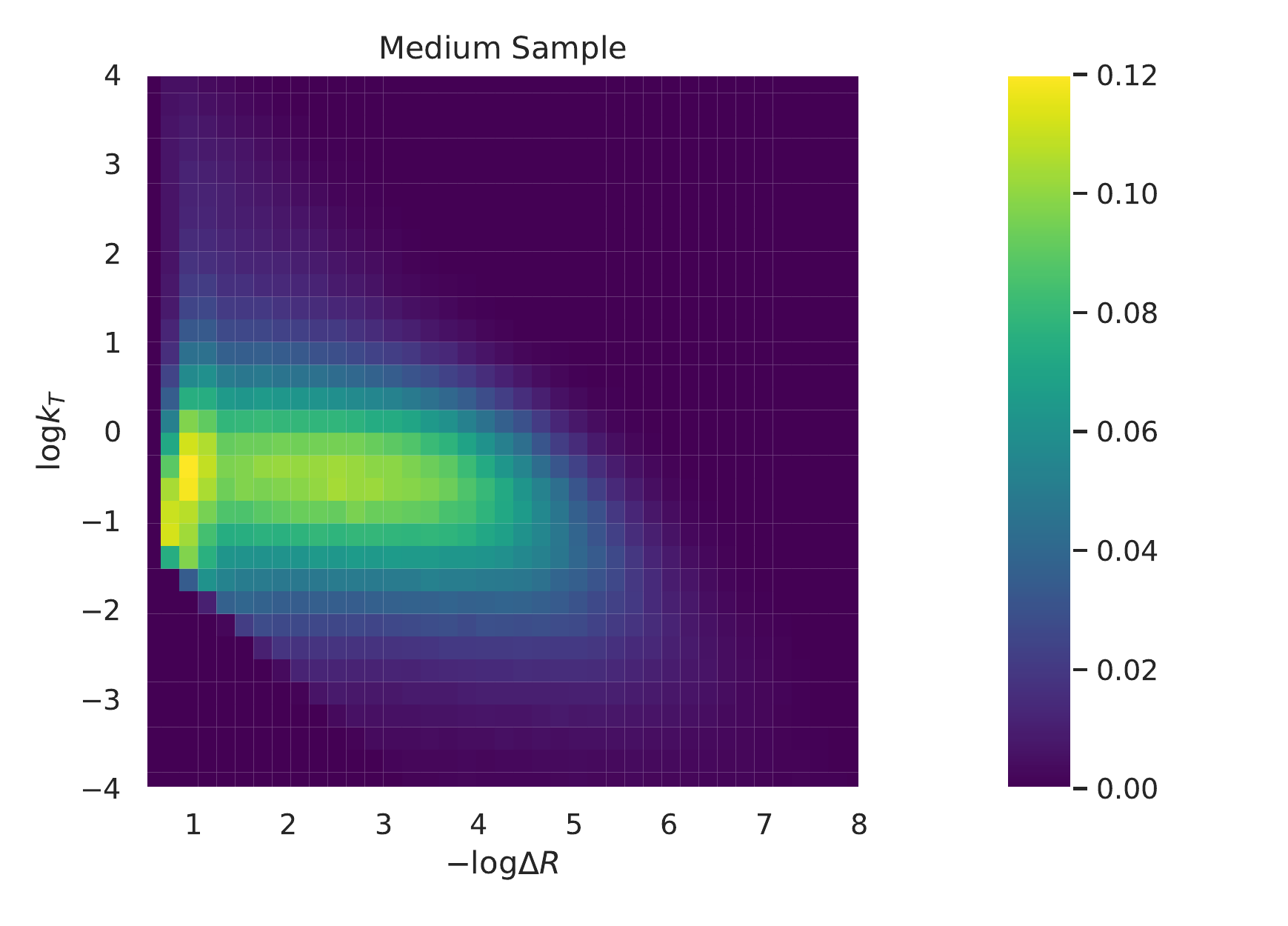} 
    \caption{\label{fig:lund-plane} Representation of the jets in the primary Lund plane  $(\mathrm{log}k_T,-\mathrm{log}\Delta R)$ for the JEWEL+PYTHIA (left) Vacuum and (right) Medium samples.}
\end{figure}

\subsubsection*{Lund plane coordinates}
The second jet representation considered is the primary sequence of Lund plane coordinates of the Cambridge-Aachen (C/A) jet clustering sequence~\cite{dokshitzer1997better}. To produce these, the jet is reclustered with the C/A clustering algorithm. The unclustering sequence, at each step, will result into two sub-jets with transverse momentum $p_{T,1}$ and $p_{T,2}$ respectively, from which we obtain the $(\log k_T, -\log \Delta R)$ coordinates~\footnote{$\Delta R$ is the distance in the rapidity $y$ and azimuthal angle $\phi$ plane between the two obtained sub-jets. $k_T = p_T \sin \Delta R$, where $p_T$ is the transverse momentum of the softest sub-jet, \emph{i.e.}, $p_T = \min (p_{T,1}, p_{T,2})$.}. The procedure goes iteratively along the primary (the hardest) branch. Since the C/A clustering algorithm produces a branching history that is angle-ordered, we retain the order of these splittings. Therefore, the jets in this format are represented by a $N \times 2$ matrix, where $N$ is the number of branches in the final clustering tree. The reclustering of the jets was performed in FastJet~\cite{Cacciari:2011ma}. 

The average representation of these primary jet Lund planes obtained from the JEWEL+\\PYTHIA Vacuum (Medium) samples is shown in \cref{fig:lund-plane} left (right). The diagonal lines with negative slope represent the kinematic cut of having a sub-jet with $p_{T,part}^{min} \leq k_T \leq p_{T,jet}/2$.

We also considered other clustering algorithms, in particular, the $\tau$ algorithm as proposed in \cite{Apolinario:2020uvt}, different settings of grooming using the Soft-Drop procedure~\cite{Larkoski:2014wba}, and different Lund plane representations and coordinates. To settle for the coordinates of the primary jet Lund planes obtained with C/A re-clustering without Soft-Drop, we performed a preliminary analysis using a non-optimised DL model to assess the dependence of its performance on these different combinations. We found that all DL networks performed similarly, and we fixed the representation that is presented above.

\subsubsection*{Tabular data - global $p_{T,jet}$ and {$n_{const}$}}

The final jet representation corresponds to tabular data containing $p_{T,jet}$ and {$n_{const}$} per jet. The purpose of this representation is to quantify the discriminating power of these two variables alone. Two of the representations above have information on both the jet $p_{T,jet}$ and its number of constituents: the unnormalised images and the Lund plane coordinate sequences. As such, we will produce a DL discriminant using only these two variables so that we can compare how much the implicit jet substructure information in the images and Lund plane coordinates improves the performance over the information on the absolute scale of these variables.

\FloatBarrier
\section{Deep Learning for jet quenching classification\label{sec:DL}}

Deep Learning provides an array of versatile models capable of performing a wide range of tasks. In addition, their capacity to learn over different data formats, including highly unstructured formats such as images, allows us to train intelligent systems in data that have not been considered before. Indeed, it is the capacity of DL models to abstract the relevant features from unstructured data that is driving many of the novel and cutting-edge DL applications.

\begin{figure}[]
    \centering
    \includegraphics[width=0.95\textwidth]{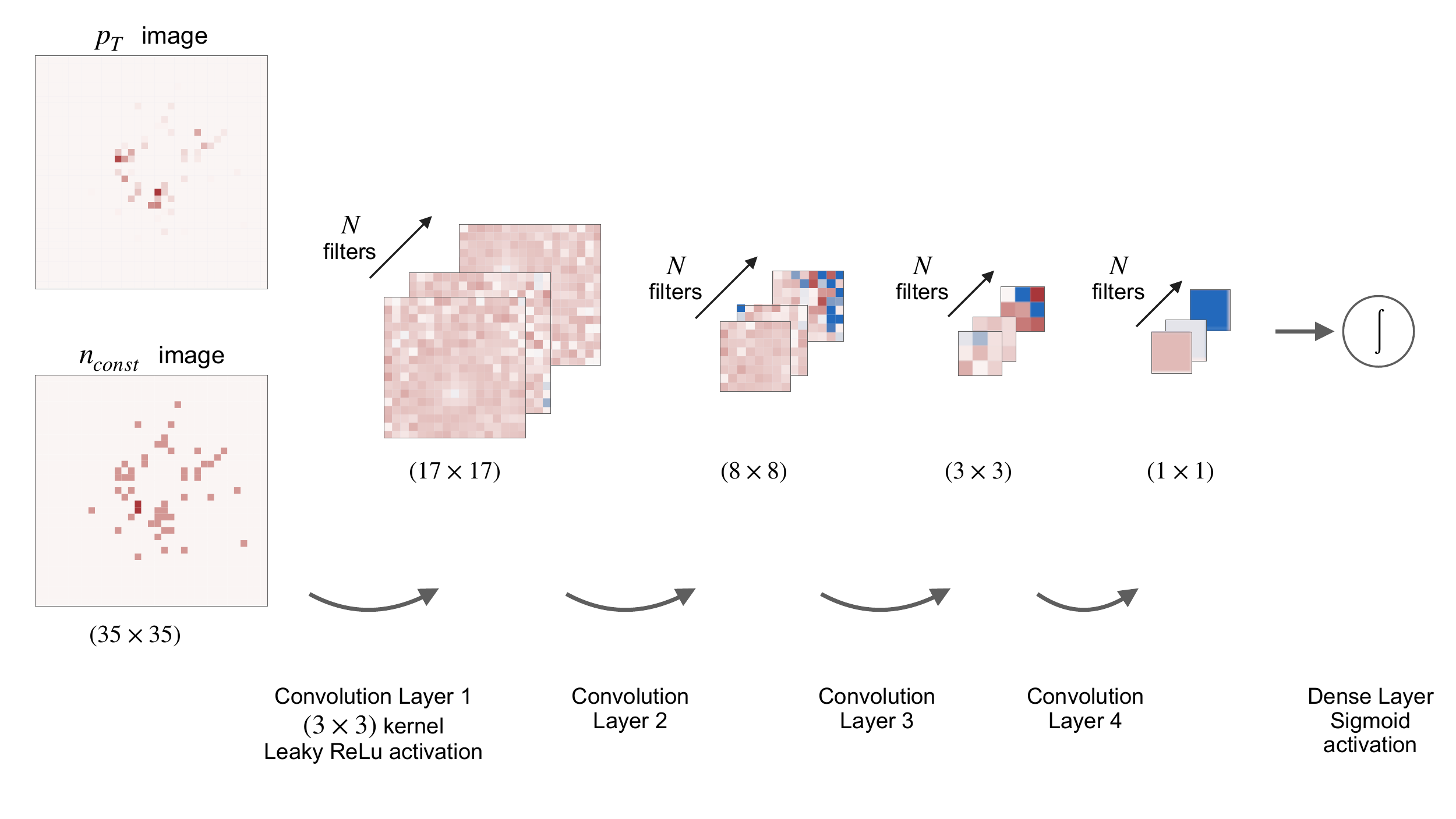}
    \caption{\label{fig:cnn} Diagram of the Convolutional Neural Network used for jet classification from image representations. The input image  corresponds to an example from the Vacuum sample.}
\end{figure}

\begin{figure}[]
    \centering
    \includegraphics[width=0.8\textwidth]{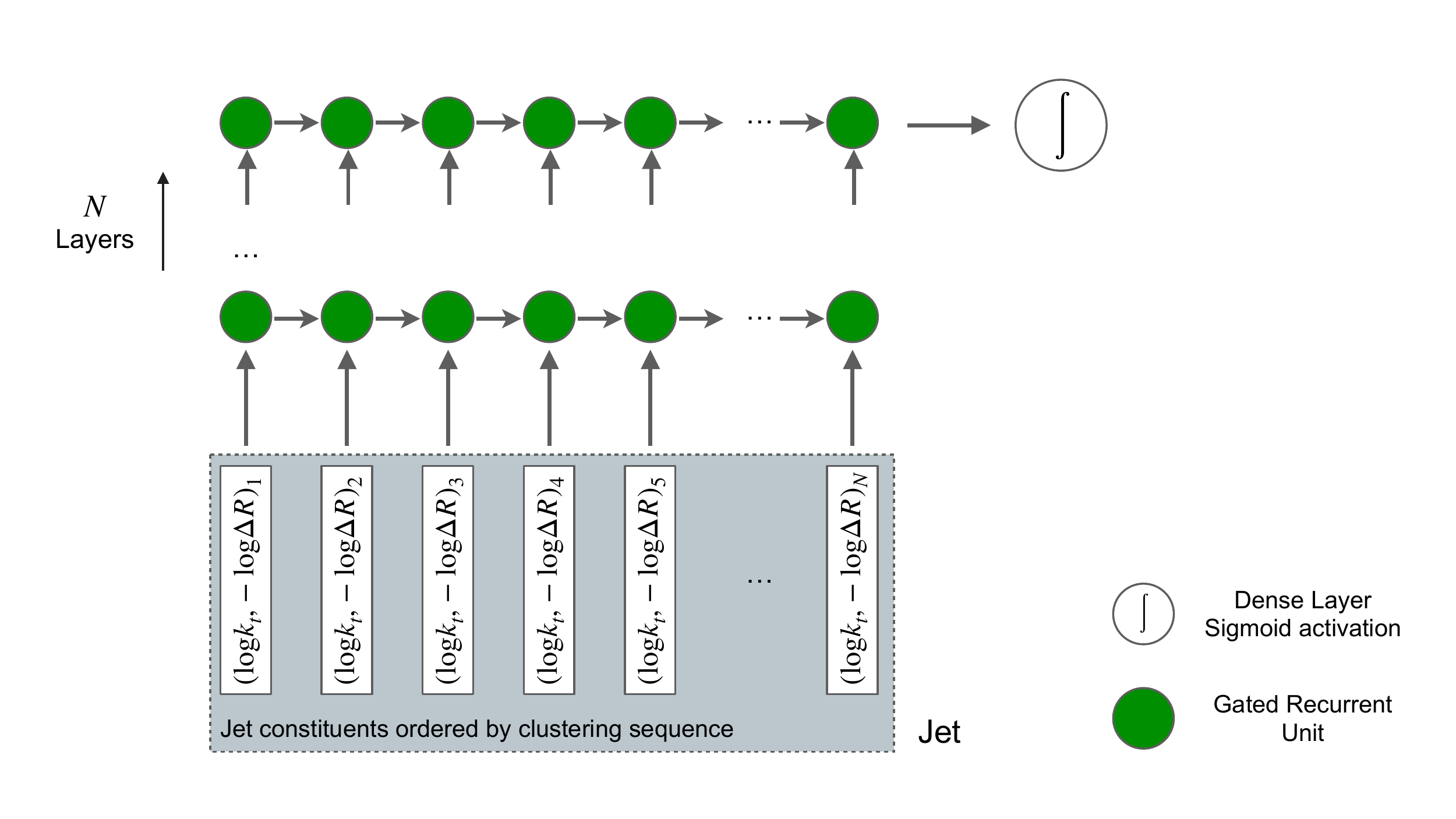}
    \caption{\label{fig:rnn} Diagram of the Recurrent Neural Network used for jet classification from sequences of coordinates of the jet constituents in the Lund plane.}
\end{figure}

In light of this, we developed three different architectures that can take the most out of the data representations that we have discussed above. These architectures were used to develop classifiers with the purpose of discriminating between vacuum and medium-modified jets, with each making use of the different implicit features in the simulated data representations:

\begin{itemize}
    \item Images: Convolutional Neural Network (CNN) for the jet $(\eta, \phi)$ images. In addition, we further considered the case that the image channels were \emph{normalised} or left \emph{unnormalised}. Schematically represented in~\cref{fig:cnn}.
    \item Lund: Recurrent Neural Network (RNN) for the sequence of the C/A re-clustered sequence of the primary Lund plane coordinates. Schematically represented in~\cref{fig:rnn}.
    \item Global: Dense Neural Network (DNN) for the tabular data of the global jet transverse momentum and the number of constituents, $(p_{T,jet}, n_{const})$.
\end{itemize}

For the jet-image representation, we use the CNN, which is the customary architecture for image-type of data, \emph{i.e.} for grid data with highly correlated localised densities (the pixels) that produce larger hierarchical relations (the textures and shapes) that also benefit from composition, which is independent of the absolute coordinates in the grid.

For the Lund plane coordinates, we produced an RNN. RNNs are sensitive to the causal order of sequential steps, for example, those also appearing in natural language or audio. Since the C/A sequence entails a physically motivated ordering in angles, we exploit this structure by using an RNN.

Finally, the Global DNN serves to set the baseline discriminating power present in the variables $(p_{T,jet}, n_{const})$ in order to disentangle the contribution of these variables from the substructure variables that the remaining networks will learn to perform the same task.

All models were developed with \verb|TensorFlow 2.3| \cite{TF}, using its internal \verb|Keras| API~\cite{Keras}. The data samples were randomly split into train, validation and test sets in a 1:1:1 proportion. This guarantees similar statistical representation for model training, selection, and physical discussion of the results. The importance of retaining the same statistical representation at each stage is understood as follows: during model training, the neural network weights are updated through the successive application of gradient descent steps which are calculated on mini-batch averages of gradients, for which having a good statistical representation is crucial to avoid biases towards kinematic regions with greater Monte Carlo statistical errors in the training set; during model selection, which is discussed in the next section, through hyper-parameter optimisation, we compare performance metrics of trained models to find the best one and, to prevent selecting a biased model, we require the validation set to have a good statistical description of the data; finally, in the application phase where we perform the analysis using the trained models, we want to have as good statistics as possible such that conclusions are statistically sound. Since we want to maximise the statistics of each of the three sets, the best solution is to split them equally as the methodology as a whole will only be as robust as the weakest link. Furthermore, at each stage, the Monte Carlo weights were used to enforce the statistical description of the kinematic distributions.

\subsection{Hyperparameter optimisation}

A crucial step in any DL application is the optimisation of the so-called hyperparameters of the model, which are the non-trainable parameters that specify the details of the architecture and its training. For each of the four cases, we performed a hyper-parameter optimisation loop using \verb|optuna|~\cite{akiba2019optuna} to tune the details of the architectures. The search space for each architecture is shown in~\cref{tab:hyperparmoptuna}. We allocated a budget of  50 trials or 12 hours, whatever came first, per \verb|optuna| loop and we evaluated the performance of each hyper-parameter combination using the validation set.

In addition, \verb|EarlyStopping| (with patience of 10 epochs) and \verb|ModelCheckpoint| callbacks were used during training to keep the network weights of the best model across all trials. To focus on the most promising combinations, we also employed the \verb|MedianPruner| pruner (with 10 warm-up epochs and evaluated every 5 epochs henceforth), which interrupts a training that does not perform better than the insofar median of the validation loss. The hyper-parameters were sampled using the built-in Tree Parzen Estimator \cite{bergstra2011algorithms}, with the \verb|multivariate| flag set to true.

\begin{table}[]
\centering
\begin{tabular}{l|l|l}
Model Type                    & Hyperparameter       & Range                               \\ \hline
\multirow{8}{*}{CNN (Images)} & Number of Filters    & $[8,128]$ in steps of $8$           \\
                              & Spatial Dropout Rate & $[0.0,0.5]$ in steps of $0.1$       \\
                              & Number of Layers     & Fixed at $4$                        \\
                              & Kernel Size          & Fixed at $3$                        \\
                              & Stride               & Fixed at $2$                        \\
                              & Padding              & Fixed at VALID                      \\
                              & Activation Function  & Fixed LeakyReLU                     \\
                              & Batch Normalisation  & After inputs and before activations \\ \hline
\multirow{3}{*}{RNN (Lund)}   & Number of Layers     & $[1,5]$                             \\
                              & Number of Units      & $[4,64]$                            \\
                              & Recurrent Unit       & Fixed GRU                           \\ \hline
\multirow{5}{*}{DNN (Global)} & Number of Layers     & $[1,10]$                            \\
                              & Number of Units      & $[4,128]$ in steps of $4$           \\
                              & Dropout Rate         & $[0.0,0.5]$ in steps of $0.1$       \\
                              & Activation Function  & Fixed LeakyReLU                     \\
                              & Batch Normalisation  & After inputs and before activations
\end{tabular}
\caption{\label{tab:hyperparmoptuna} Hyperparameter search spaces for the different Deep Learning architecture types.}
\end{table}

Furthermore, in the same loop, we also optimised the details of the learning rate scheduler used during training. For all the cases the \verb|Adam| optimiser \cite{kingma2014adam} was chosen, with an \verb|ExponentialCyclicalLearningRate| schedule as implemented by \verb|TensorFlowAddons 0.11| with \verb|initial_learning_rate=1e-5|, \verb|maximal_learning_rate=1e-2|, \verb|scale_mode="cycle"|, \verb|step_size| set to half the total number of batches, and \verb|gamma| to be optimised by the \verb|optuna| loop in the range $[0.925, 0.975]$ in steps of $0.005$. For all cases, the batch size was set to $1024$, as well as a maximum of $100$ epochs. Both the hyper-parameter bounds in~\cref{tab:hyperparmoptuna} and the training details discussed above were defined after an initial round of manual trials to determine reasonable configurations within the hardware and time constraints. The final best hyperparameters are shown in~\cref{tab:hyperparmfinal}, where we observe that no optimal configuration is set at the boundaries defined in~\cref{tab:hyperparmoptuna}, which reinforces the initial choice of the hyperparameter space. 

In~\cref{tab:hyperparmoptuna} some hyperparameters can be seen as explicitly fixed, whereas others are implicitly fixed through the default values of the relevant classes implemented in the \verb|Keras| API. The explicitly fixed values are justified in virtue of the DL architecture as follows.

For the CNN architecture, the values of the Kernel Size, Stride, and Padding fix the maximum depth to $4$ for $35\times35$ images without the use of pooling layers after the convolutions, making this architecture purely convolutional. Consequently, the network does not lose information from the data through pooling operations and can still progressively reduce the representation size (c.f.~\cref{fig:cnn}) to the point where the output of the last convolution is already of $N$ filters of dimension $1$. In turn, this means that the head of the network is a linear classifier over patterns learned by the filters, which will allow us to better interpret what the network learned to perform the classification, as further discussed in~\cref{app:CNNinterpretation}. Finally, the usage of \verb|LeakyReLU| and Batch Normalisation are standard recommended practices for Deep Neural Networks.

For the RNN architecture, we fixed the recurrent unit to be the Gated Recurrent Unit (GRU). In early experiments, we did not observe any variation in performance between GRU and the Long Short-Term Memory (LSTM) unit, for which we fixed the choice on GRU before the \verb|optuna| loop as this unit has fewer parameters than the LSTM. In addition, we did not optimise the inner hyper-parameters of the GRU since only a few combinations allow for \verb|Keras| optimised CUDA implementation that significantly increases the training speed. No inter-layer Batch Normalisation or Dropout was used as it is common in RNN since these are meant to be applied on the outputs of layers, whereas in an RNN the learning process step is performed \emph{step}-wise across the sequence jointly across all layers.

Finally, for the DNN architecture, the implementation of \verb|LeakyReLU| and Batch Normalisation was fixed to simplify the hyperparameter optimisation loop and to allow for deep, \emph{i.e.} many layers, configurations.

\begin{table}[]
\centering
\begin{tabular}{ll|l|l}
\hline
\multicolumn{2}{l|}{Model Type}                               & Hyperparameter       & Value   \\ \hline
\multirow{6}{*}{CNN (Images)} & \multirow{3}{*}{Normalised}   & Number of Filters    & $104$   \\
                              &                               & Spatial Dropout Rate & $0.3$   \\
                              &                               & Gamma                & $0.925$ \\ \cline{2-4} 
                              & \multirow{3}{*}{Unnormalised} & Number of Filters    & $88$    \\
                              &                               & Spatial Dropout Rate & $0.0$   \\
                              &                               & Gamma                & $0.970$ \\ \hline
\multicolumn{2}{l|}{\multirow{3}{*}{RNN (Lund)}}              & Number of Layers     & $2$     \\
\multicolumn{2}{l|}{}                                         & Number of Units      & $15$    \\
\multicolumn{2}{l|}{}                                         & Gamma                & $0.935$ \\ \hline
\multicolumn{2}{l|}{\multirow{4}{*}{DNN (Global)}}            & Number of Layers     & $6$     \\
\multicolumn{2}{l|}{}                                         & Number of Units      & $116$   \\
\multicolumn{2}{l|}{}                                         & Dropout Rate         & $0.1$   \\
\multicolumn{2}{l|}{}                                         & Gamma                & $0.93$ 
\end{tabular}
\caption{\label{tab:hyperparmfinal} Best hyper-parameter configurations for each Deep Learning architecture type.}
\end{table}

\subsection{Performance of the Deep Learning Architectures}

\begin{figure}[h]
    \centering
    \includegraphics[width=0.40\textwidth]{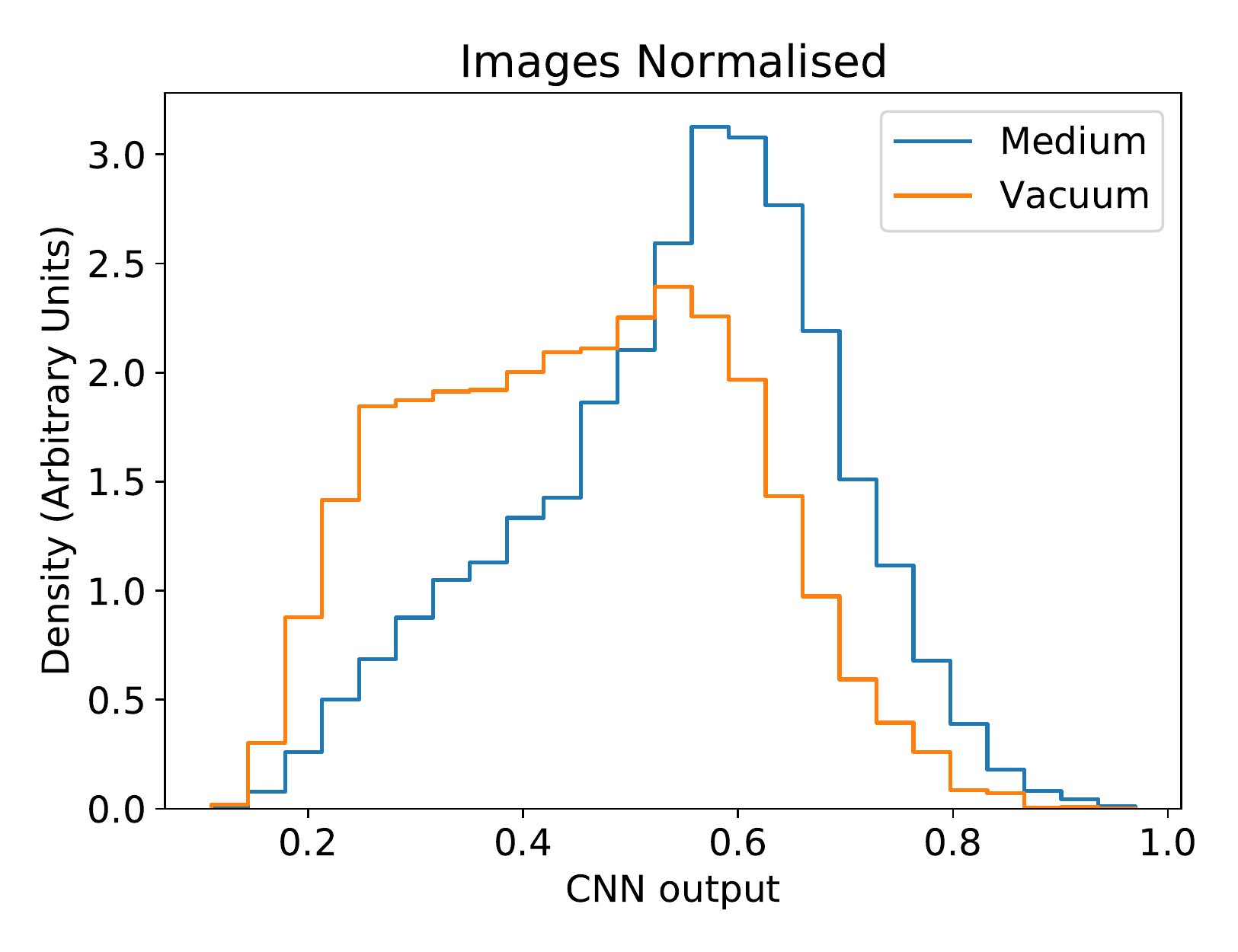}
    \includegraphics[width=0.40\textwidth]{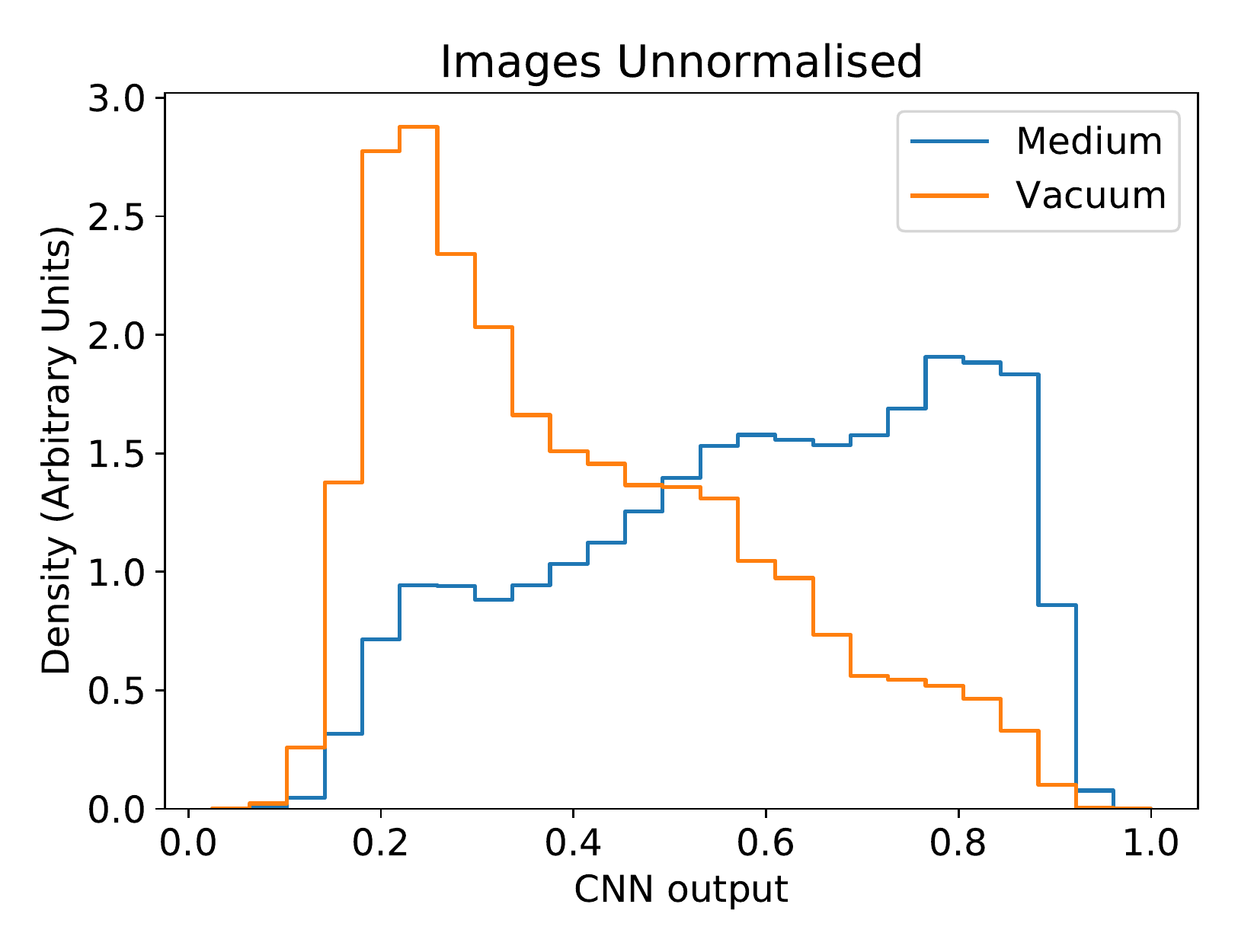}
    \includegraphics[width=0.40\textwidth]{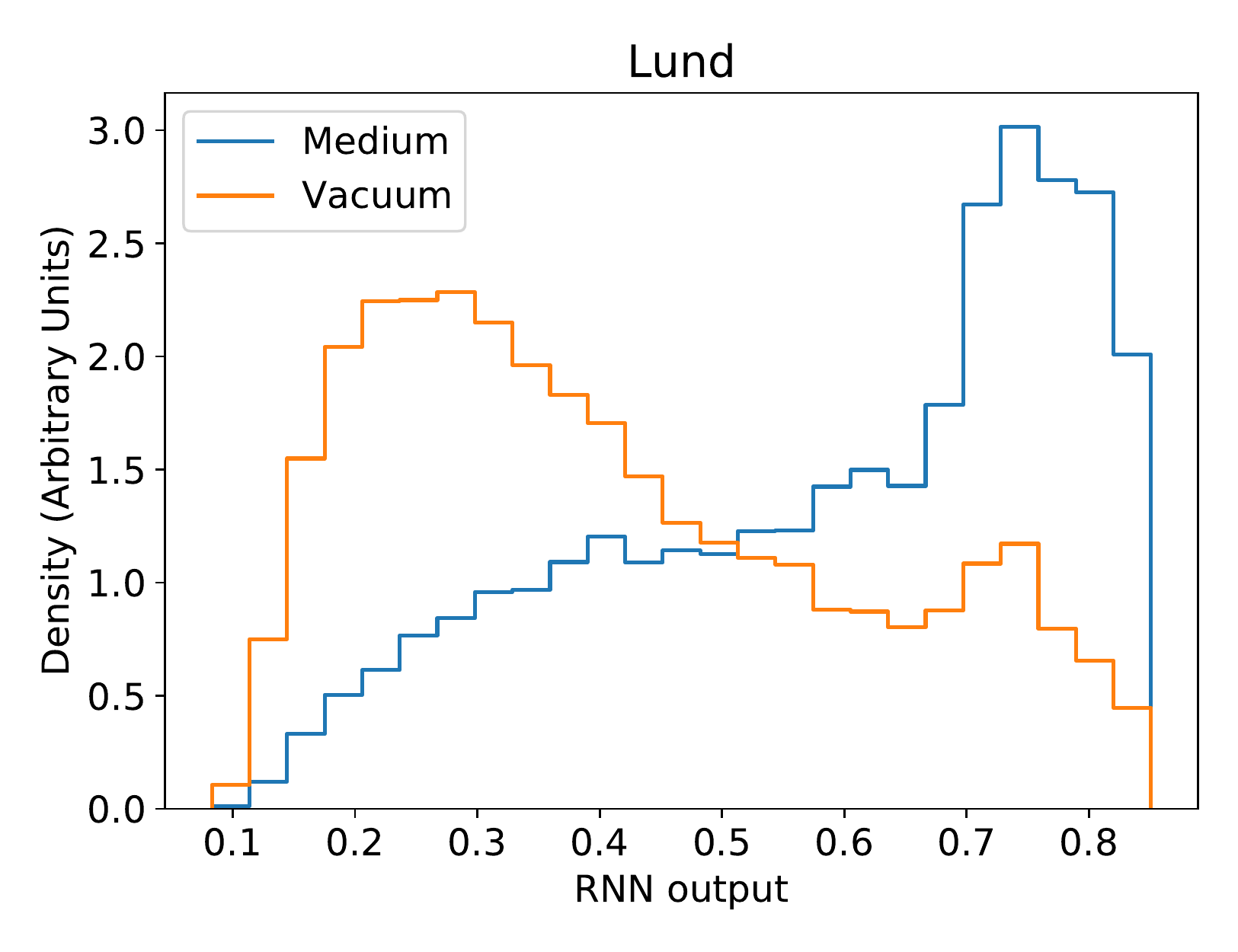} \includegraphics[width=0.40\textwidth]{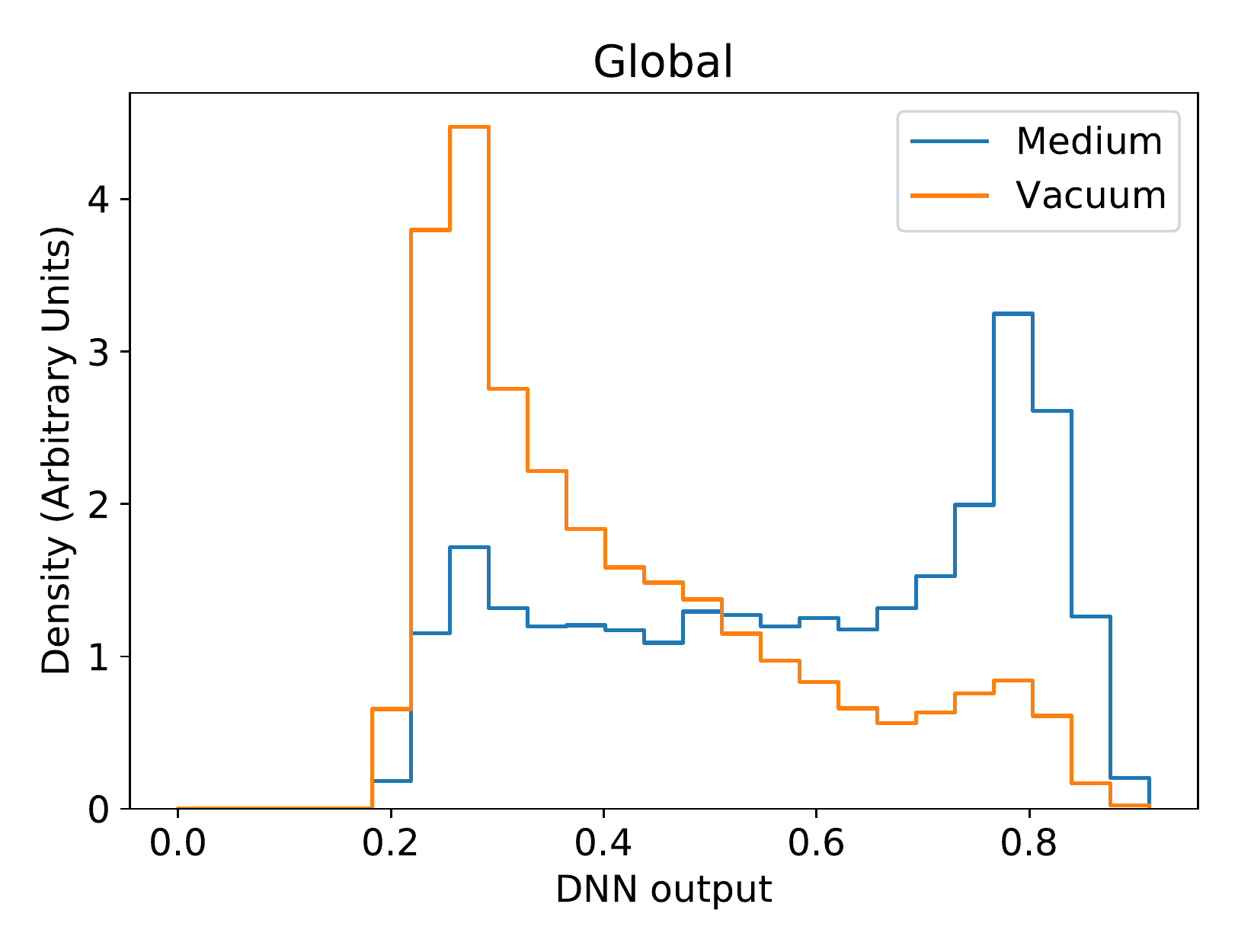}
    \caption{\label{fig:nnoutputs} Distribution of the different Deep Learning outputs for the Vacuum and Medium samples.}
\end{figure}

The outputs of the DL networks are shown in \cref{fig:nnoutputs} for the validation data set. During network training, the Vacuum sample is identified with a true target value of 0 and the Medium sample with 1. Thus, the distribution of the predicted labels should be closer to 1 for jets obtained from the Medium sample and closer to 0 for jets obtained from the Vacuum simulation. This is observed for all DL architectures. 

The final goal of these classifiers is to identify jets that experienced strong jet quenching effects. However, the Medium sample does not yield a pure sample of medium-modified jets, containing also a collection of reconstructed jets that, probabilistically, did not experience strong energy loss modifications (events for which $x_{jZ}\sim1$). Nevertheless, while learning to distinguish between the Vacuum and Medium samples, part of the network will learn the effects of jet quenching on each data representation type. At the same time, this fact limits the capacity of the models to discern between the pure \textit{vacuum-like} jets (proton-proton collisions) and \textit{medium-like} jets (whose fragmentation pattern was modified by the presence of in-medium scatterings and in-medium radiation). 

\begin{figure}[h]
    \centering
    \includegraphics[width=0.40\textwidth]{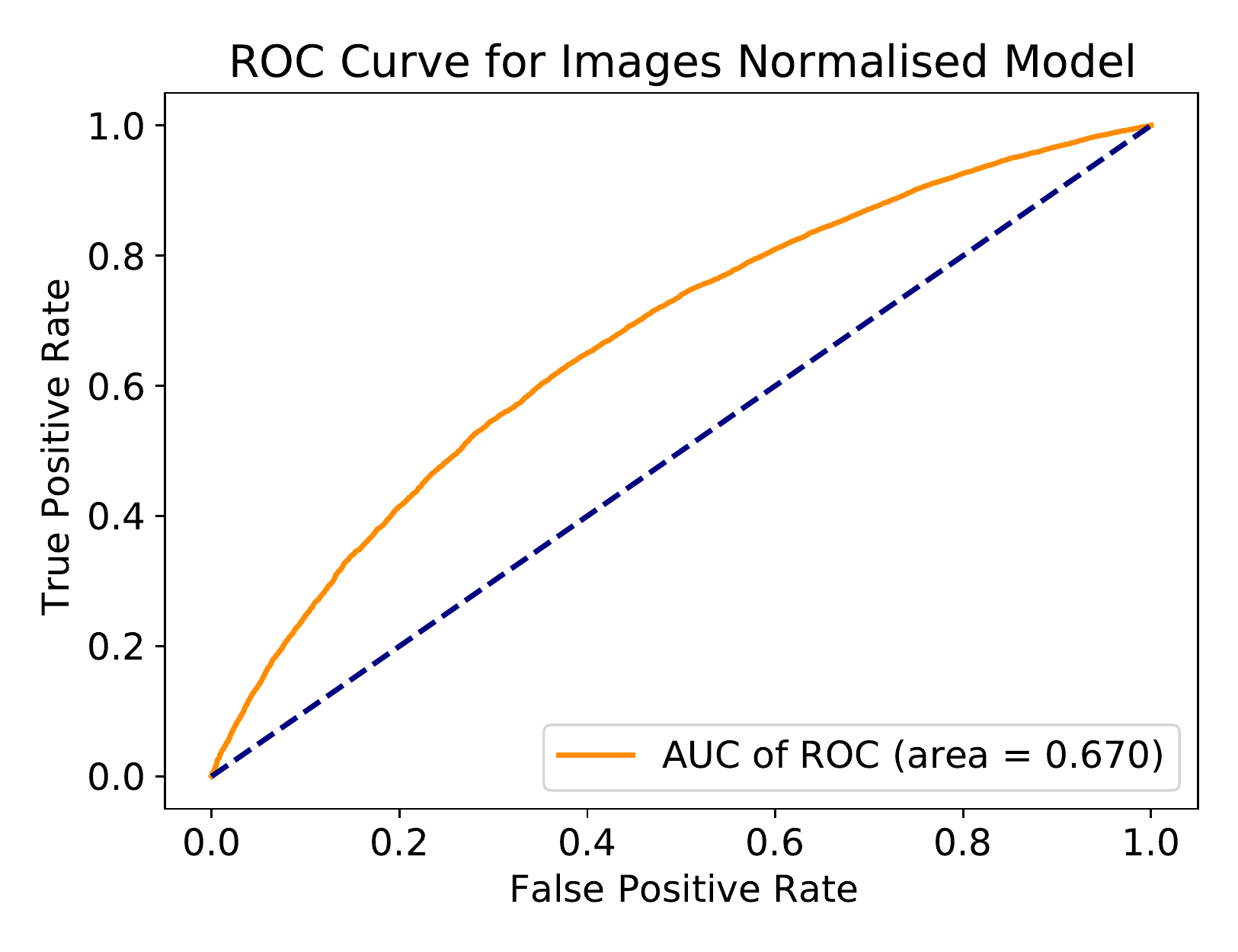} 
    \includegraphics[width=0.40\textwidth]{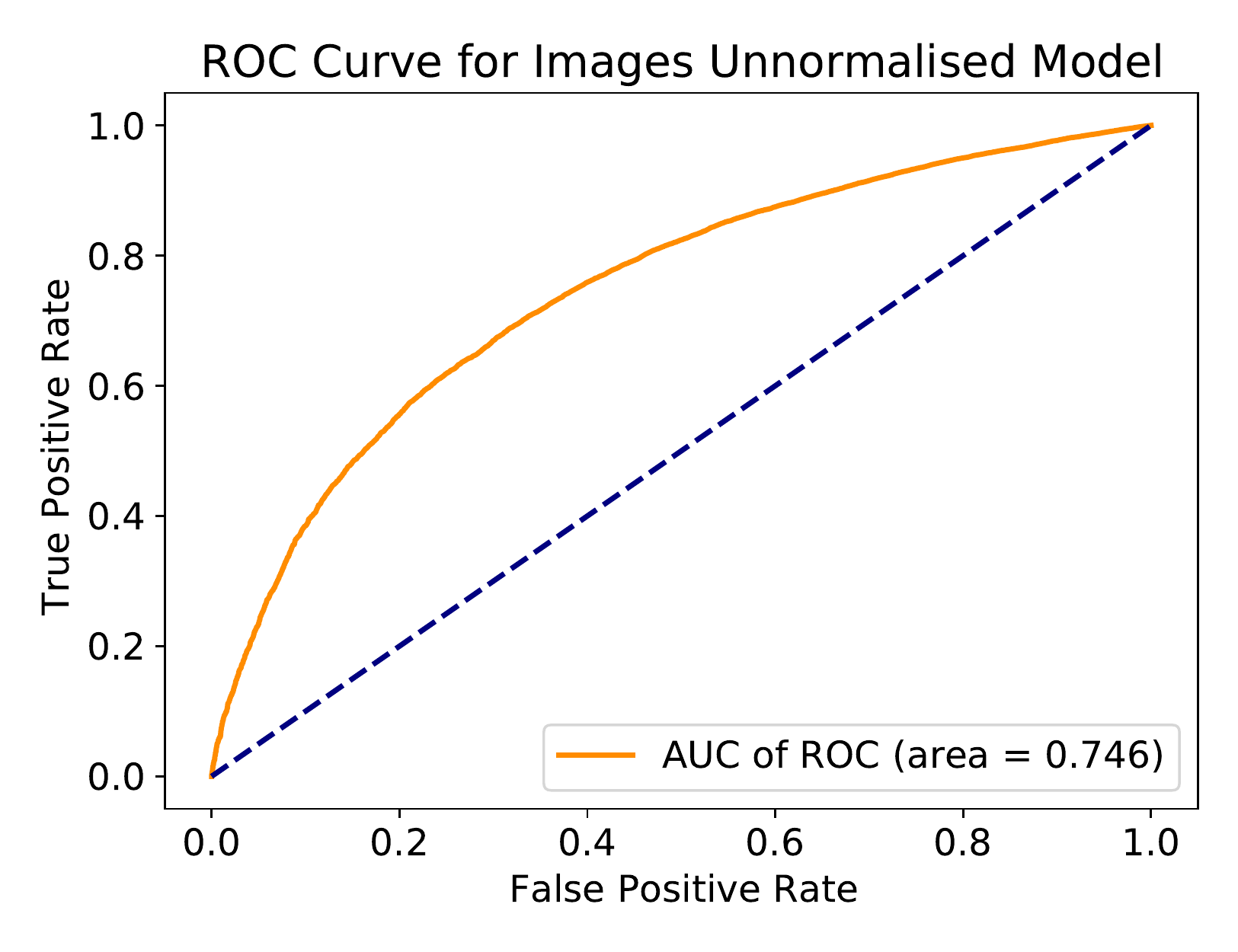}
    \includegraphics[width=0.40\textwidth]{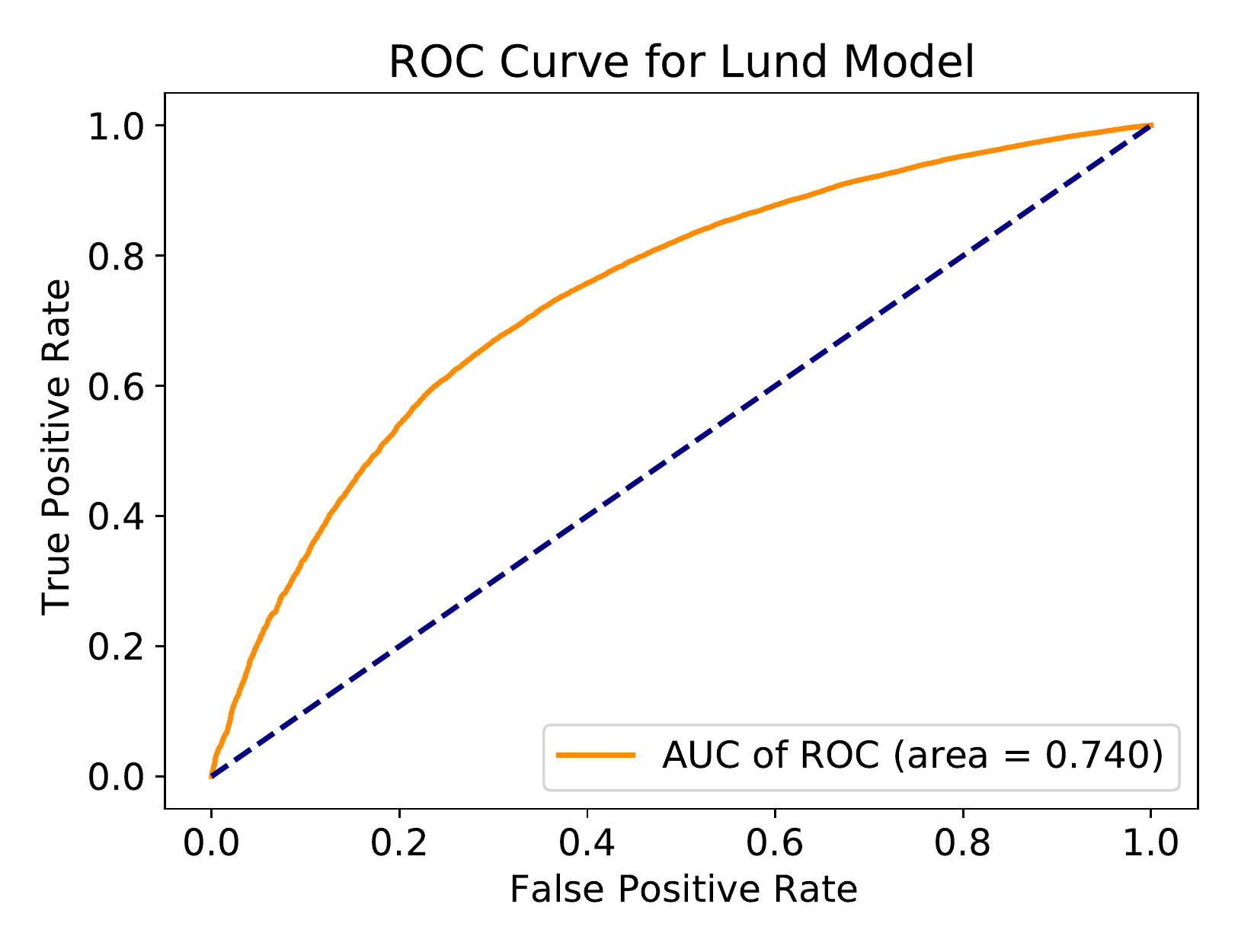} \includegraphics[width=0.40\textwidth]{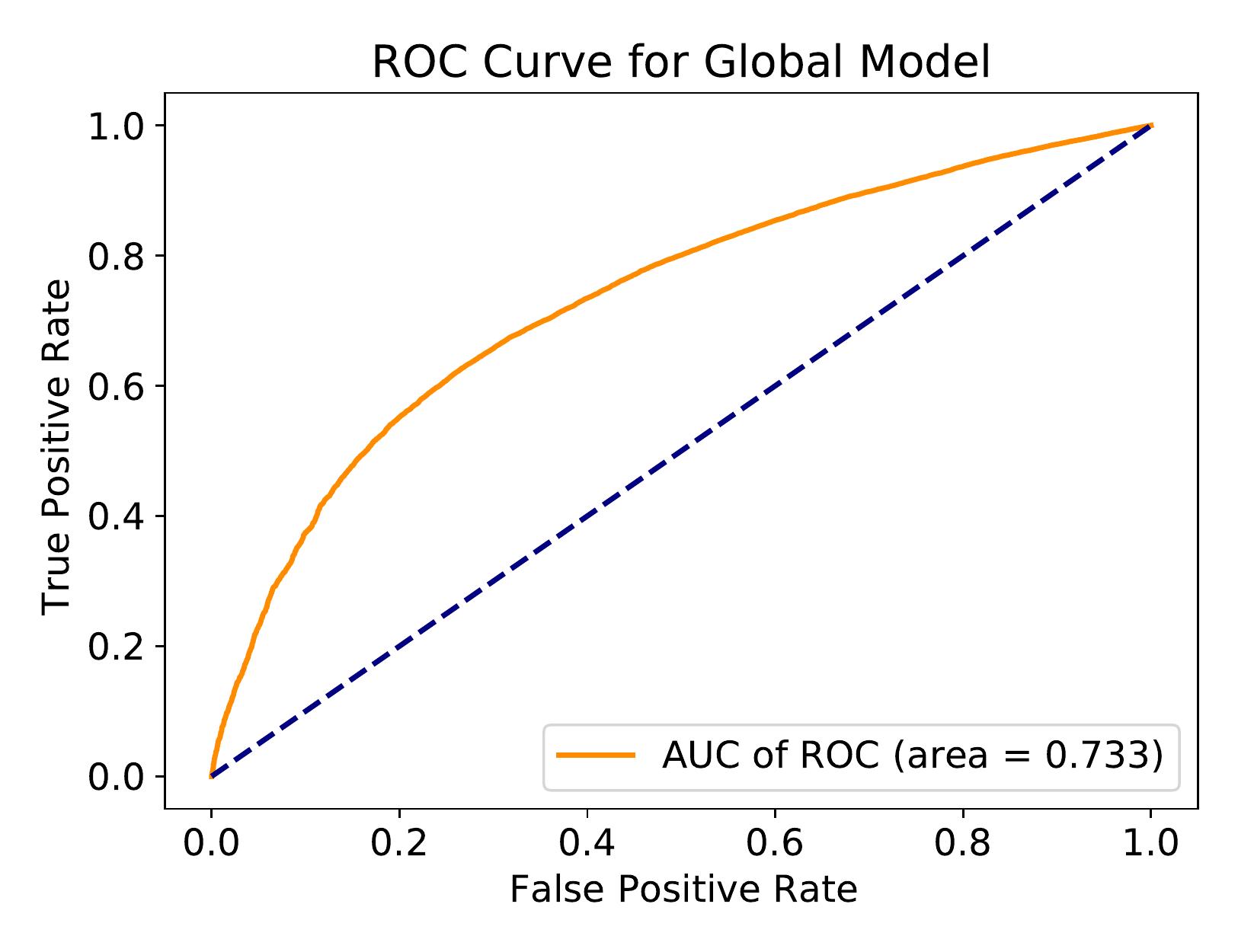}
    \caption{\label{fig:rocs} ROC curve for the separation of the Vacuum and Medium samples using the different Deep Neural Network models.}
\end{figure}
\begin{table}[h]
\centering
\begin{tabular}{l | c | c}
\hline
Model                        &  $p_{T,jet}>$30~GeV & $p_{T,jet}>$125~GeV \\
\hline 
Normalised jet images CNN    & 0.67 & 0.65 \\
Unnormalised jet images CNN  & 0.75 & 0.68 \\
Lund sequences RNN           & 0.74 & 0.69 \\
Global DNN                   & 0.73 & 0.64 \\
\end{tabular}
\caption{\label{tab:roc_auc} Area under the ROC curve of the different Deep Learning architectures for the separation of the Vacuum and Medium samples in the pre-defined case ($p_{T,jet}>30$~GeV) and in the large jet transverse momentum regime ($p_{T,jet}>$125~GeV).}
\end{table}

The outputs provided by the RNN, DNN and CNN trained on unormalised images show the best separation between the Medium and Vacuum samples generated by JEWEL+PYTHIA. The effect shows up on the corresponding Receiver Operating Characteristic (ROC) curves represented in \cref{fig:rocs}, where the area under the ROC curve (AUC) is also reported. The CNN for normalised images has the poorer AUC, 0.67, while the remaining models achieve an AUC around 0.74. This is an indication that the jet absolute $p_T$ and number of constituents play an important role on distinguishing between the Vacuum and Medium samples. In Section~\ref{sec:Results}, we further investigate the outputs provided by the DL architectures to understand if the two classes of jets identified by the networks are compatible with the desired \textit{medium-} $versus$ \textit{vacuum-like} jets separation.

Moreover, in \cref{tab:roc_auc}, we also present the AUCs obtained for the different DL models over the same samples after performing a $p_T>125$~GeV cut. The reason to do this is that by increasing the minimum $p_{T,jet}$, while keeping the same cut on $p_{T,Z}$, we are discarding most of the events with $p_{T,Z} <$125~GeV on both samples (the few vacuum events that will pass this cut will be the ones with a large ISR contamination; in the presence of a medium, those will fall below the cut). Most of the selected events will then have a $Z$-boson with a $p_{T,Z}$ that is near the momentum threshold for the jet. As such, while jet quenching effects will still be present, the magnitude of those will be highly reduced by definition, since those should come from the high end of the $p_T$ distribution. We observe that the AUCs obtained with the DNN, RNN and CNN with unnormalised images decrease around 10\% for jets with $p_T>$125~GeV, where the $p_T$ spectra are identical between the medium and vacuum categories. Contrarily, the performance of CNNs trained on normalised images are only slightly affected by the jet $p_T$.

\section{Results and interpretation of the Deep Learning architectures}
\label{sec:Results}

\begin{figure}[p]
    \centering 
    \includegraphics[width=0.35\textwidth]{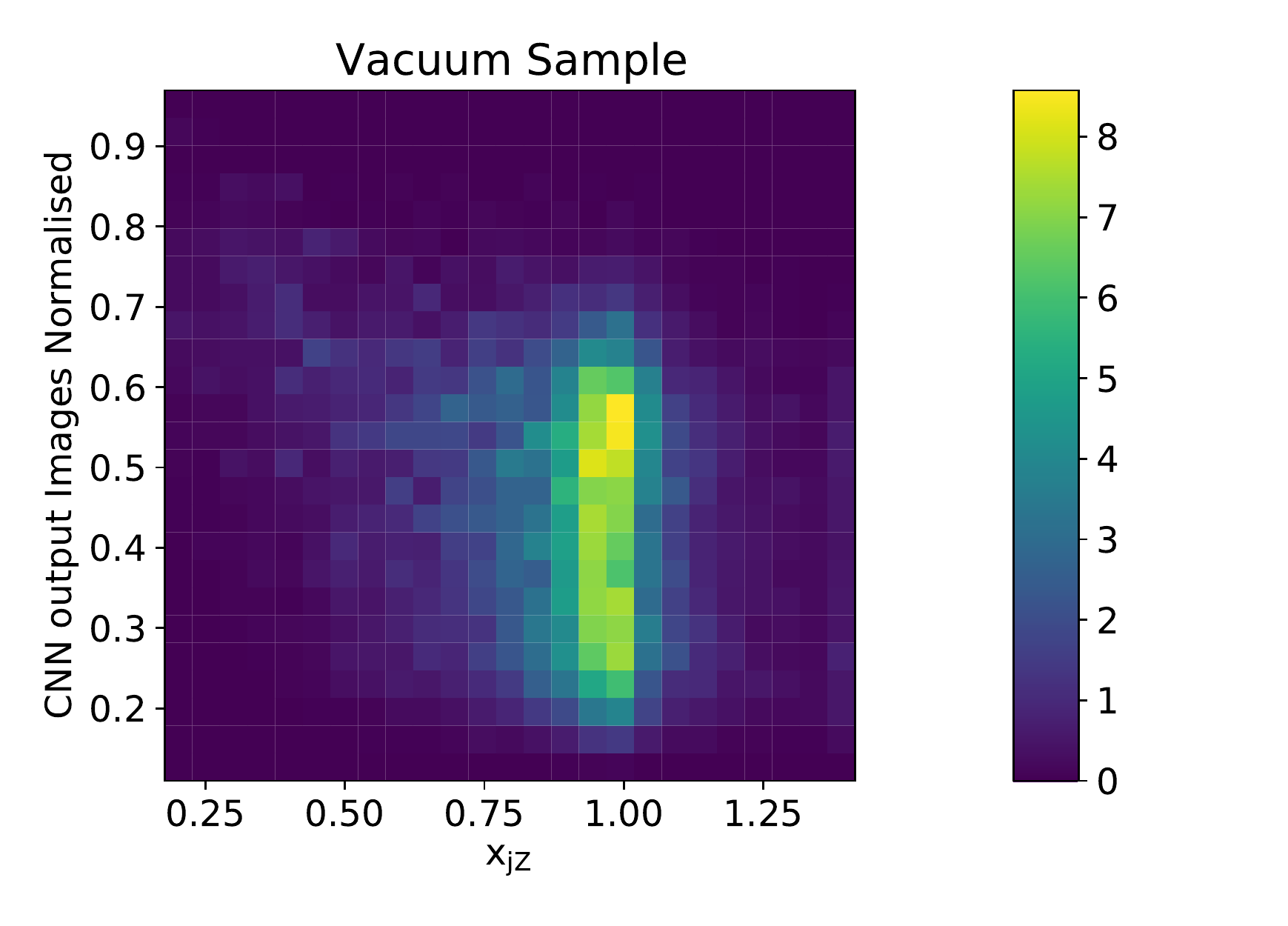}
    \includegraphics[width=0.35\textwidth]{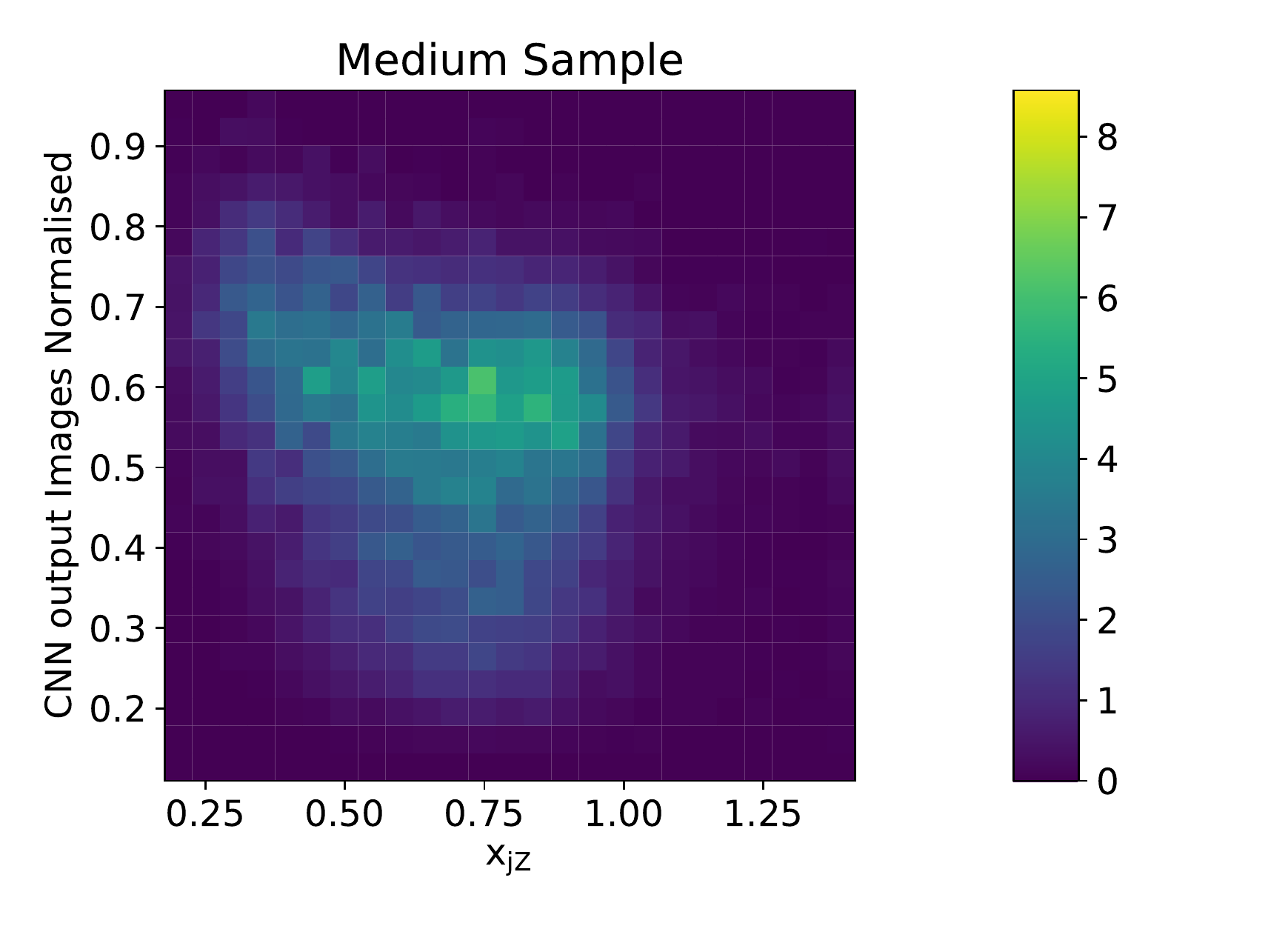}
    \includegraphics[width=0.35\textwidth]{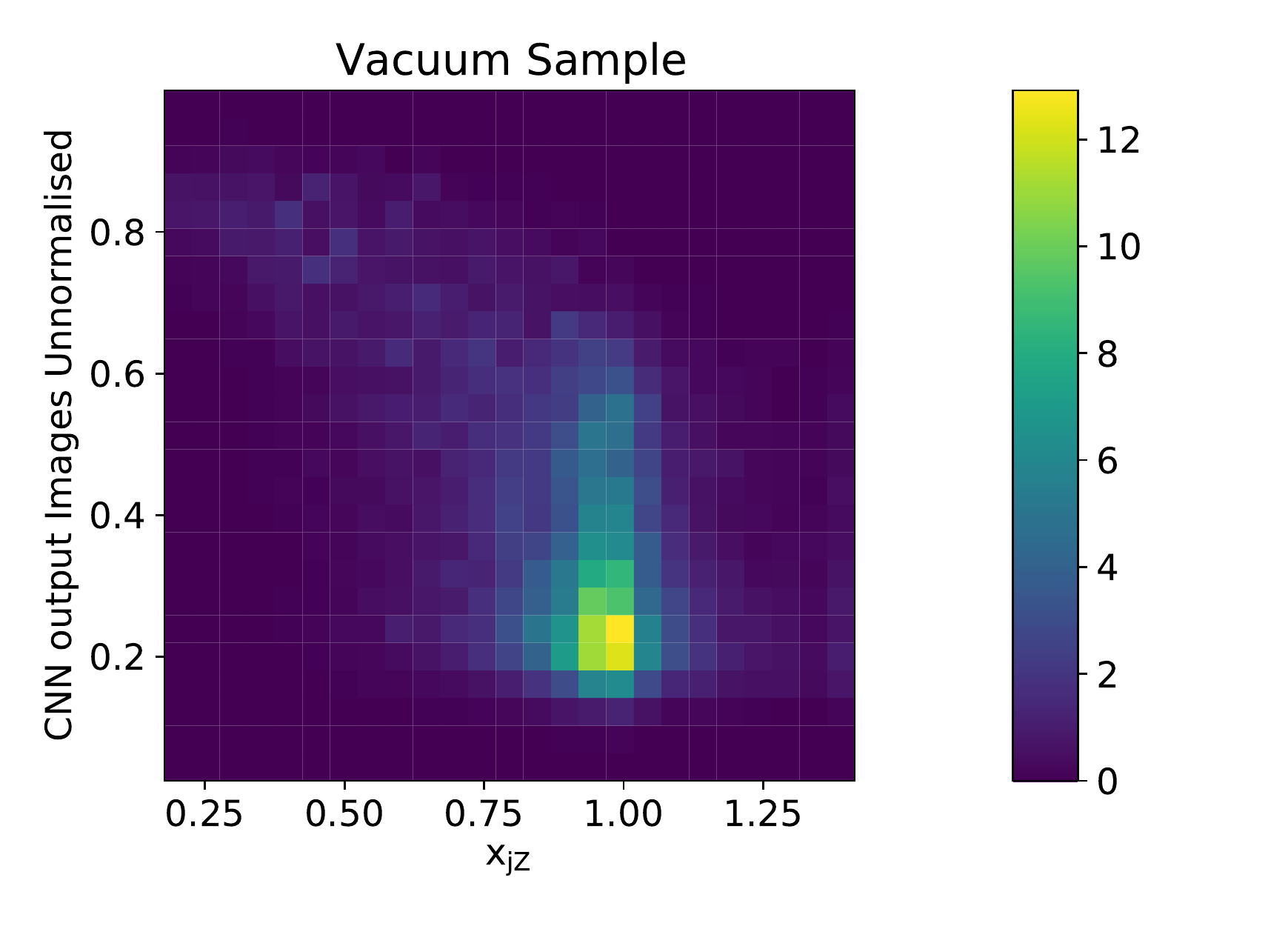} 
    \includegraphics[width=0.35\textwidth]{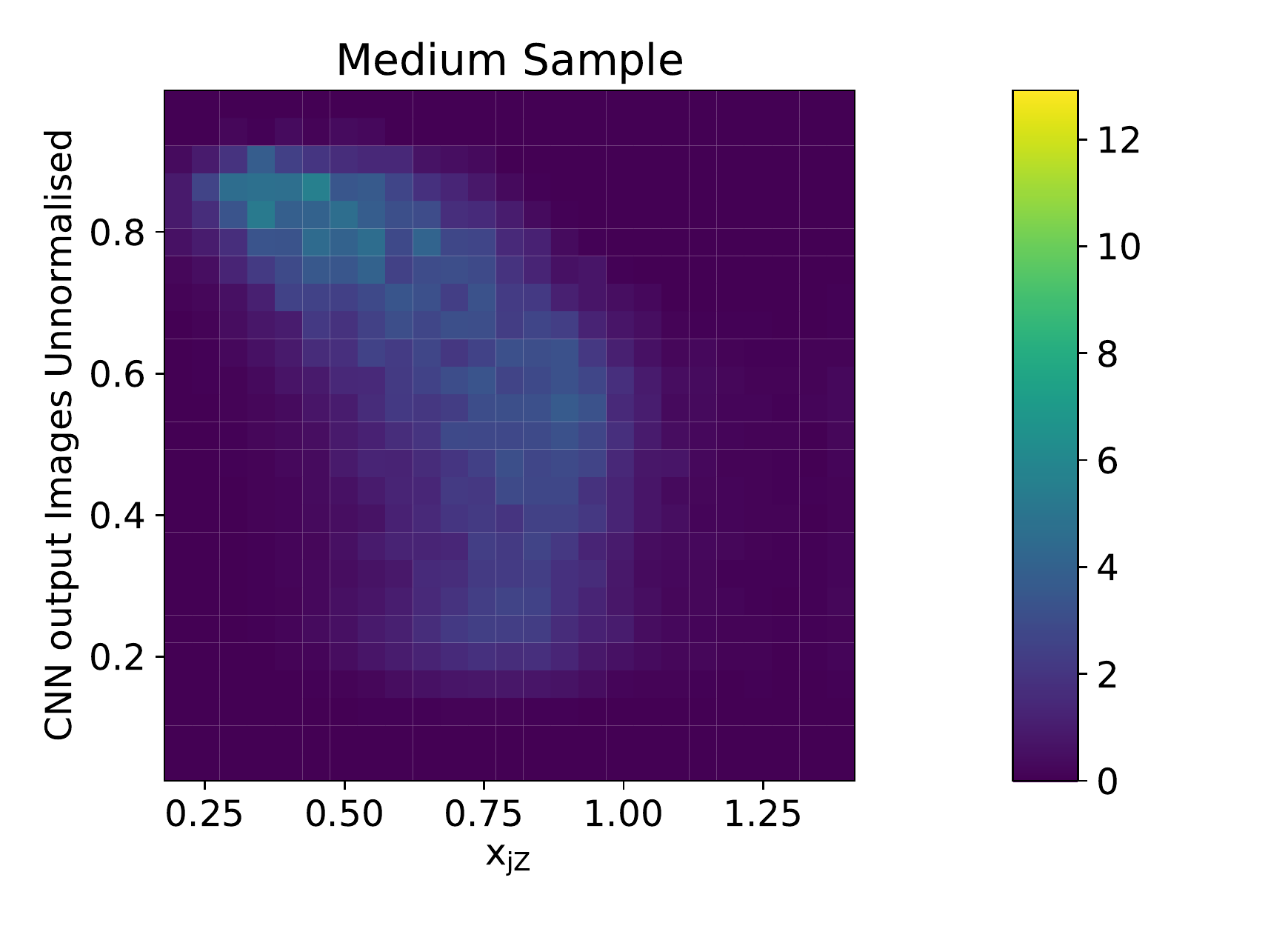}
    \includegraphics[width=0.35\textwidth]{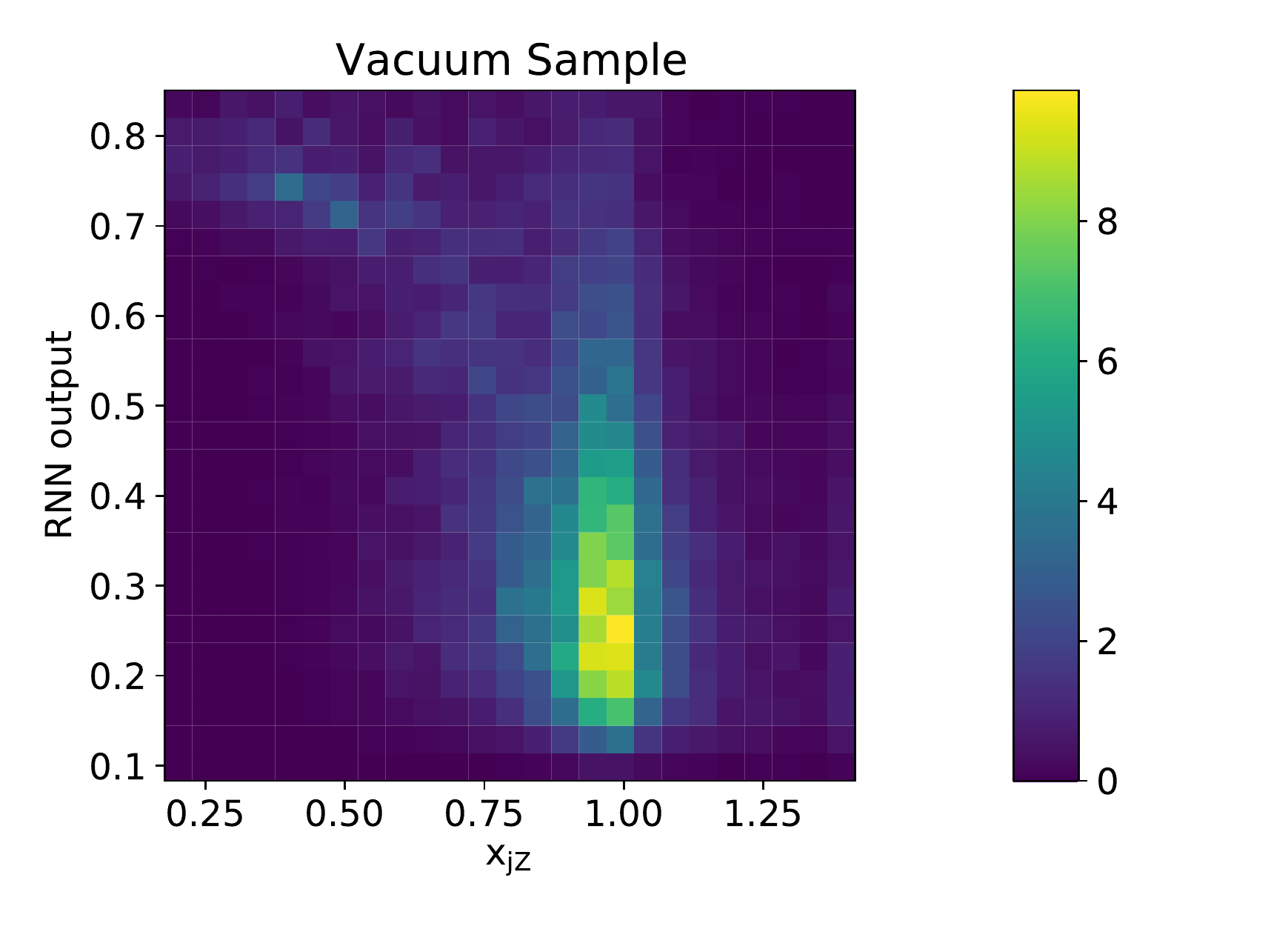} \includegraphics[width=0.35\textwidth]{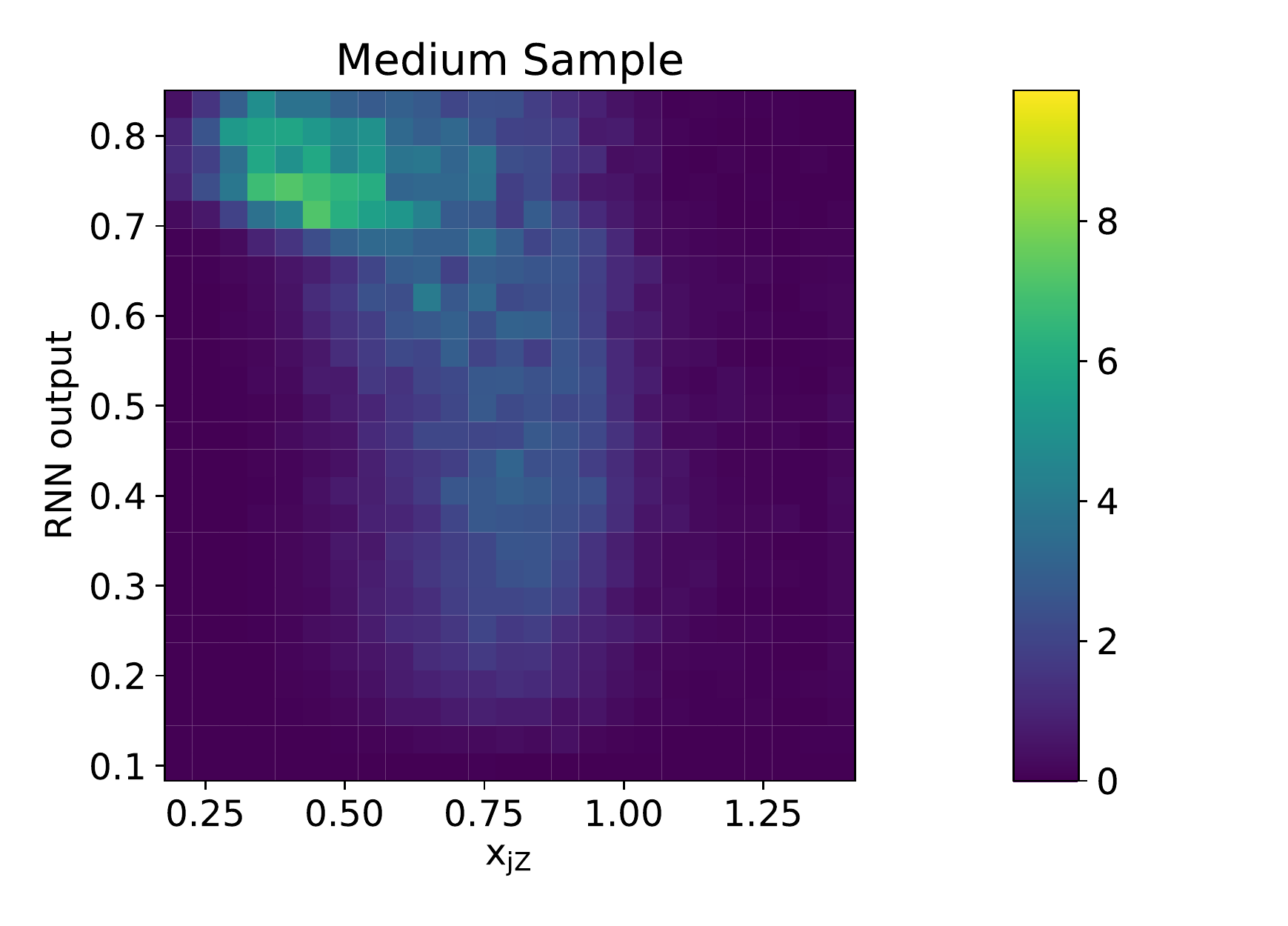}
   \includegraphics[width=0.35\textwidth]{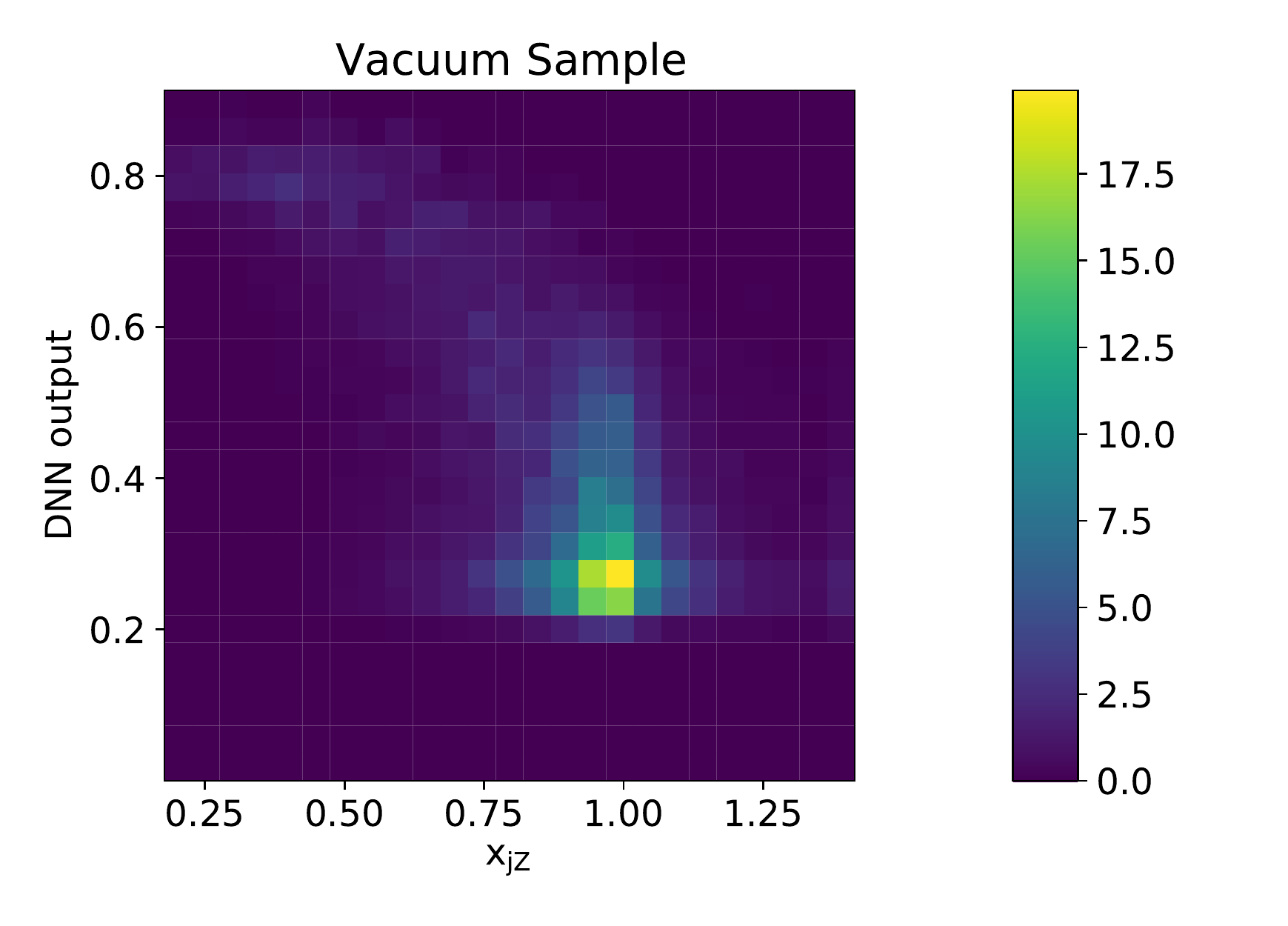} \includegraphics[width=0.35\textwidth]{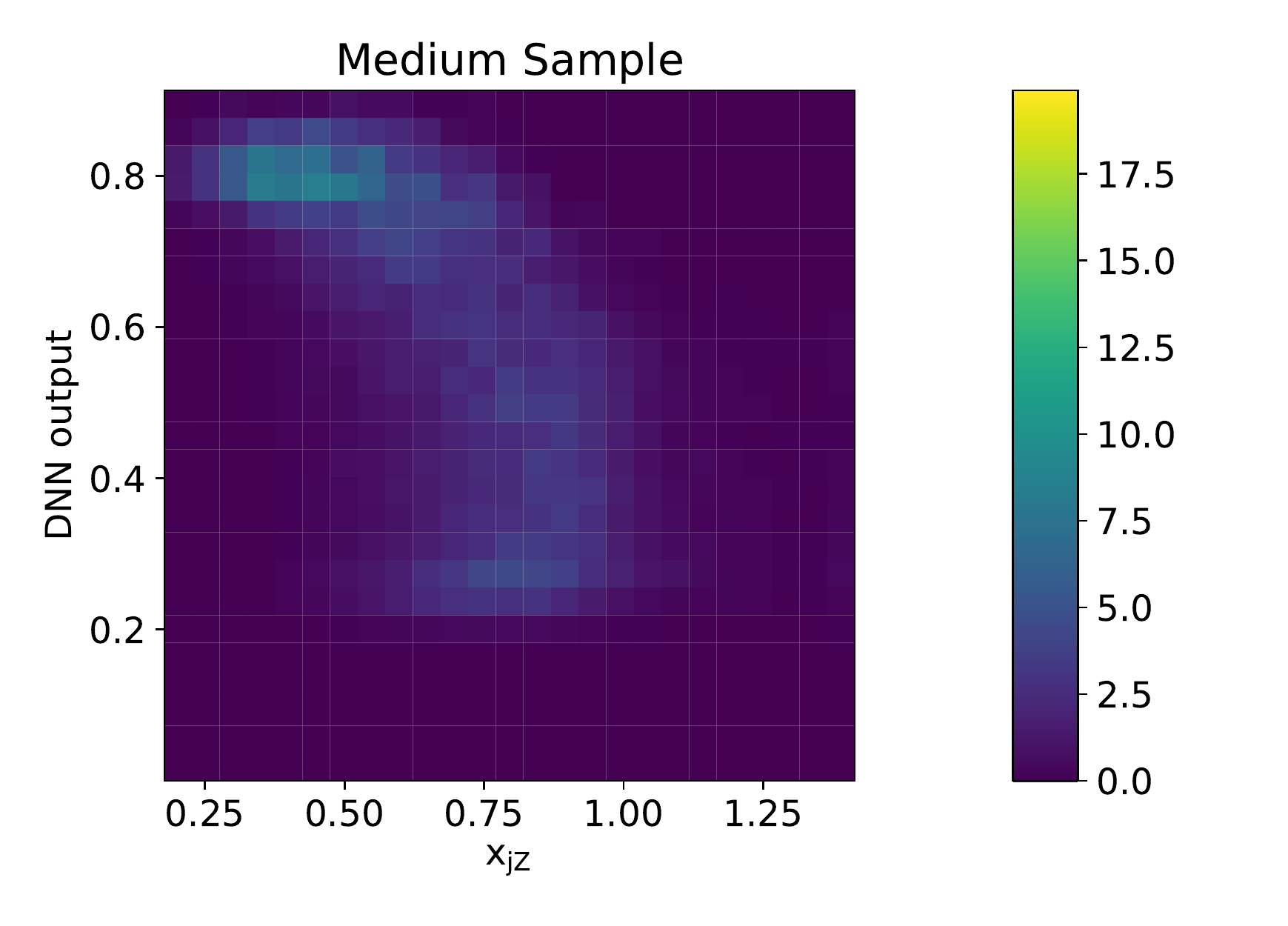}
    \caption{\label{fig:nnoutputsxjcorr} Distribution of the Deep Learning network output as a function of $x_{jZ}$ for the (left) Vacuum sample and the (right) Medium sample.}
\end{figure}

In order to investigate how the DL networks separate between jets reconstructed from the Vacuum and Medium sample, we plot the predicted DL outputs versus $x_{jZ}$ in ~\cref{fig:nnoutputsxjcorr}. Simultaneously, since $x_{jZ}$ is a good proxy for the quenching phenomenon at the jet level, this allows evaluating the potential of the networks for a jet quenching tagging application. The outputs of the different DL architectures are nearly uncorrelated with $x_{jZ}$ for vacuum (see ~\cref{app:DNNcorrelation}), which is a desired property for the tagger since events for which $x_{jZ}$ differs from 1 in the vacuum result from spurious effects, independent of jet quenching through interaction with the QGP. On the other hand, the DNN, RNN and CNN from unnormalised images have larger predictions for smaller values of $x_{jZ}$, \emph{i.e.} when the jet modification by the medium is also larger on average. Therefore, these networks are predicting better the labels of jets which are quenched and misidentifying as vacuum jets with lower $x_{jZ}$, effectively behaving as a jet quenching classifier. Using normalised images, the CNN seems only slightly correlated with $x_{jZ}$, which means that in principle the decision boundary of the model is not the most adequate for tagging quenched jets. Furthermore, in~\cref{app:DNNcorrelation}, we inspect the correlations between the DL discriminants.

\begin{figure}[t]
    \centering 
    \includegraphics[width=0.45\textwidth]{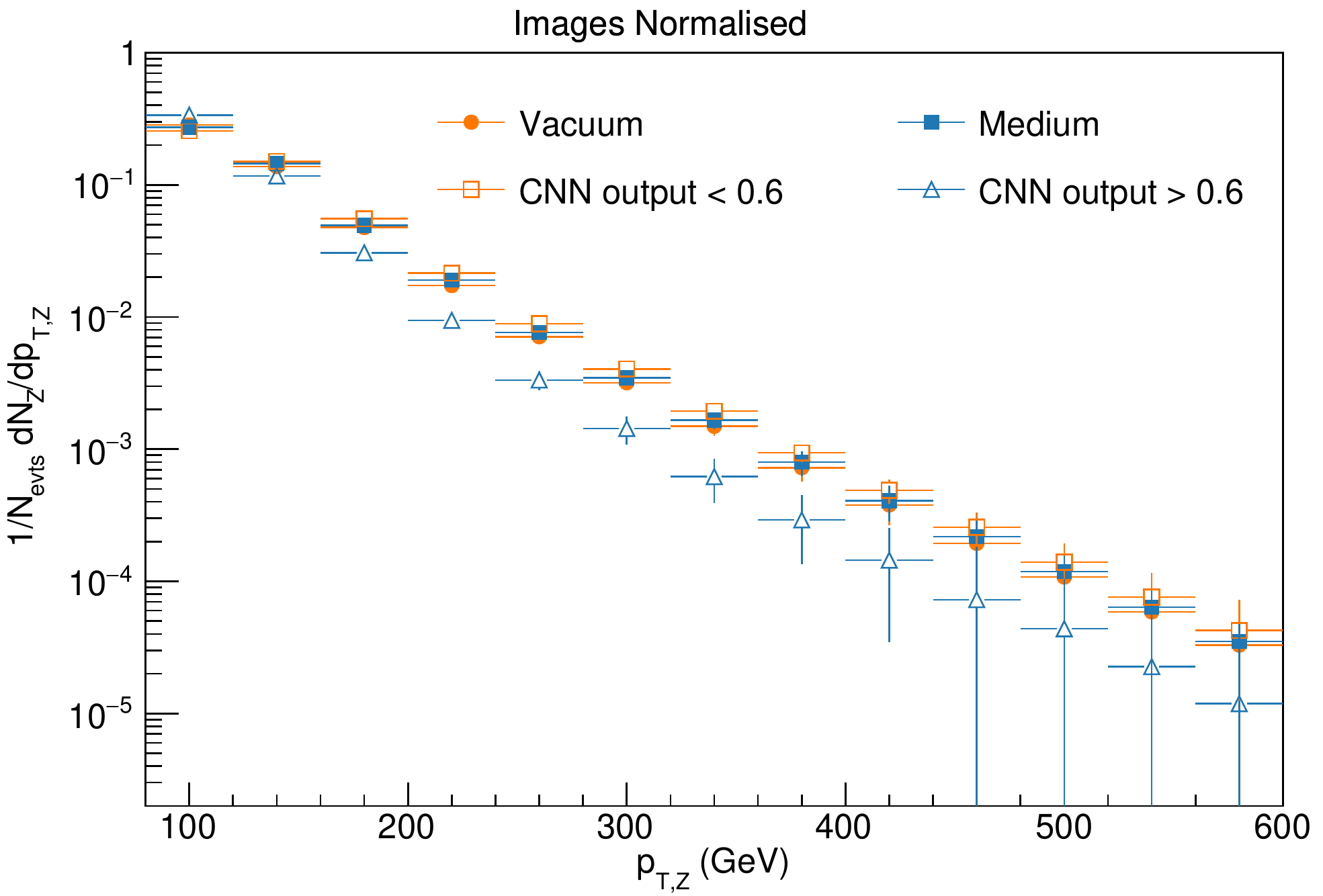}
    \includegraphics[width=0.45\textwidth]{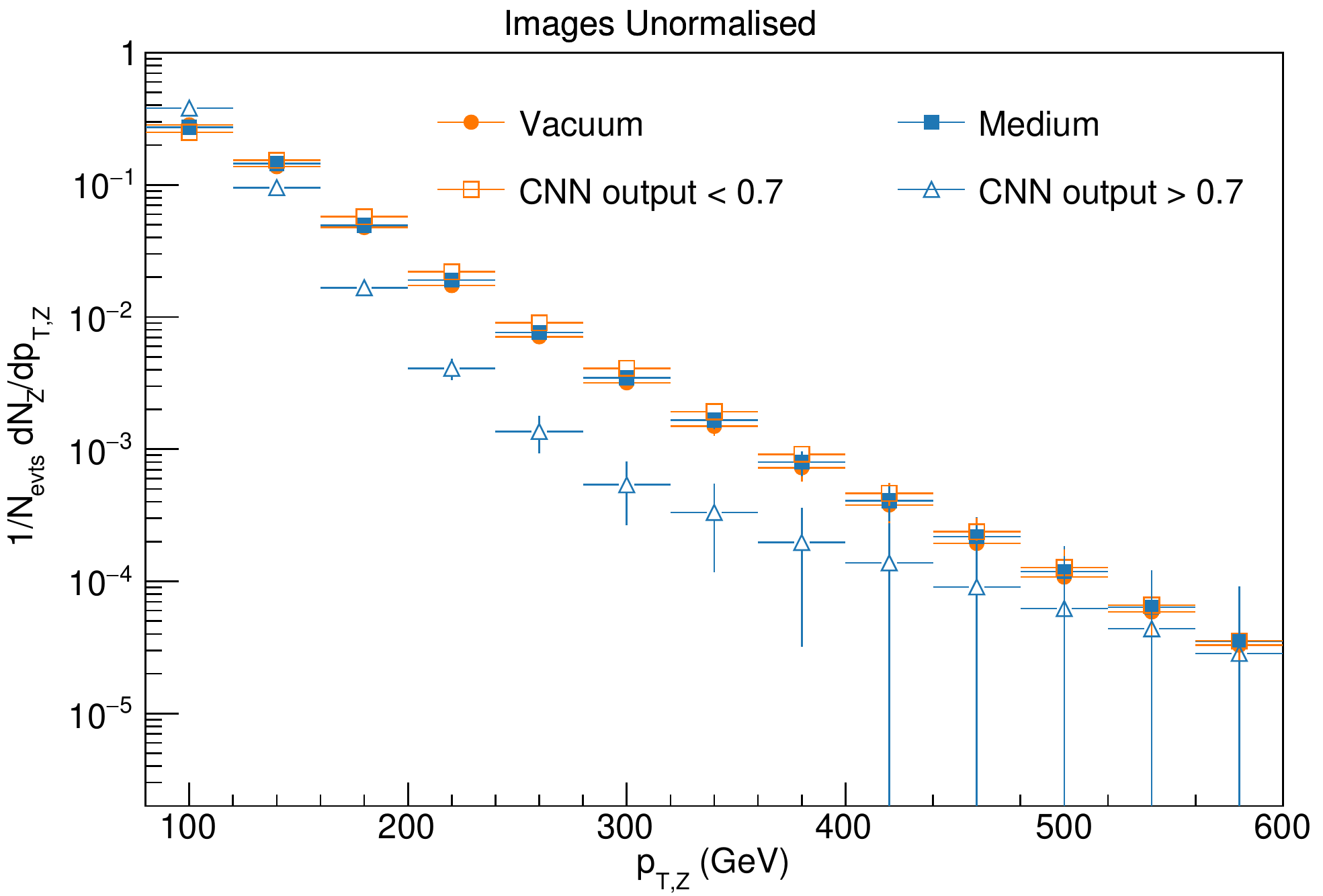}
    \includegraphics[width=0.45\textwidth]{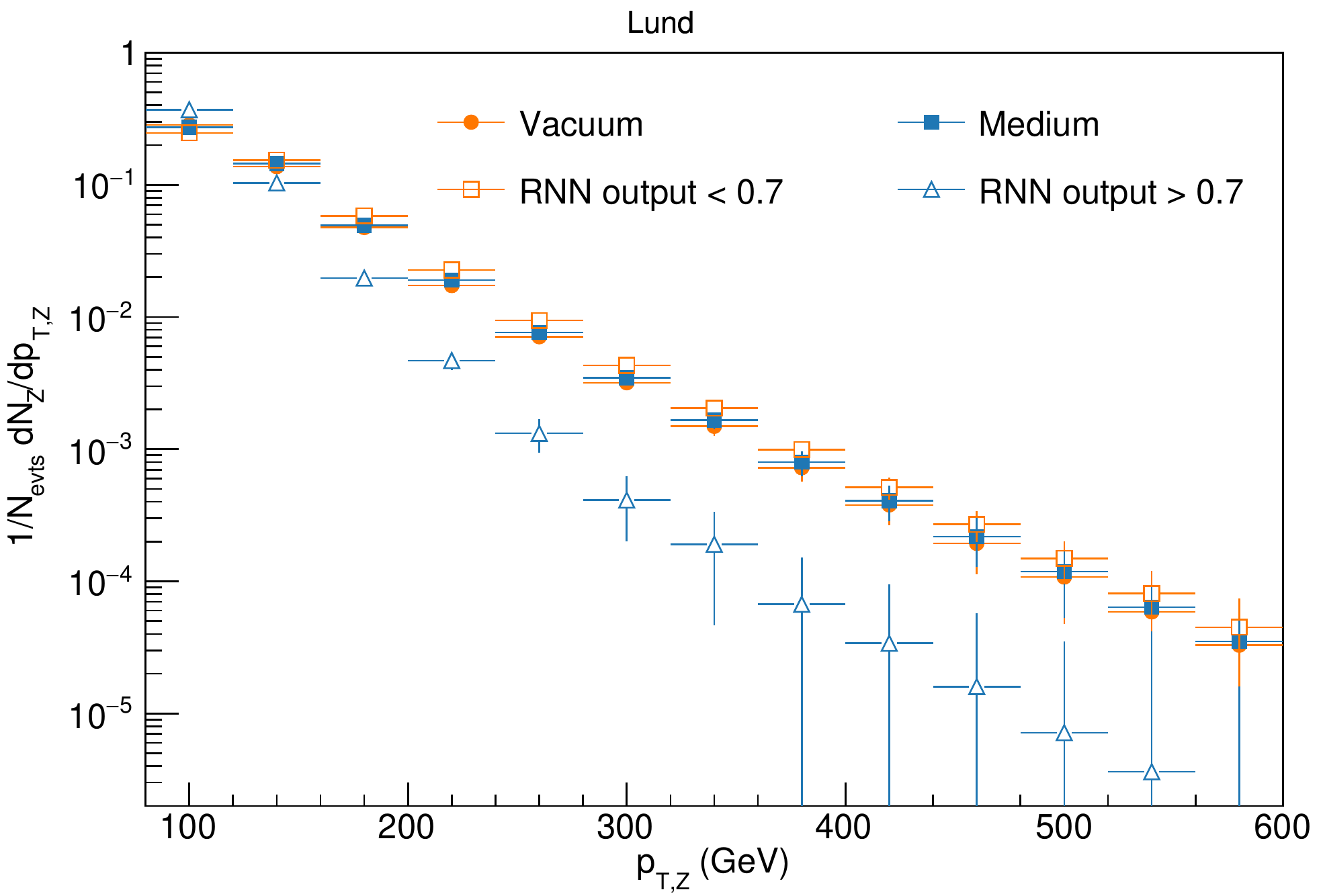} 
    \includegraphics[width=0.45\textwidth]{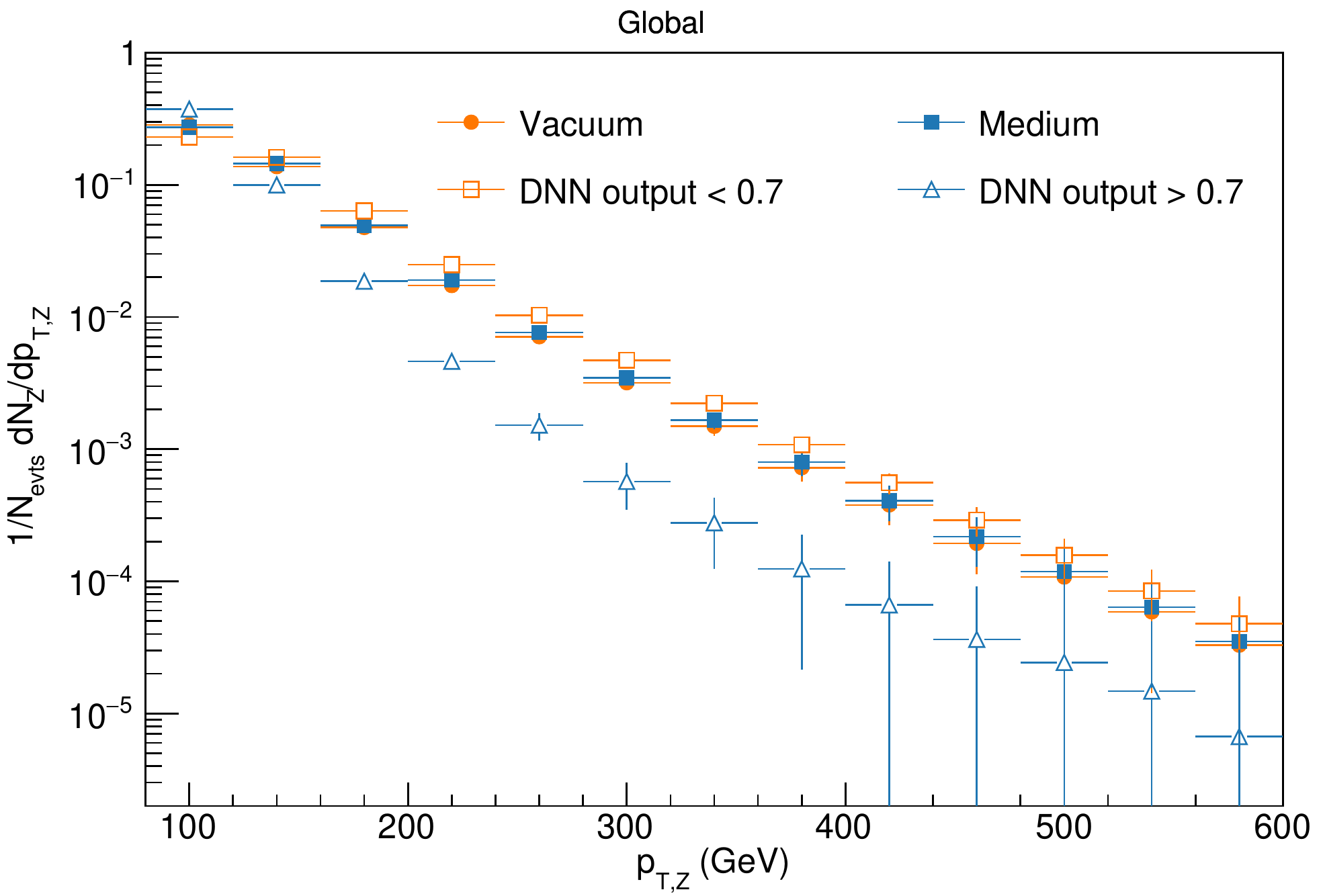}
    \caption{\label{fig:ptZ} Transverse momentum spectra of the reconstructed $Z$-boson $p_{T,Z}$, for the different Deep Learning architectures. Monte Carlo truth from JEWEL+PYTHIA is provided in solid symbols for the Vacuum and Medium samples and a subset of events selected by the DL discriminant appears in open symbols. The DL output selection employed to identify \textit{vacuum-like} jets (open blue) and \textit{medium-like} jets (open orange) is made explicit in the legend of each plot.}
\end{figure}

To test the results of the different architectures, we created two samples of \textit{medium-like} and \textit{vacuum-like} jets as identified by the output of each DL network. On both samples generated by JEWEL+PYTHIA (Vacuum and Medium), we classified jets as quenched (if the DL discriminant was above a given reference value) or vacuum (if the result was below). This reference value was not optimised and it was chosen for illustration purposes only. Taking the results of~\cref{fig:nnoutputs}, we set this reference cut to $0.7$ except for the CNN trained on normalised images, which was set to $0.6$. A comparison of the resulting $Z$-boson spectra contrasting the Monte Carlo truth is shown in~\cref{fig:ptZ}. We kept the solid lines representing the Vacuum (orange) and Medium (blue) simulations withdrawn from JEWEL+PYTHIA, while the open symbols reflect the selection identified by each network as being Vacuum (orange) and Medium (blue). In all DL architectures, it is possible to fully recover the vacuum expectations, a good indication of the ability of DL to identify \textit{vacuum-like} jets. We note that a better agreement of the \textit{vacuum-like} to the Vacuum sample could in principle be achieved by optimizing the reference cut. Nevertheless, our purpose is to distinguish \textit{medium-} from \textit{vacuum-like} jets. We thus expect some events from the Medium sample to be identified by the DL models as \textit{vacuum-like}, thus slightly distorting the resulting distributions. The \textit{medium-like} $p_{T,Z}$ spectra obtained through DL selection is always suppressed with respect to the Medium sample, thus indicating that all DL models are making a separation that does not follow our Vacuum vs Medium simulations. To confirm if the DL classifiers were not misidentifying \textit{medium-like} jets out of the JEWEL+PYTHIA Vacuum simulation, we checked the percentage of the test events categorised as being \textit{medium-like}. This amounts to $9\%$ for the CNN trained on unnormalised images, $11\%$ for the Global DNN, $13\%$ for the RNN, and $18\%$ for the CNN trained on normalised images, thus confirming that, overall, the obtained DL discriminants can correctly identify jets whose fragmentation pattern follows the same as vacuum physics. The CNN trained on normalised images is the one where misidentification can potentially impact the interpretation of the results. It is well known that the $p_{T,jet}$ is a fairly good discriminant of non-quenching. Selecting higher transverse momentum jets (higher $p_{T,Z}$) will likely bias our sample towards lower fragmentation patterns, and as such, subject to smaller energy loss effects. Since this quantity is related to the $p_{T,Z}$, if this information was used by the DL network as a discriminant, we expect a different $p_{T,Z}$ dependence from the Monte Carlo truth. From~\cref{fig:ptZ}, we observed that the resulting transverse momentum dependence of the \textit{medium-like} jets provided by the DL architecture vary between them. The Lund sequences (RNN) are the ones that show a larger $p_{T,Z}$ dependence, followed by unnormalised images (CNN) and the global information (DNN). By construction, the CNN trained on normalised images shows a very weak dependence on the $p_{T,Z}$. This is in agreement with the results in Table~\ref{tab:roc_auc}, where the performance of the Global DNN was the most affected when increasing the minimum cut on $p_{T,jet}$. Moreover, we also see that the RNN and unnormalised CNN are using additional information from the jet fragmentation pattern since they have a different $p_{T,Z}$ dependence when compared to the Global DNN, trained solely on $p_{T,jet}$ and $n_{const}$.

\begin{figure}[t]
    \centering 
    \includegraphics[width=0.45\textwidth]{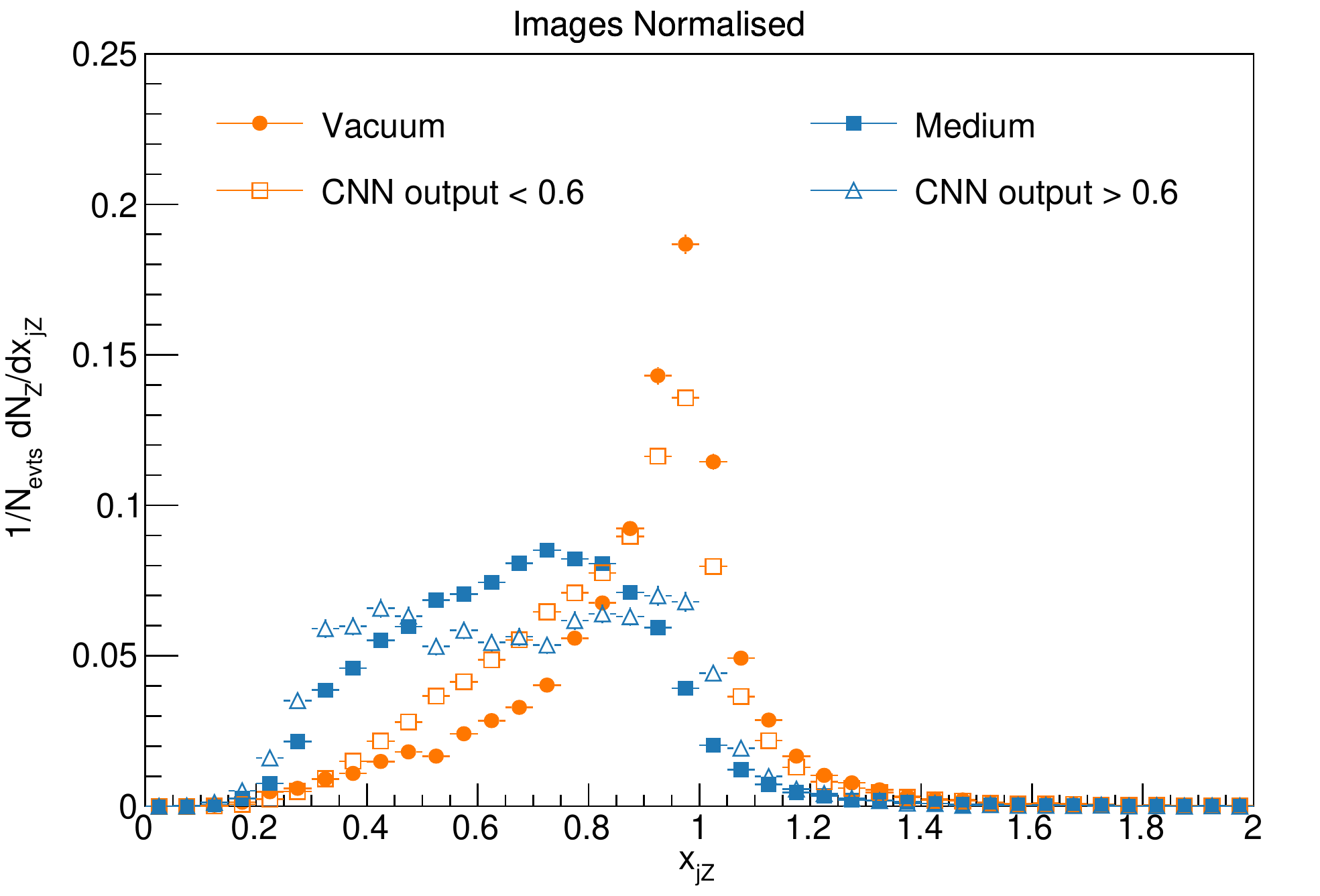}
    \includegraphics[width=0.45\textwidth]{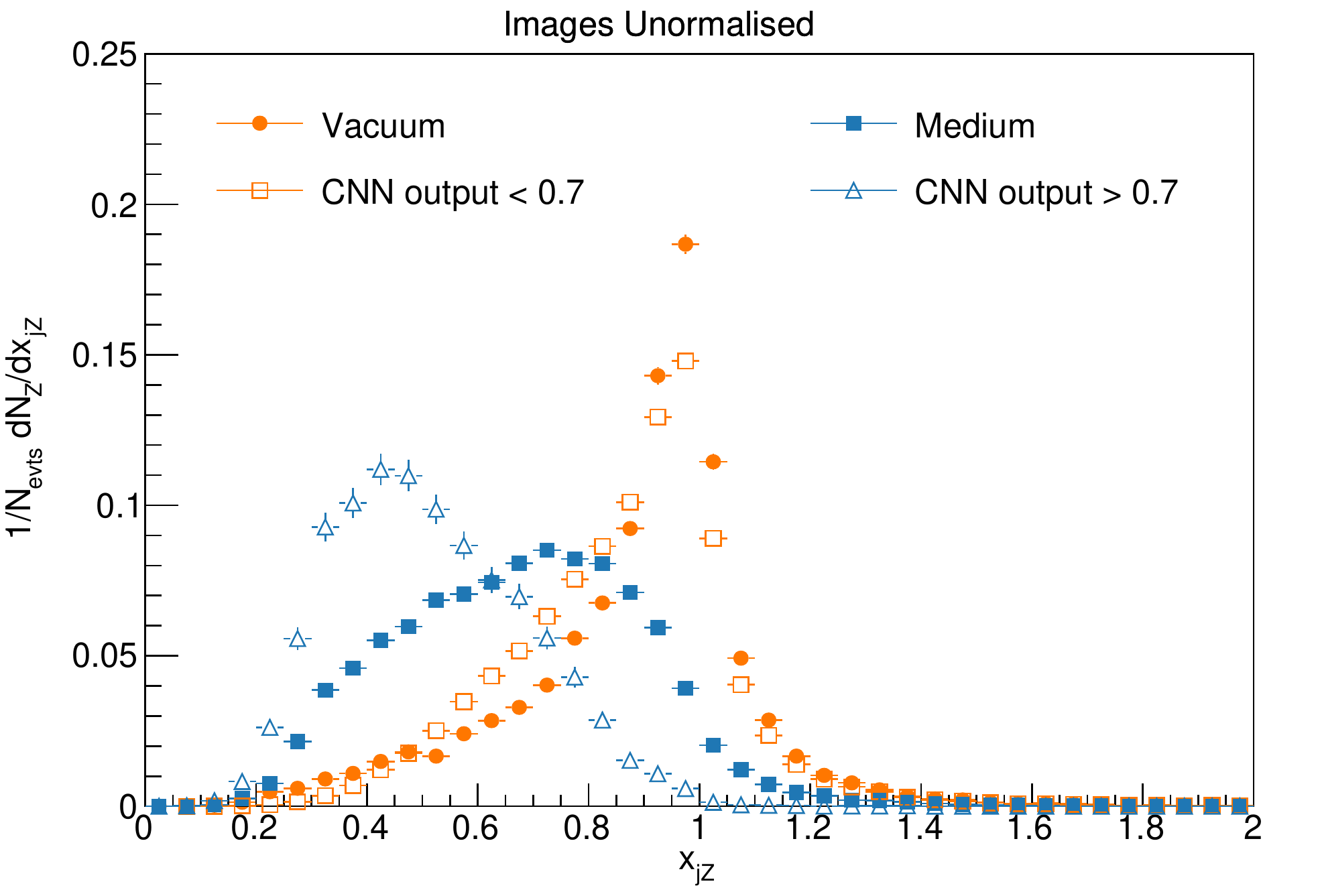}
    \includegraphics[width=0.45\textwidth]{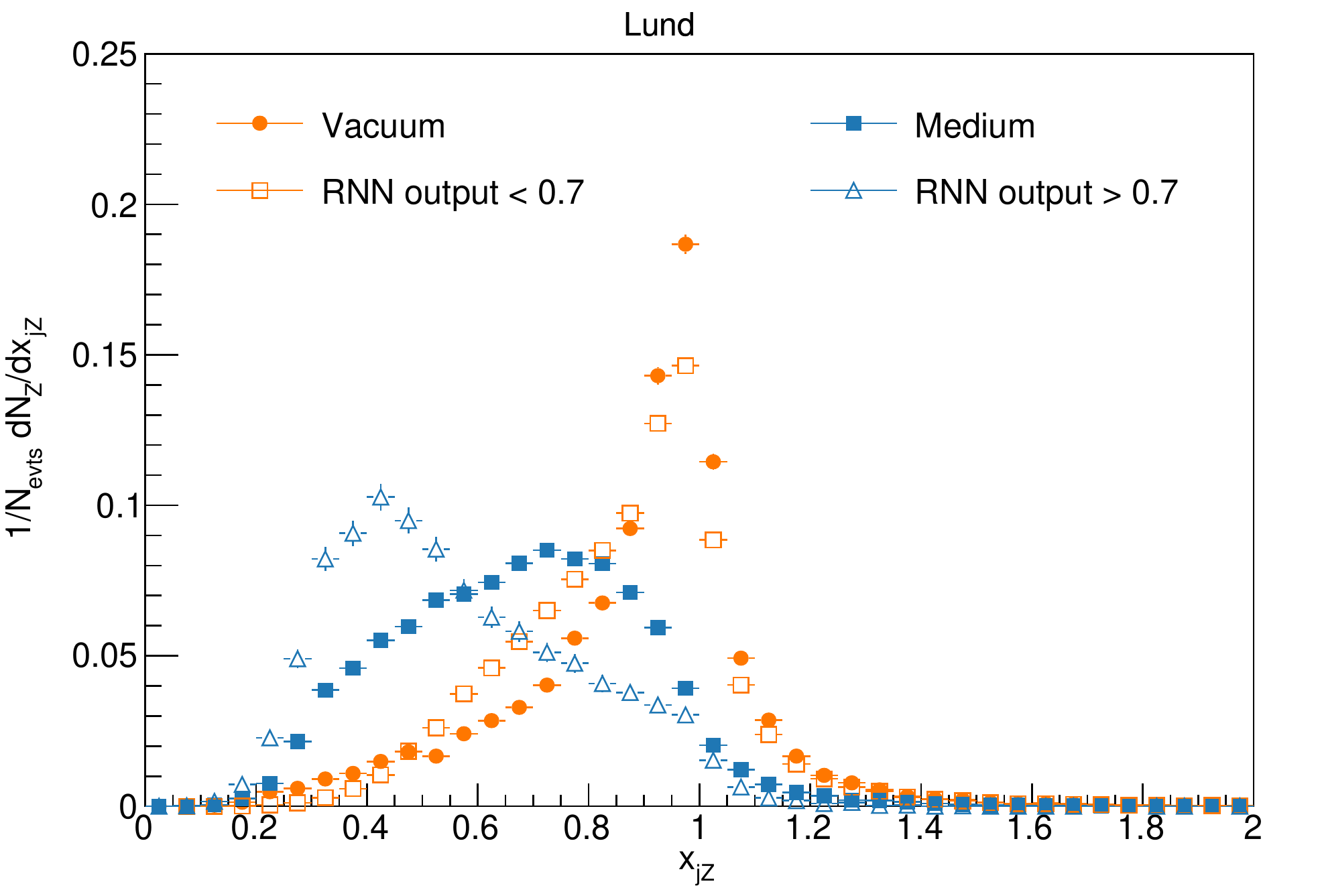} 
    \includegraphics[width=0.45\textwidth]{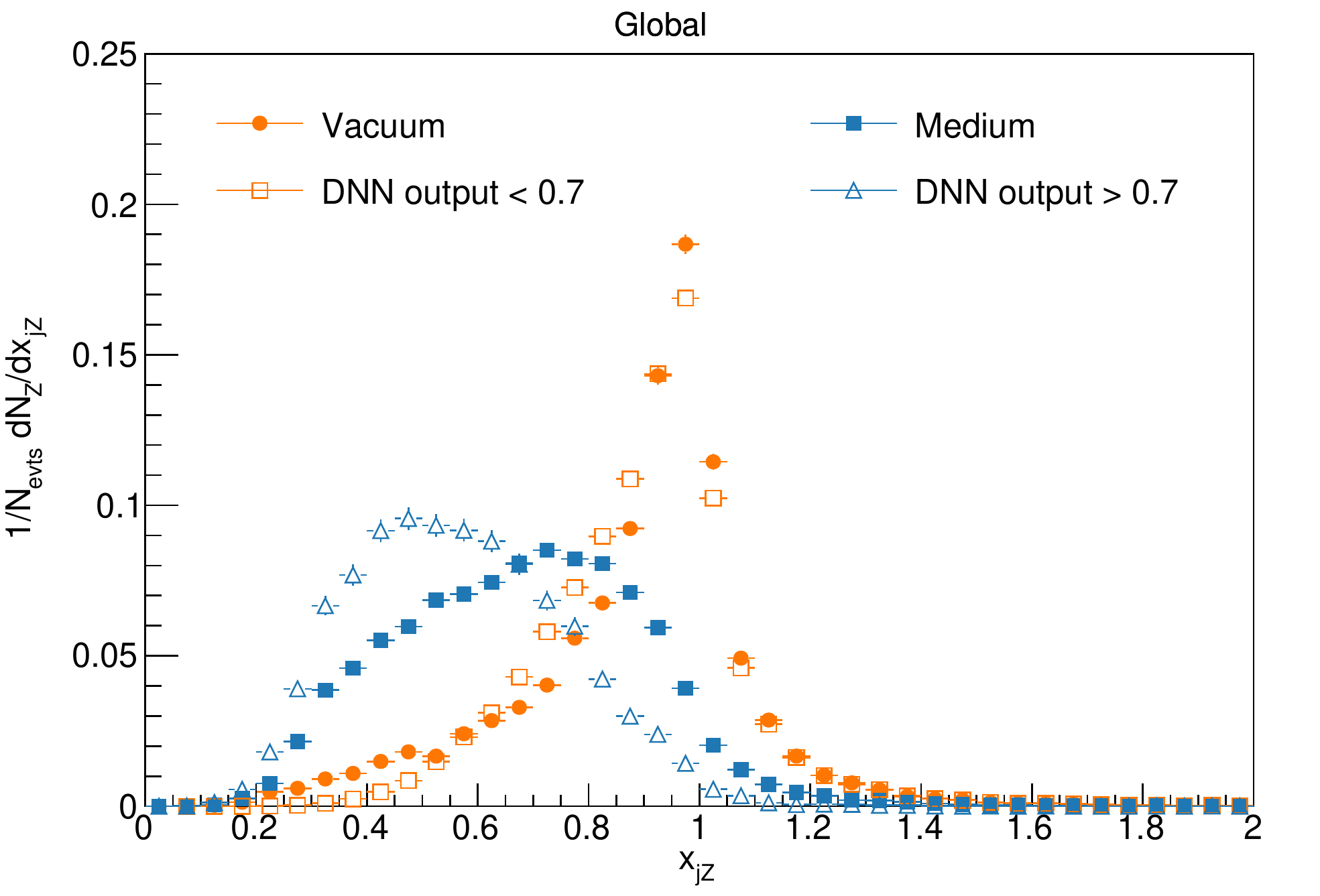}
    \caption{\label{fig:dijet_xj} Transverse momentum imbalance $x_{jZ}$, for the different Deep Learning architectures. Monte Carlo truth from JEWEL+PYTHIA is provided in solid symbols for the Vacuum and Medium samples and a subset of events selected by the DL discriminant appears in open symbols. The DL output selection employed to identify \textit{vacuum-like} jets (open blue) and \textit{medium-like} jets (open orange) is made explicit in the legend of each plot.}
\end{figure}

We now proceed to analyse the transverse momentum imbalance of the \textit{medium-} and \textit{vacuum-like} event sample. This observable is only sensitive to the fraction of transverse momentum that is captured inside the jet area, with respect to the $p_{T,Z}$. However, energy loss induced by jet quenching effects is associated with a change in the fragmentation pattern of a jet. As such, we expect some differences between the Global and the DL architectures that do use clustering information as input. On the opposite end, we have the CNN trained on normalised images whose input is the relative differences in the fragmentation pattern. The resulting distributions are illustrated in~\cref{fig:dijet_xj}, keeping the same symbol (open symbols - DL output; full symbols - JEWEL+PYTHIA) and colour notation (orange - Vacuum; blue - Medium) as before. Clearly, the CNN that is trained on normalised images shows the most different results. It is not able to recover so well the $x_{jZ}$ distribution obtained from the Vacuum sample (the Global DNN seems to excel in this sense) and the $x_{jZ}$ of \textit{medium-like} jets is also more flat when compared to the Medium sample. The distribution of the output of this network (\cref{fig:nnoutputs}) for the Medium and Vacuum sample overlaps significantly, making it more difficult to select a suitable reference value. Nonetheless, the \textit{medium-like} $x_{jZ}$ provided by this CNN seems to enhance \textit{medium-like} features with respect to the medium sample as its $x_{jZ}$ distribution is displaced towards smaller values. Training only on jet-wise variables, such as the Global DNN, provides an excellent description of the vacuum $x_{jZ}$. As expected, using $p_{T,jet}$ during the training helps to describe observables that are exclusively sensitive to energy loss effects. The \textit{medium-like} $x_{jZ}$ provided by the Global DNN is shifted towards the left and has approximately the same shape as the Medium Monte Carlo truth. By using a more complete set of jet information - unnormalised images or Lund planes - we see that the \textit{medium-like} distribution selected by the corresponding DL architectures is even more peaked at lower $x_{jZ}$. The \textit{vacuum-like} distribution is slightly displaced from the Monte Carlo truth Vacuum sample. This might also hint that these networks can identify jets in the Medium JEWEL+PYTHIA sample that did not experience major interactions with the medium, thus categorizing them as \textit{vacuum-like} jets.

After checking the $p_{T,Z}$ dependence and the results on $x_{jZ}$ we move to observables that require information from jet substructure: the average jet radial profile, that keeps track of the number of particles in bins of $\Delta R$ inside of the jet, and the jet mass, $m_j$, that weights distance and transverse momentum of the particles inside the jet.

\begin{figure}[t]
    \centering 
    \includegraphics[width=0.45\textwidth]{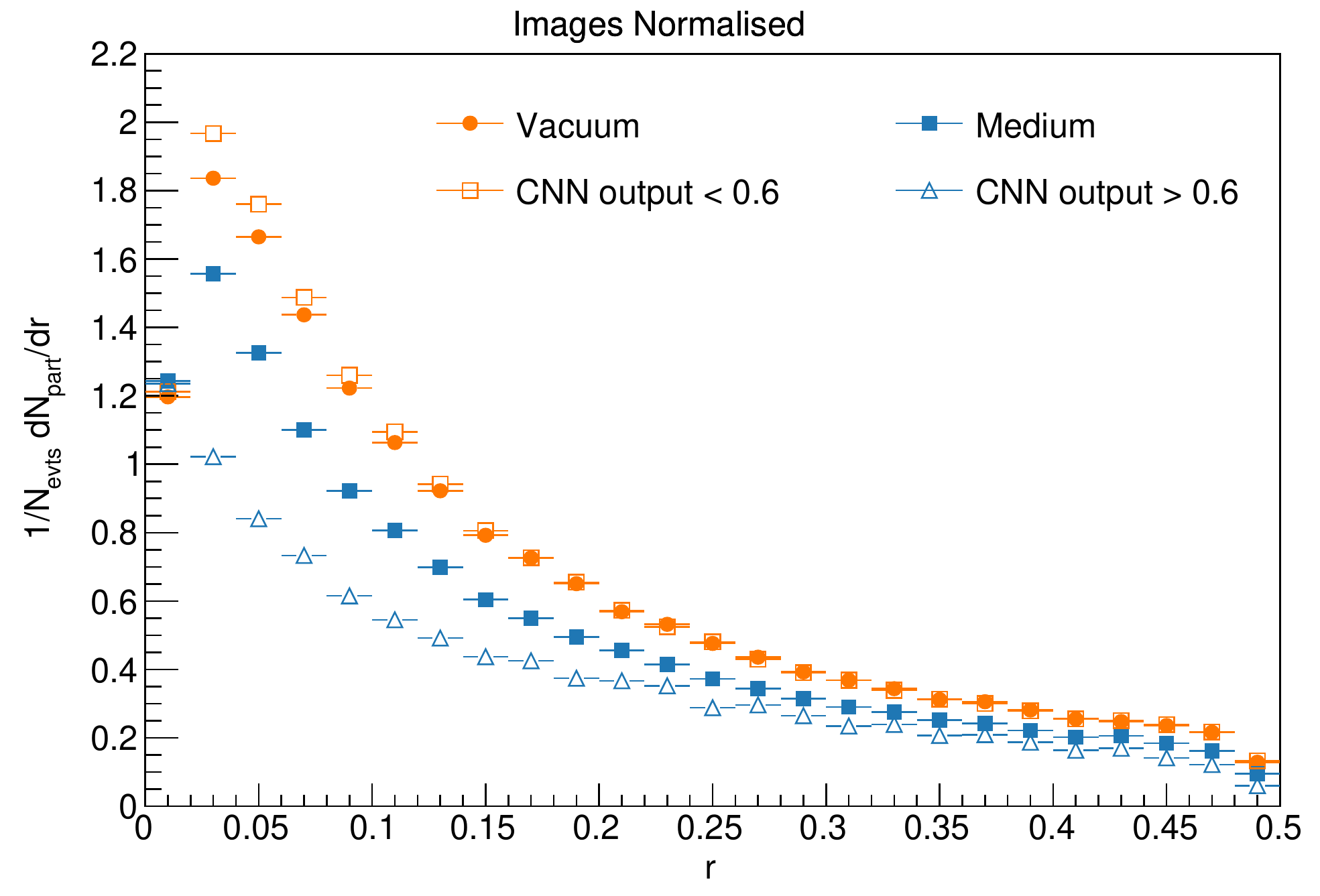}
    \includegraphics[width=0.45\textwidth]{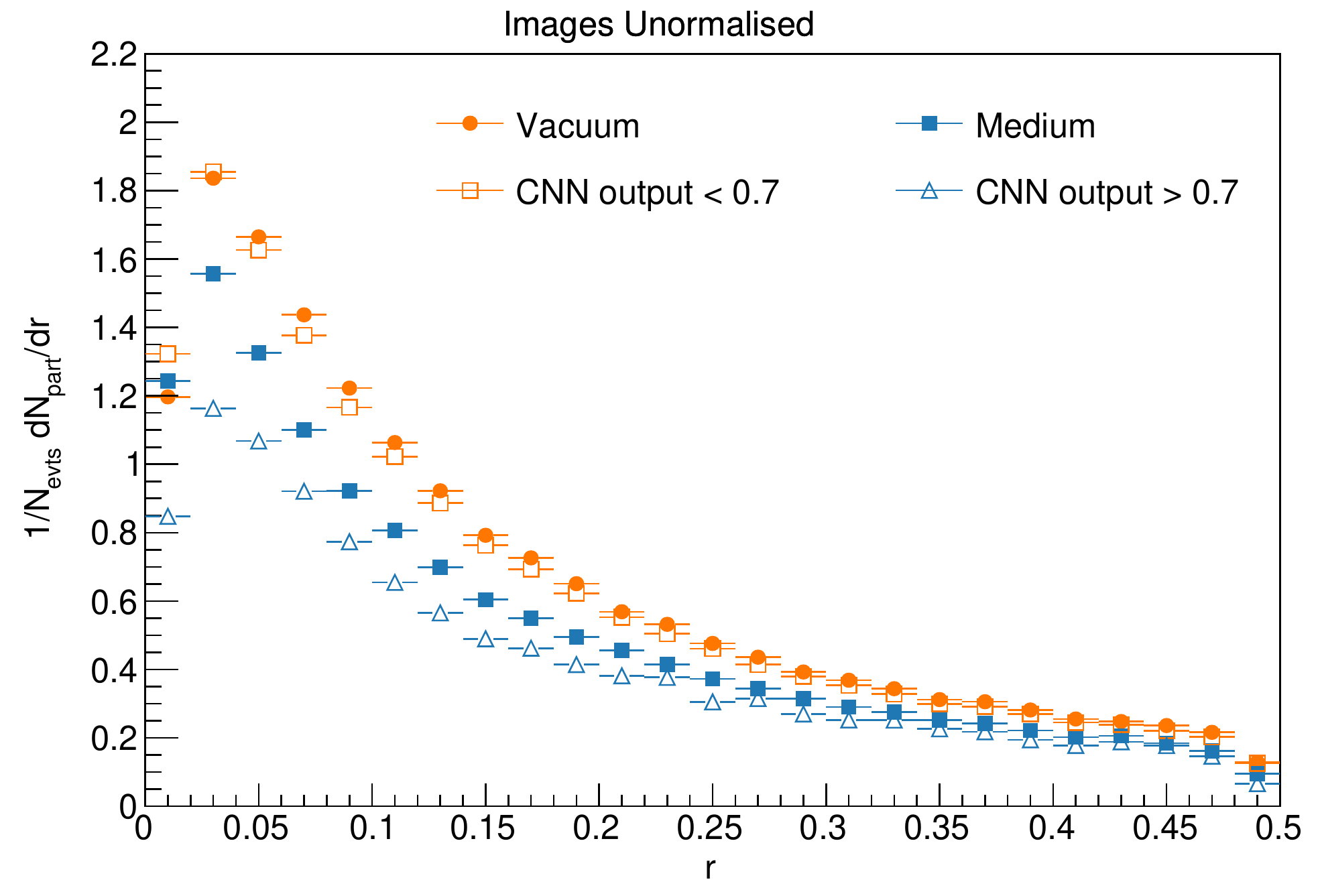}
    \includegraphics[width=0.45\textwidth]{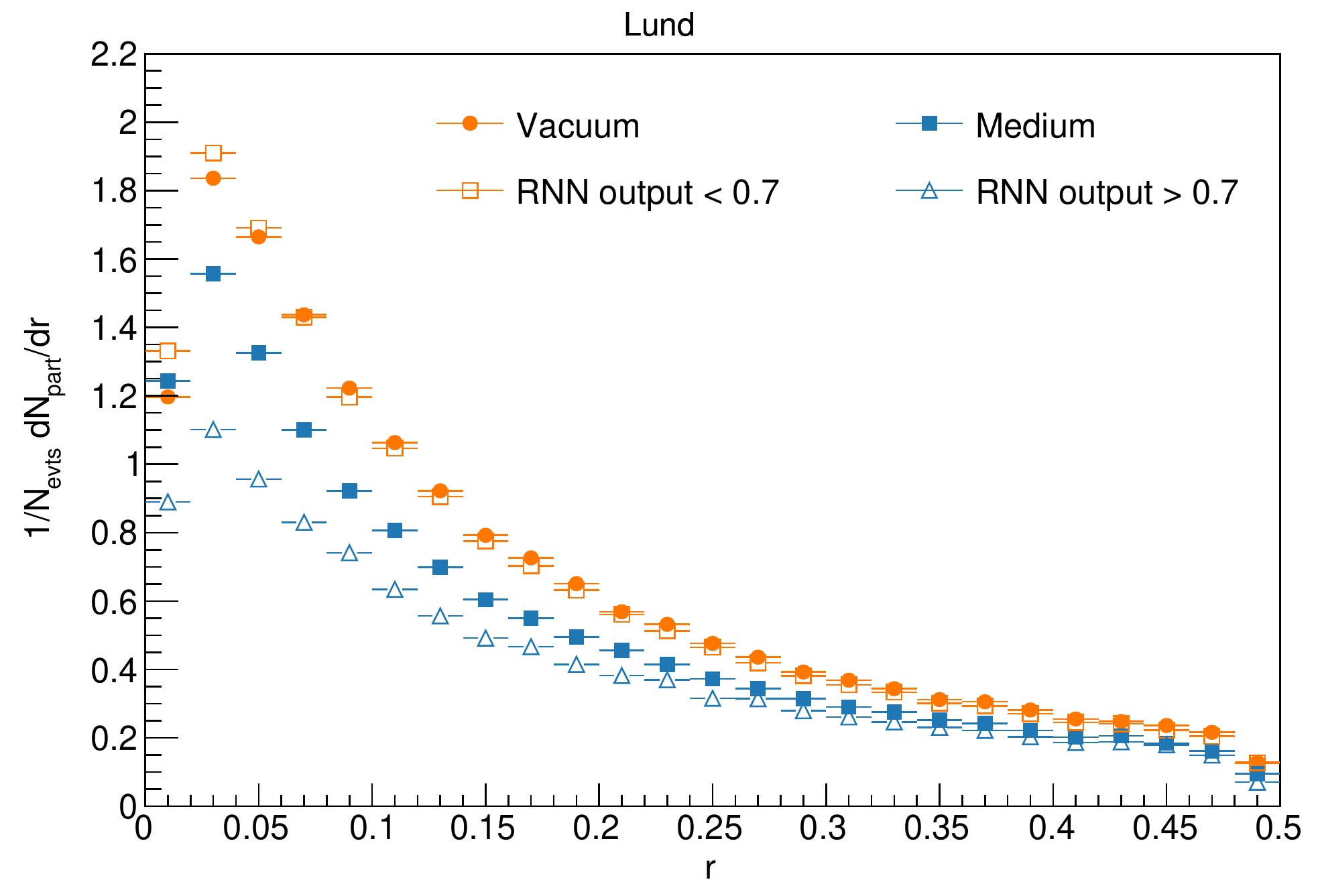} 
    \includegraphics[width=0.45\textwidth]{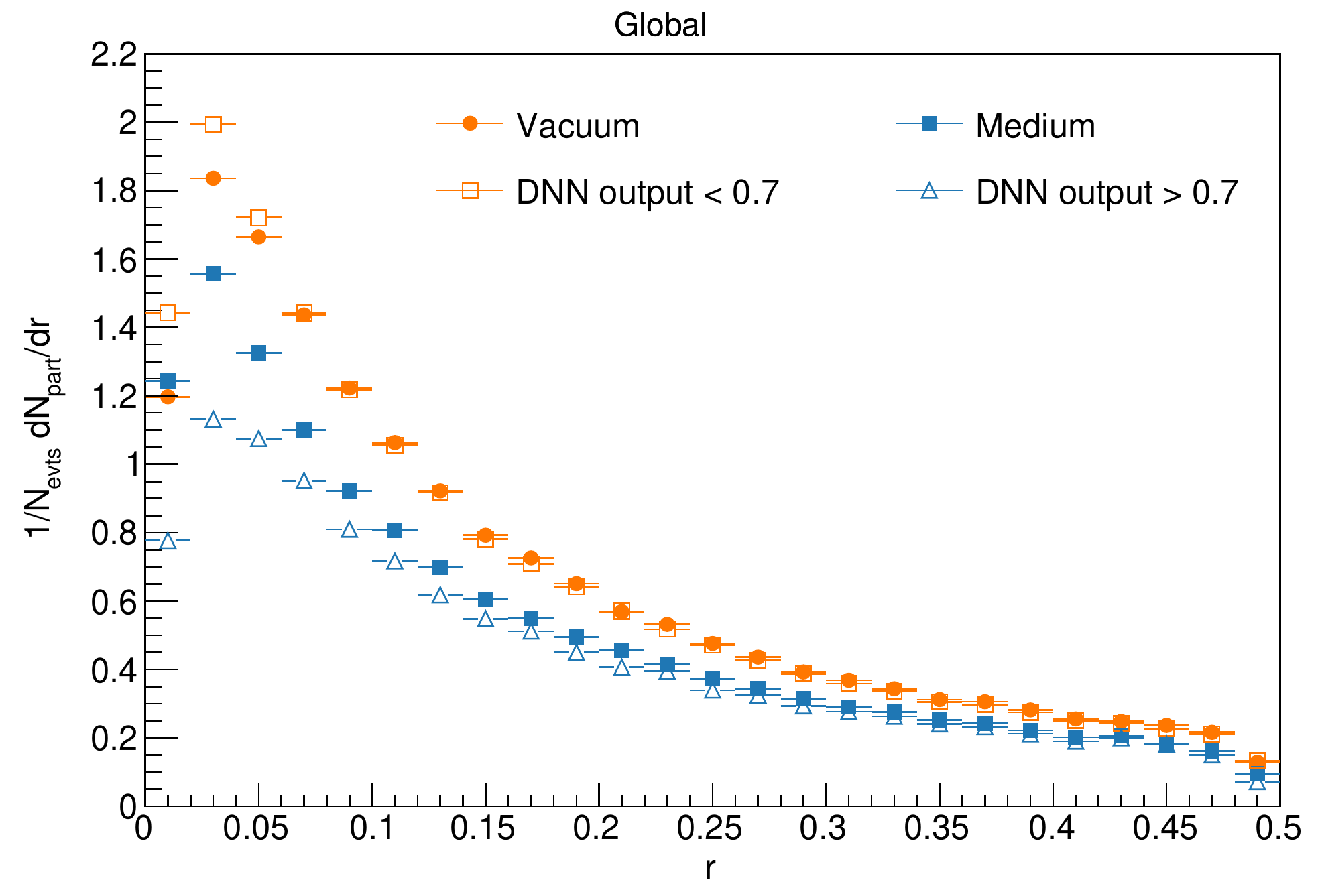}
    \caption{\label{fig:radial_profile} Reconstructed jet radial profile (average number of constituents) $r$, for the different Deep Learning architectures. Monte Carlo truth from JEWEL+PYTHIA is provided in solid symbols for the Vacuum and Medium samples and a subset of events selected by the DL discriminant appears in open symbols. The DL output selection employed to identify \textit{vacuum-like} jets (open blue) and \textit{medium-like} jets (open orange) is made explicit in the legend of each plot.}
\end{figure}

The results for the average jet radial profile are shown in~\cref{fig:radial_profile}. Overall, all DL architectures select the same type of \textit{vacuum-like} pattern jets as the Vacuum sample, even though the resulting $x_{jZ}$ distribution can vary. The Global DNN, that did not receive information from the jet fragmentation during training, shows the same trend, but this can be a consequence of providing an exceptional good agreement on the highly peaked vacuum $x_{jZ}$ distribution. It is also possible to see that all but the Global DNN identify \textit{medium-like} jets as being narrower than the ones within the Medium JEWEL+PYTHIA sample. Since the latter sample contains a mixture of different levels of quenching, it is thus expected that a more pure sample of \textit{medium-like} jets will be even narrower. The details between the DNNs results differ, nonetheless. The CNN trained on normalised images is the one that shows the highest deviation because it is trained only on the relative fragmentation. It follows the Lund planes and unormalised jet images. We note that while the presence of jet quenching will induce a narrower average jet radial profile, the opposite is not necessarily verified. For this reason, the CNN trained on normalised images results into a more flat $x_{jZ}$ distribution despite showing a selection of very narrow jets. On the other hand, the DL networks exploring unnormalised images or Lund planes identify a not so narrow jet, but that indeed lost a significant amount of energy relative to its initial momentum ($p_{T,Z}$). The Global DNN, whose training did not contain any information on the jet substructure, still selects jets whose centre is depleted concerning the Medium sample. These jets are more evenly populated, and thus likely to contain medium-induced radiation that travelled along the jet direction. While retaining this energy, these jets continue to experience collisional energy loss as its absolute multiplicity continues to be smaller than the Medium Monte Carlo reference.

\begin{figure}[t]
    \centering 
    \includegraphics[width=0.45\textwidth]{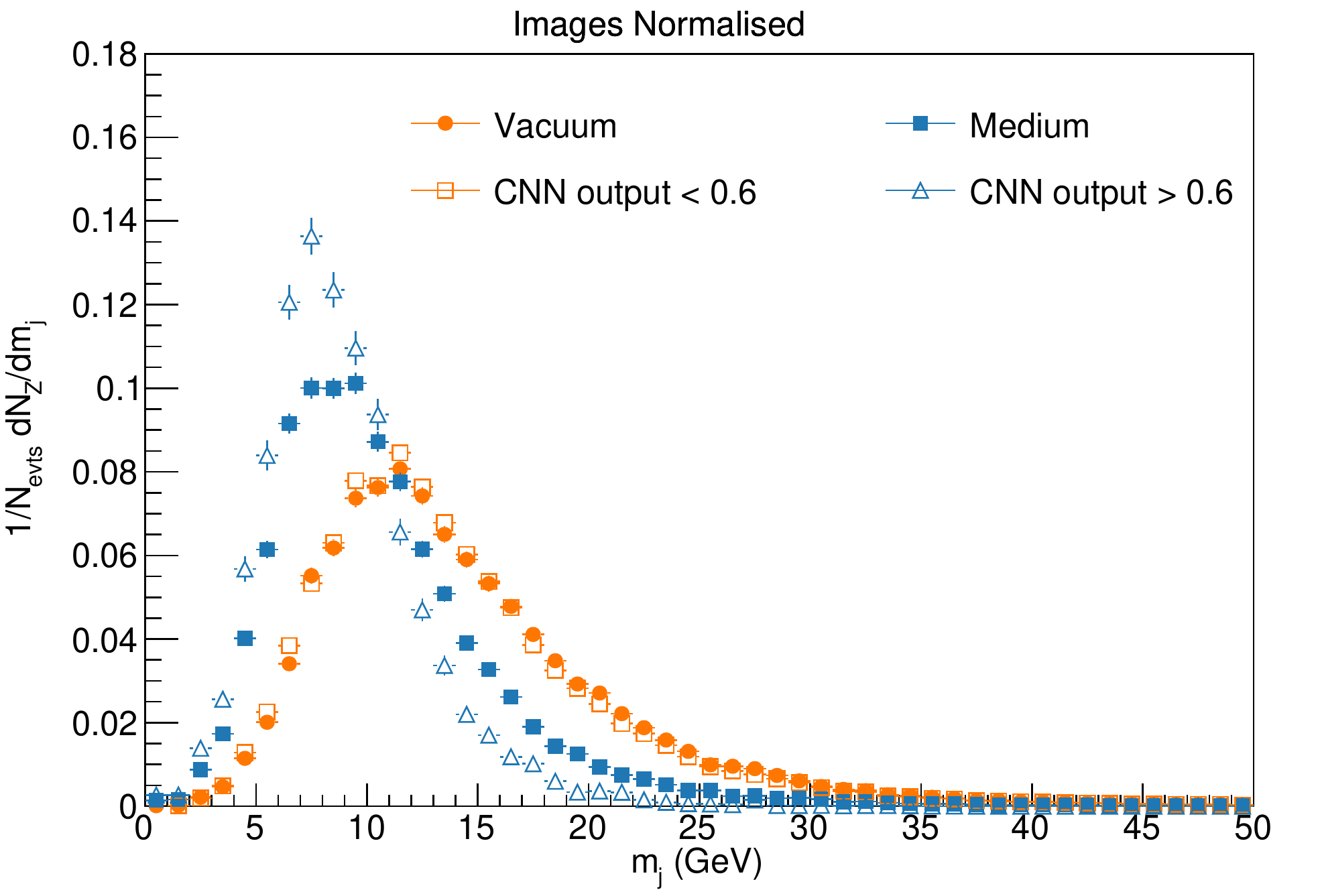}
    \includegraphics[width=0.45\textwidth]{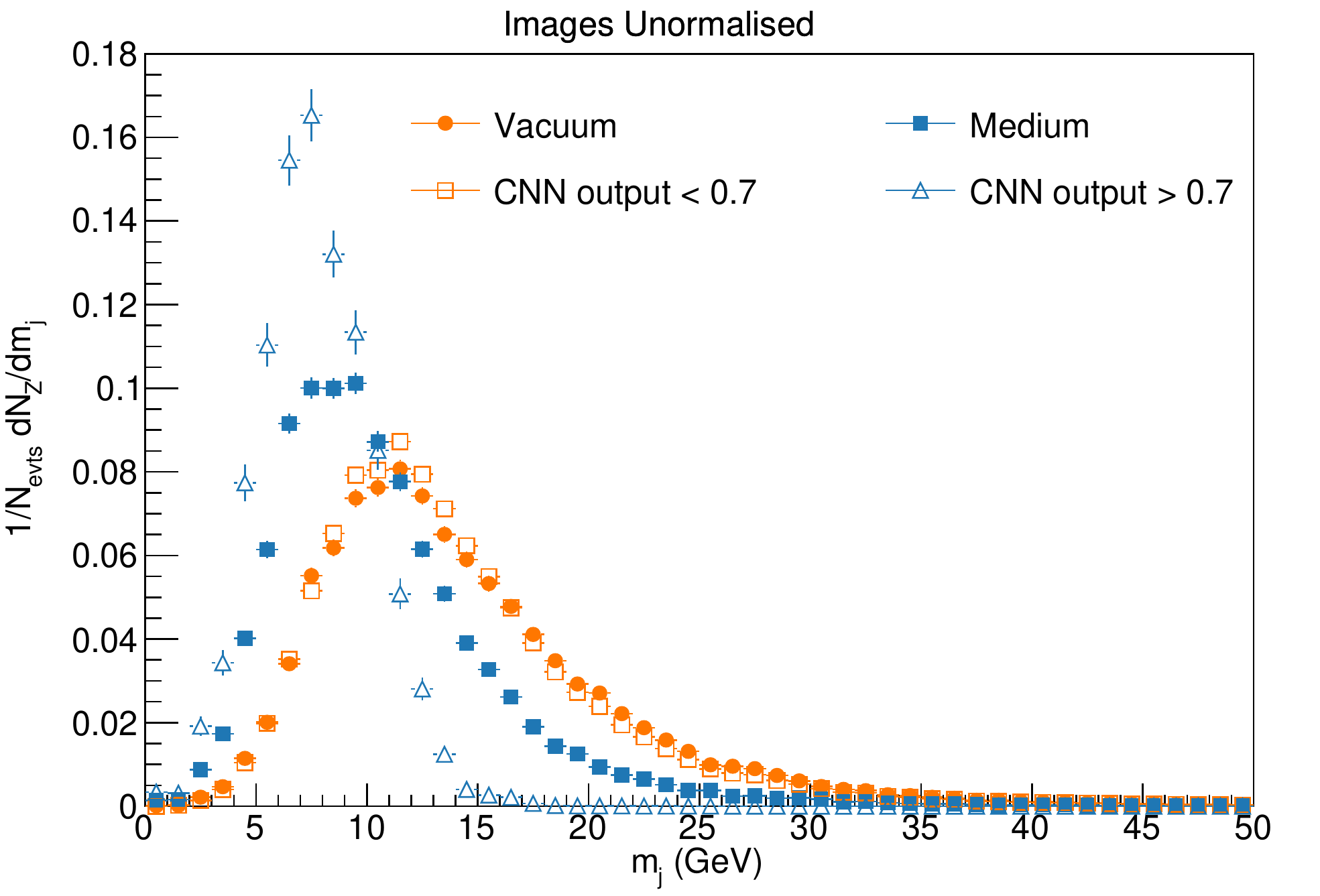}
    \includegraphics[width=0.45\textwidth]{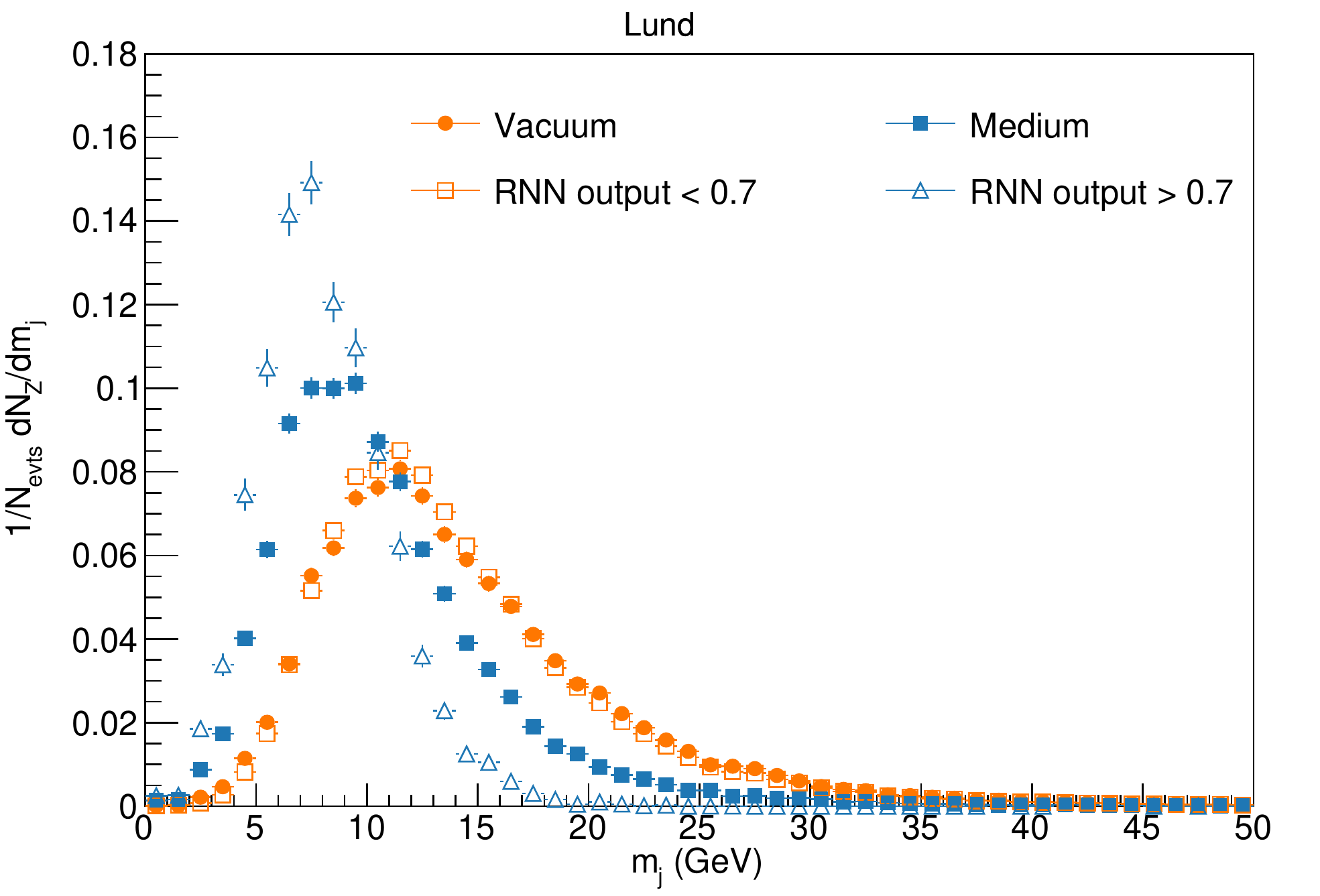} 
    \includegraphics[width=0.45\textwidth]{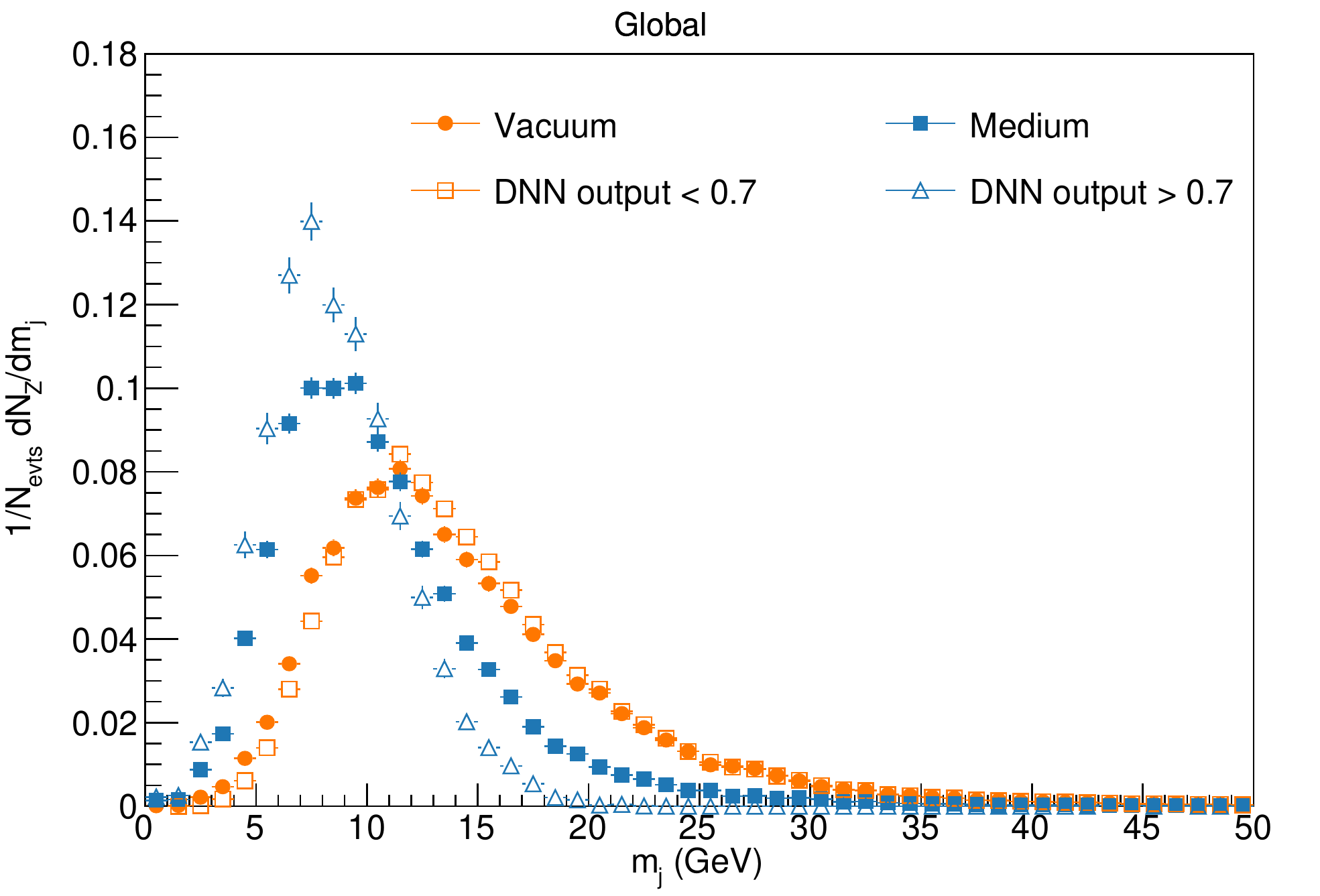}
    \caption{\label{fig:mass_jet} Reconstructed jet mass $m_{j}$, for the different Deep Learning architectures. Monte Carlo truth from JEWEL+PYTHIA is provided in solid symbols for the Vacuum and Medium samples and a subset of events selected by the DL discriminant appears in open symbols. The DL output selection employed to identify \textit{vacuum-like} jets (open blue) and \textit{medium-like} jets (open orange) is made explicit in the legend of each plot.}
\end{figure}

Finally, the results on the jet mass are shown in ~\cref{fig:mass_jet}. As mentioned before, this observable keeps track of all jet input variables used in this training by definition. Thus, all DL architectures used in this study were given (partial) input about this observable, and as such, all of them are able to identify \textit{vacuum-like} as being the same as the Monte Carlo Vacuum sample. Simultaneously \textit{medium-like} jets show a jet mass that is smaller than the Medium sample. Nonetheless, we see that the two networks trained with all information (unormalised images and Lund planes) tag a jet population with smaller jet mass induced by jet quenching effects.

As such, relative differences on the jet pattern alone or jet-wise variables ($p_{T,jet}$ and $n_{const}$) can be used to identify energy loss effects or differences in the jet fragmentation function, independently if a particular observable is insensitive to one effect or the other. However, it seems that alone they do not suffice. A combination of both seems to work better to emphasise jet quenching features across a wider range of observables. In this regard, both the CNN trained on unnormalised images (final state particles only) or the RNN in Lund planes (declustering information) seem to perform equally well. In~\cref{app:CNNinterpretation} we scrutinise further which jet features are the CNNs triggering on.

\FloatBarrier
\section{Conclusions}
\label{sec:conclusions}

In this work we set out to explore how different DL architectures learn to discriminate between \textit{medium-like} and \textit{vacuum-like} jets. For this purpose, we used JEWEL as our Monte Carlo event generator to produced Vacuum and Medium $Z$+jet samples. The different architectures presented were chosen as to utilise different data representations of the jets: Convolutional Neural Networks for jet-images, Dense Neural Networks for jet-wise observables, and Recurrent Neural Networks for Lund plane paths. Since each data format carries different explicit and implicit information, this comparison allowed us to further understand how DL can help isolate medium-induced effects in jets.

By looking at how a DL-based classification affect the distributions of the jet observables, we observed that while all DL networks seem to identify the same \textit{vacuum-like} distributions, they did not always produced the same \textit{medium-like} distributions. More specifically, we observed that CNN on unnormalised images and the RNN on Lund plane paths identified \textit{medium-like} jets to be more different from the results provided by the Monte Carlo (Medium sample). These two architectures have access to the absolute scale of the jet transverse momentum and to its number of constituents, but their discriminant power is not based on the two observables alone, since they outperform the Global DNN establishing the ground performance of the transverse momentum and multiplicity of constituents. As such, the result indicates that the CNN on unnormalised images and the RNN are also learning from the fragmentation pattern that will yield different jet profiles (CNN) as well as different jet Lund sequences (RNN).

The samples that include jet quenching effects were produced with JEWEL, a widely used jet quenching Monte Carlo event generator. While our results are biased towards this Monte Carlo truth, the agreement of this model over a wide range of jet observables~\cite{Zapp:2013vla, KunnawalkamElayavalli:2017hxo} provides a robust baseline to establish the first step towards using Deep Learning techniques to identify in-medium modifications. Therefore, the effect of medium-induced recoils, the most model-dependent feature of current jet quenching descriptions, was neglected. As an outlook, we plan to use the most performant networks from this study (RNN and CNN for unnormalised images) in different jet quenching Monte Carlo event generators and with the presence of recoiling scattering centres. This will further probe the capability of the proposed DNN methodology to identify jet quenching effects induced by the presence of a QGP.

\section{Acknowledgments}

We acknowledge the support from FCT Portugal,
Lisboa2020, Compete2020, Portugal2020 and FEDER under project
PTDC/FIS-PAR/29147/2017, CERN/FIS-PAR/0024/2019 and contract DL57/2016/CP1345/CT0004. This work was also supported by the European research Council project ERC-2018-ADG-835105 YoctoLHC. The computational part of this work was supported by
INCD (funded by FCT and FEDER under the project 01/SAICT/2016 nr.
022153) and by FCT under project CPCA/A2/4075/2020. Part of the computation work was done in the Research Centre in Digitalization and Intelligent Robotics  -- CeDRI -- at Instituto Politécnico de Bragança. This project has received funding from the European Union’s Horizon 2020 research and innovation programme under grant agreement No 824093.

\newpage

\appendix

\section{Correlation between Deep Neural Networks}\label{app:DNNcorrelation}

\begin{figure}[h]
    \includegraphics[width=0.32\textwidth]{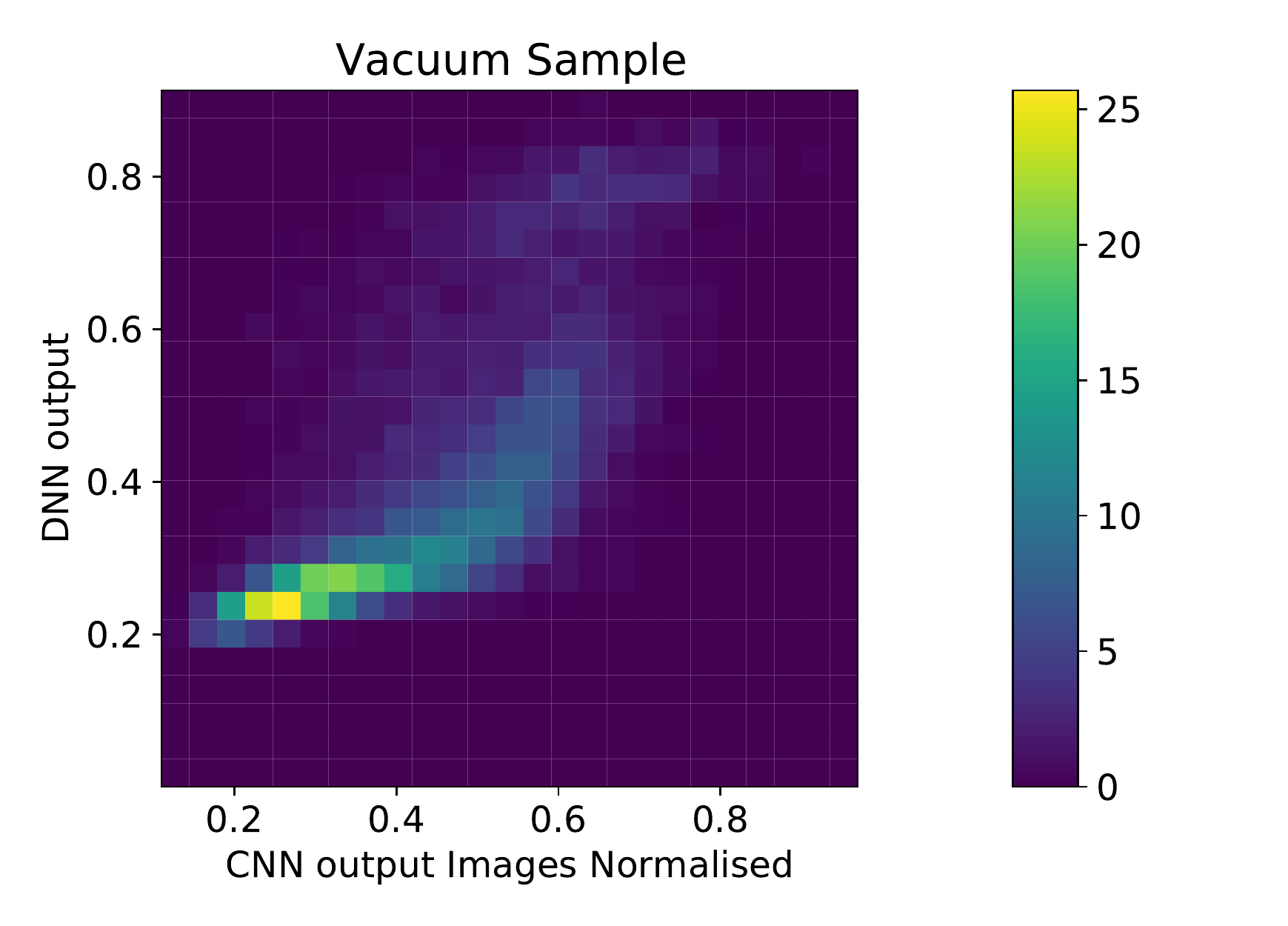}
    \includegraphics[width=0.32\textwidth]{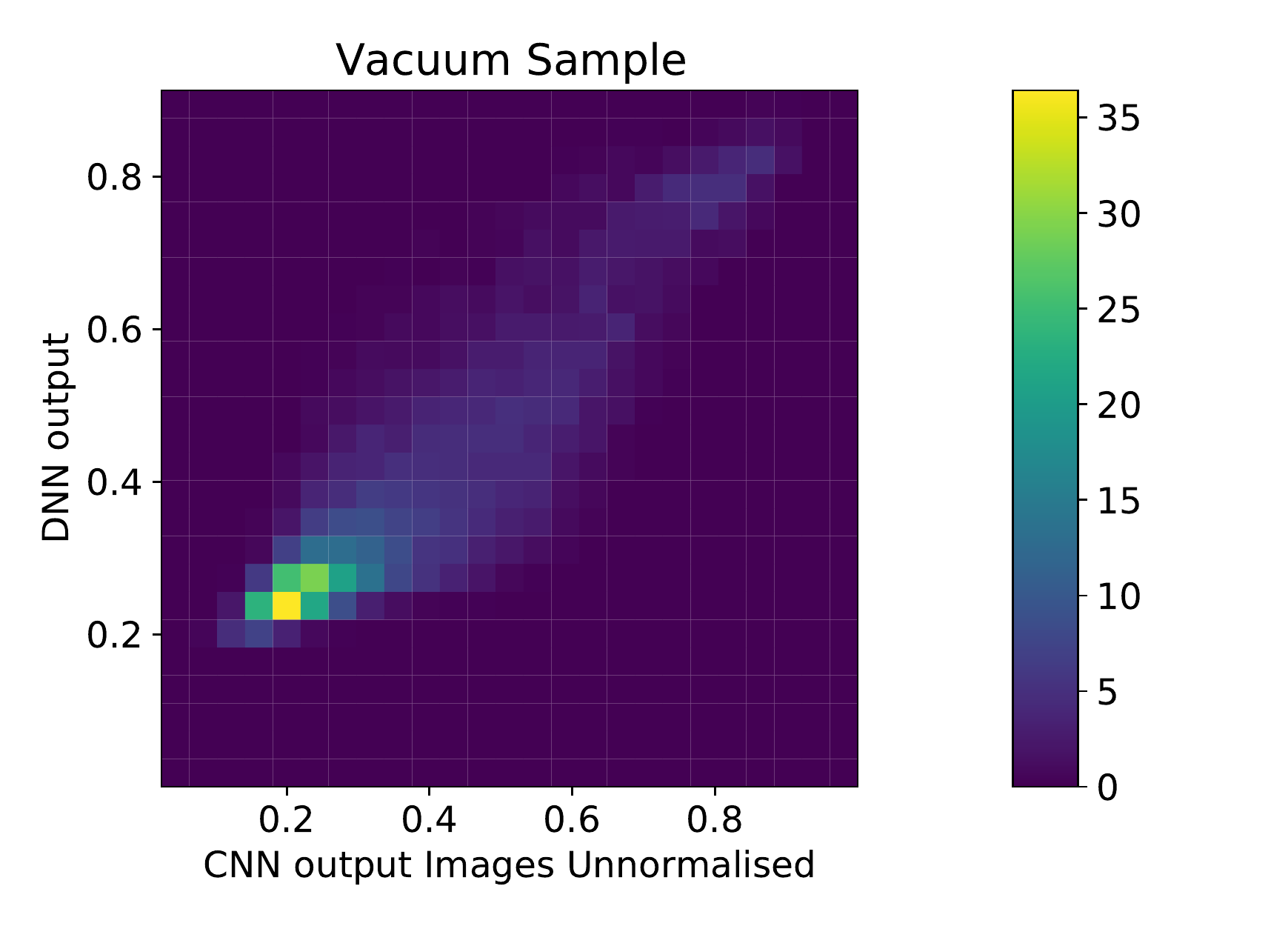}
    \includegraphics[width=0.32\textwidth]{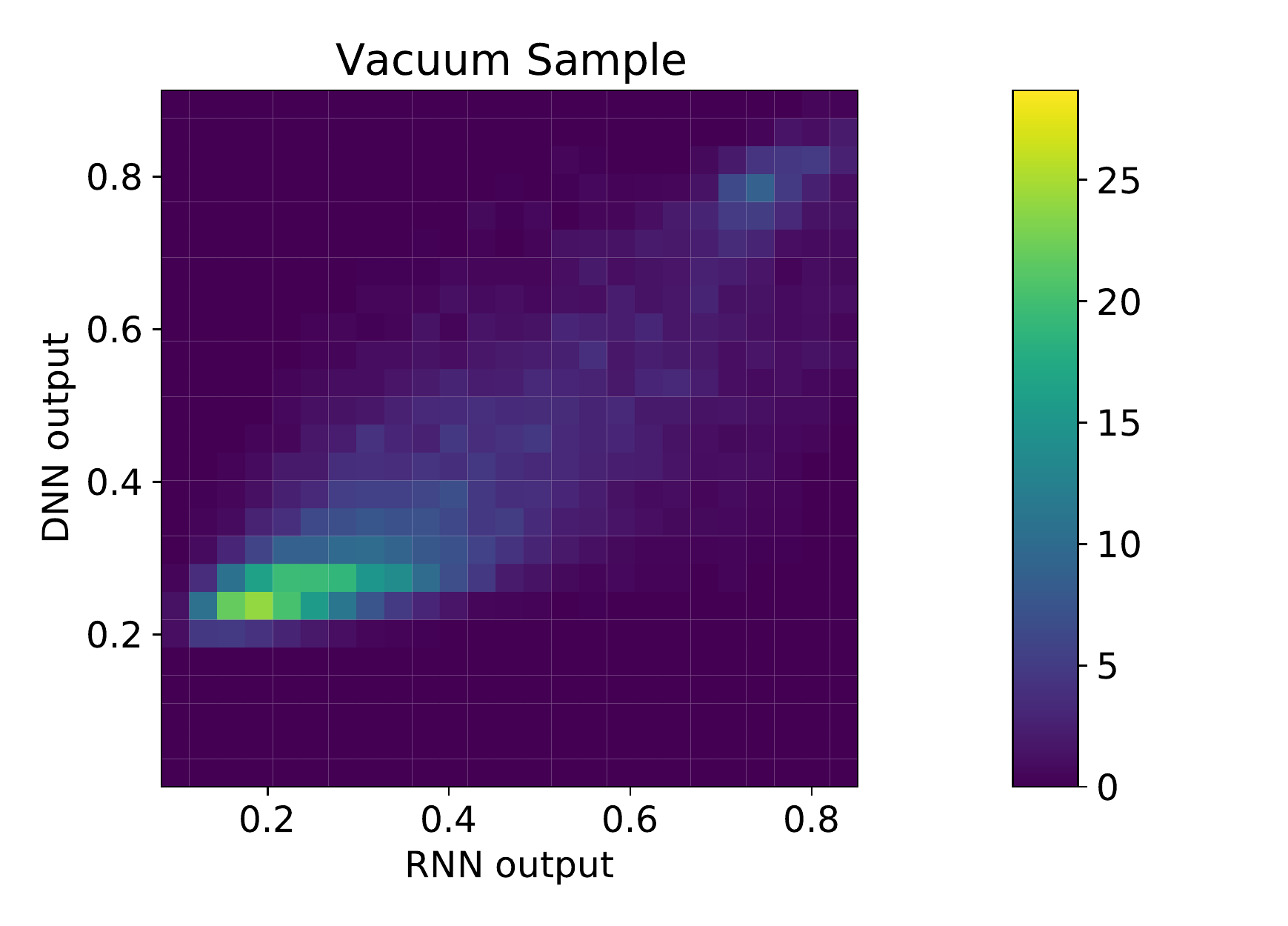}\\
    \includegraphics[width=0.32\textwidth]{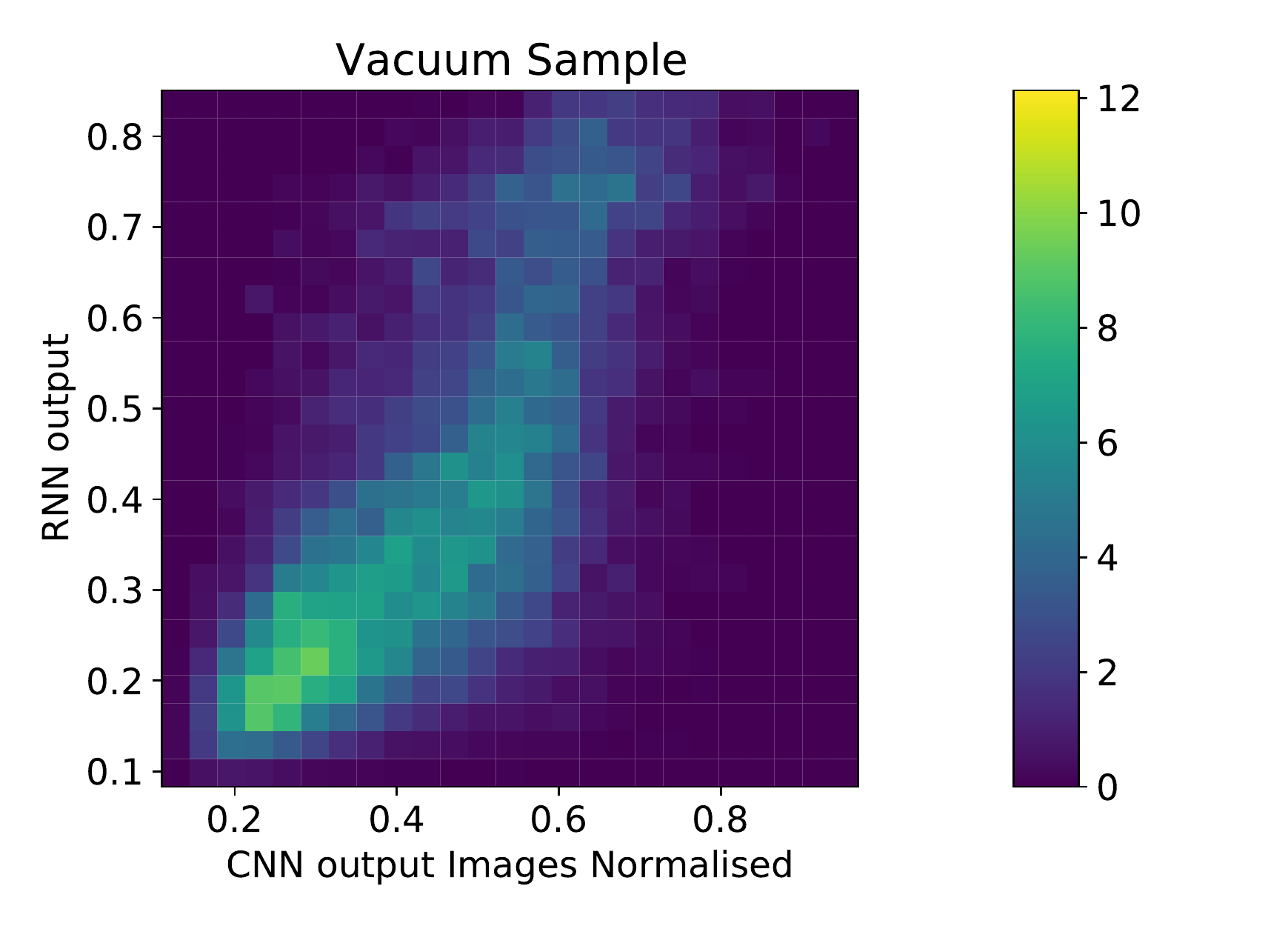}
    \includegraphics[width=0.32\textwidth]{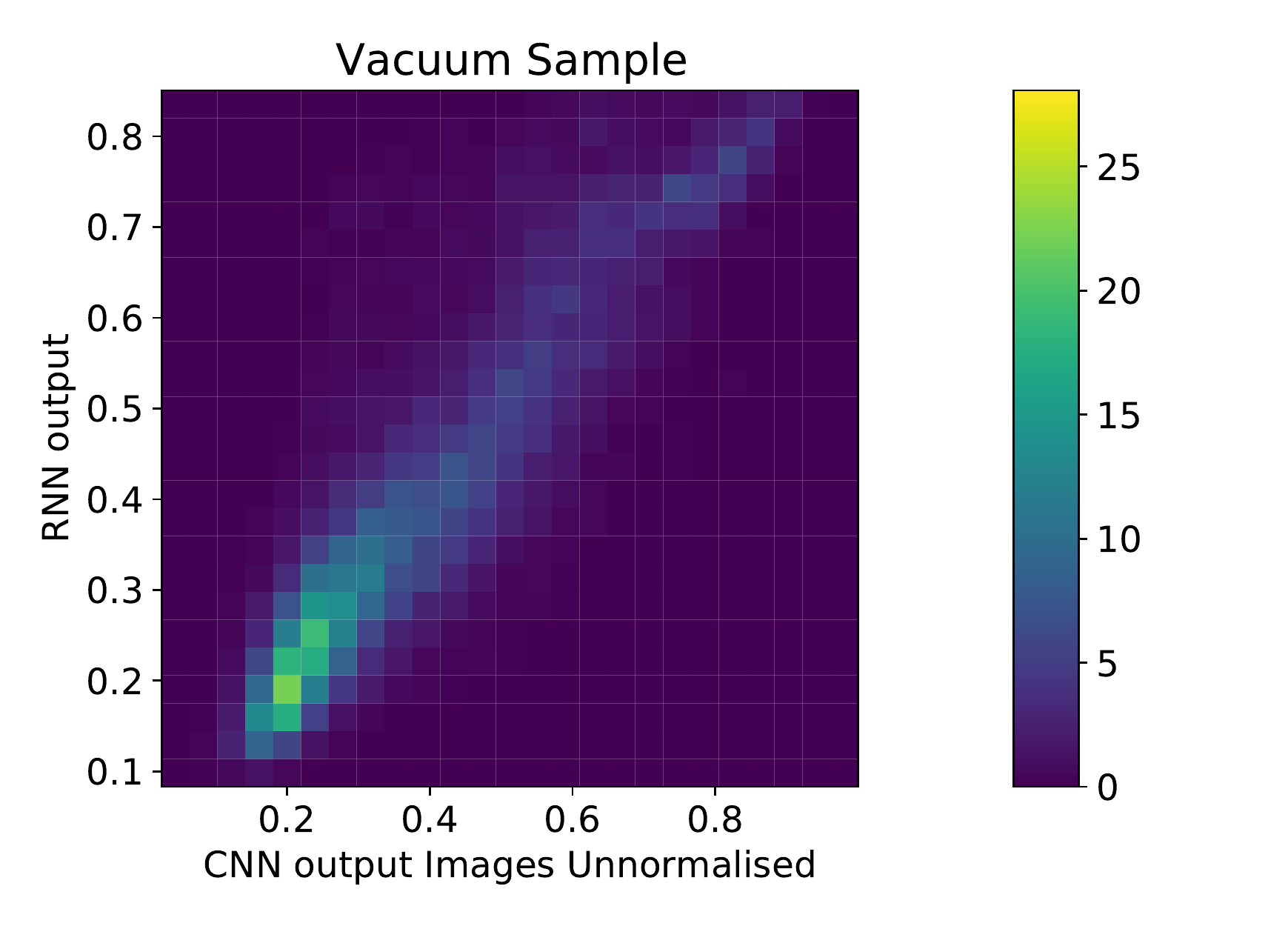}\\
    \includegraphics[width=0.32\textwidth]{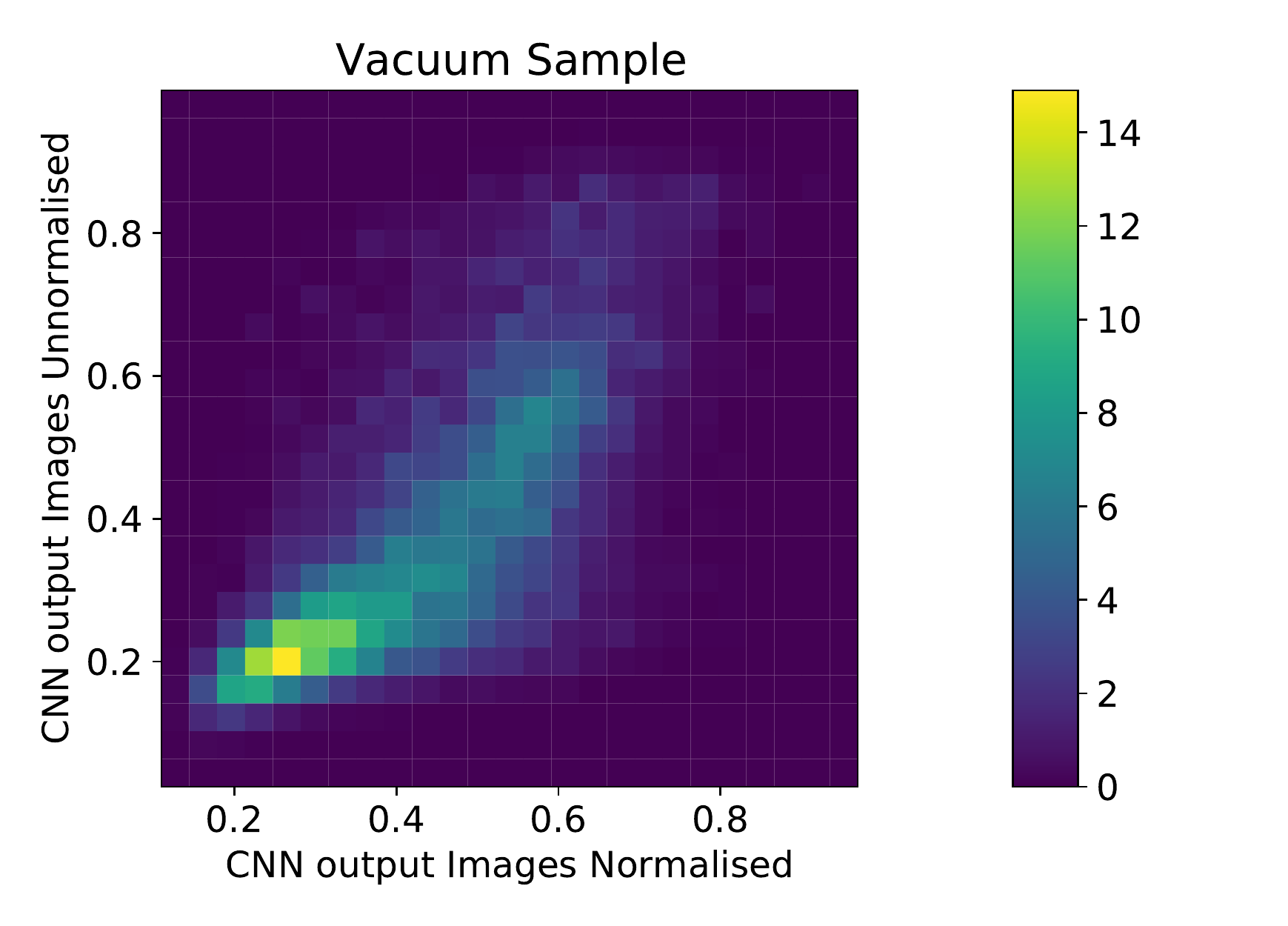}
    \caption{\label{fig:nnoutputscorr_vac} Bi-dimensional distributions of the Deep Neural Network outputs for the Vacuum sample.}
\end{figure}

\begin{figure}[]
    \includegraphics[width=0.32\textwidth]{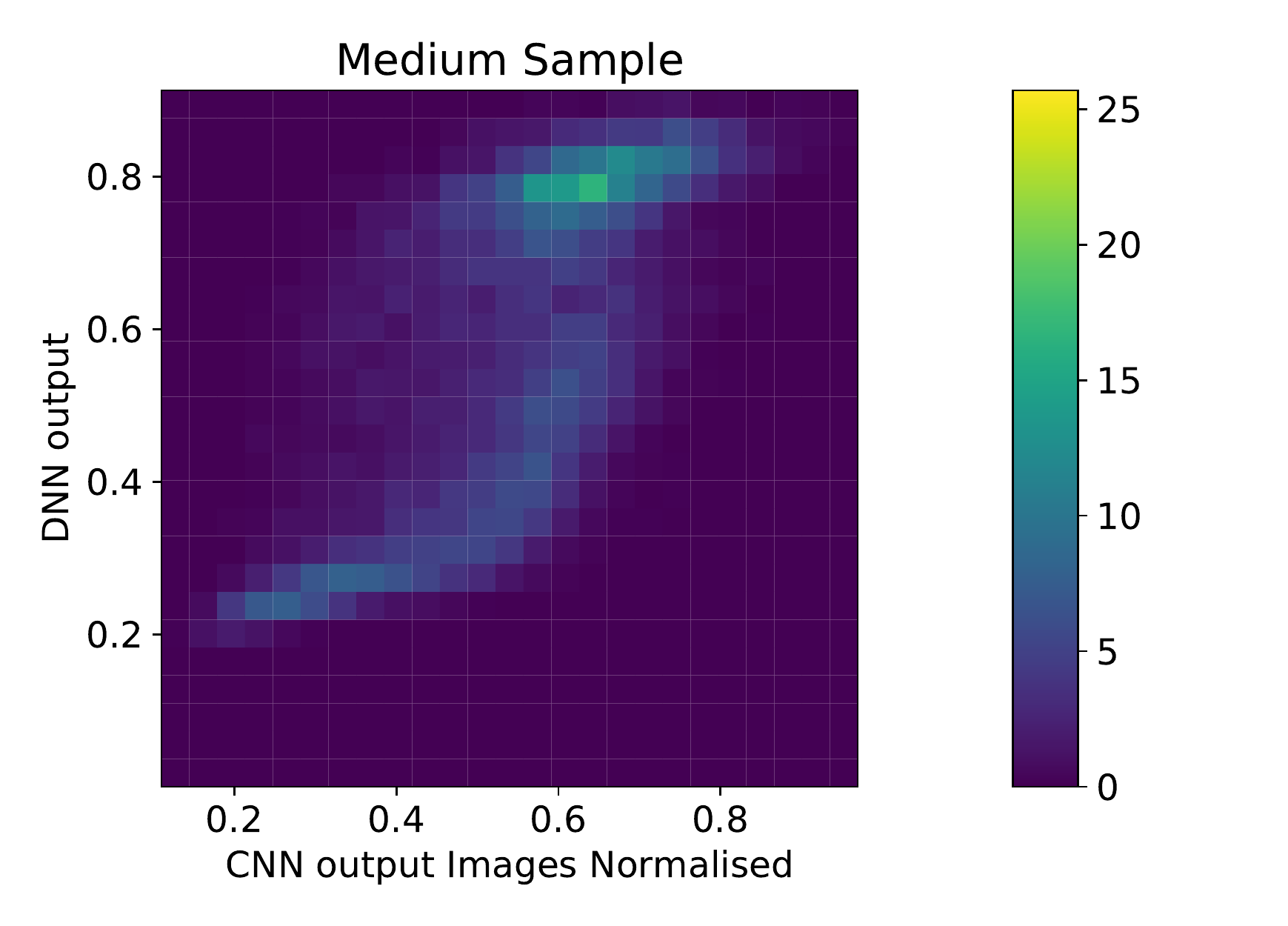}
    \includegraphics[width=0.32\textwidth]{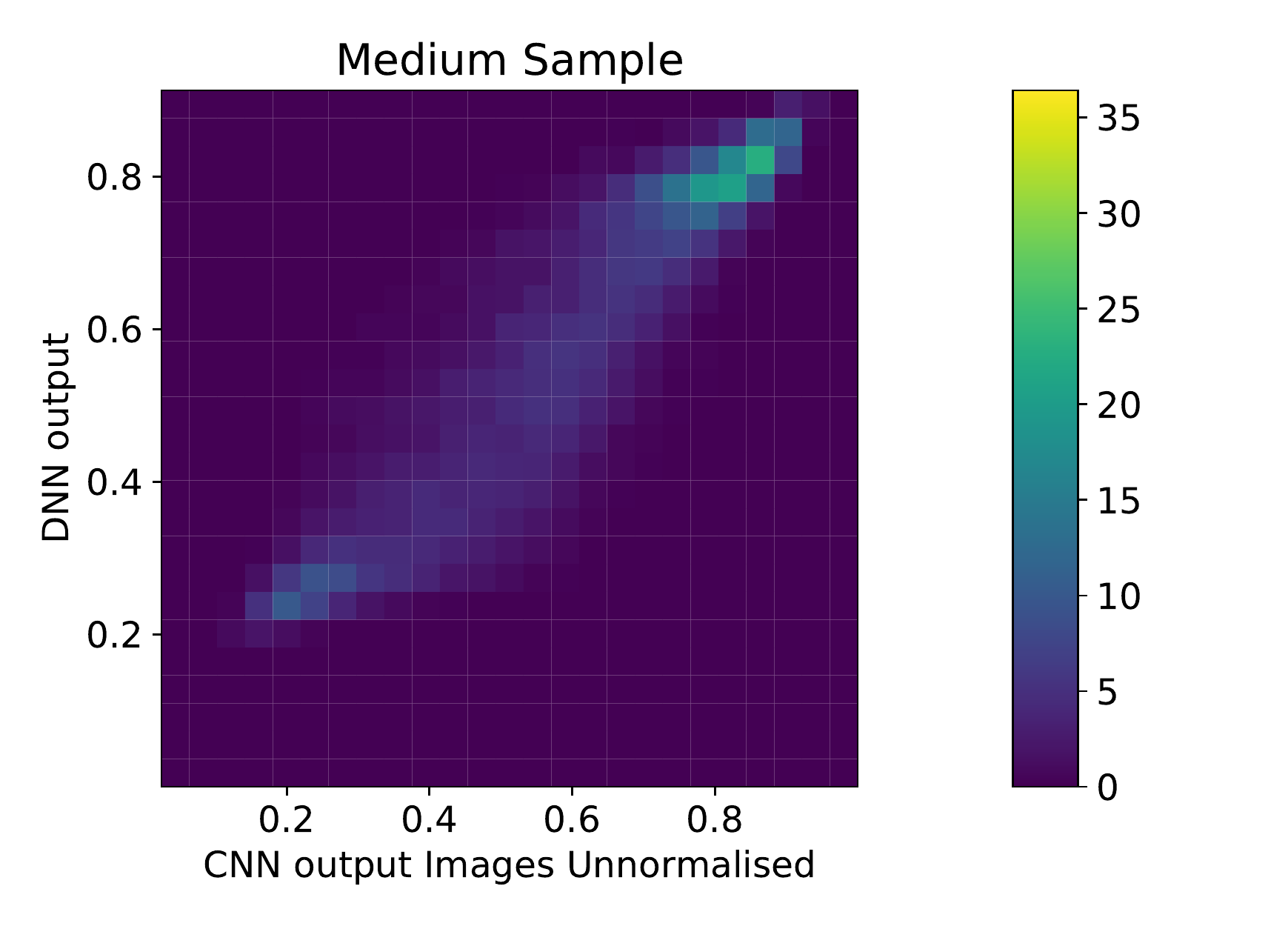}
    \includegraphics[width=0.32\textwidth]{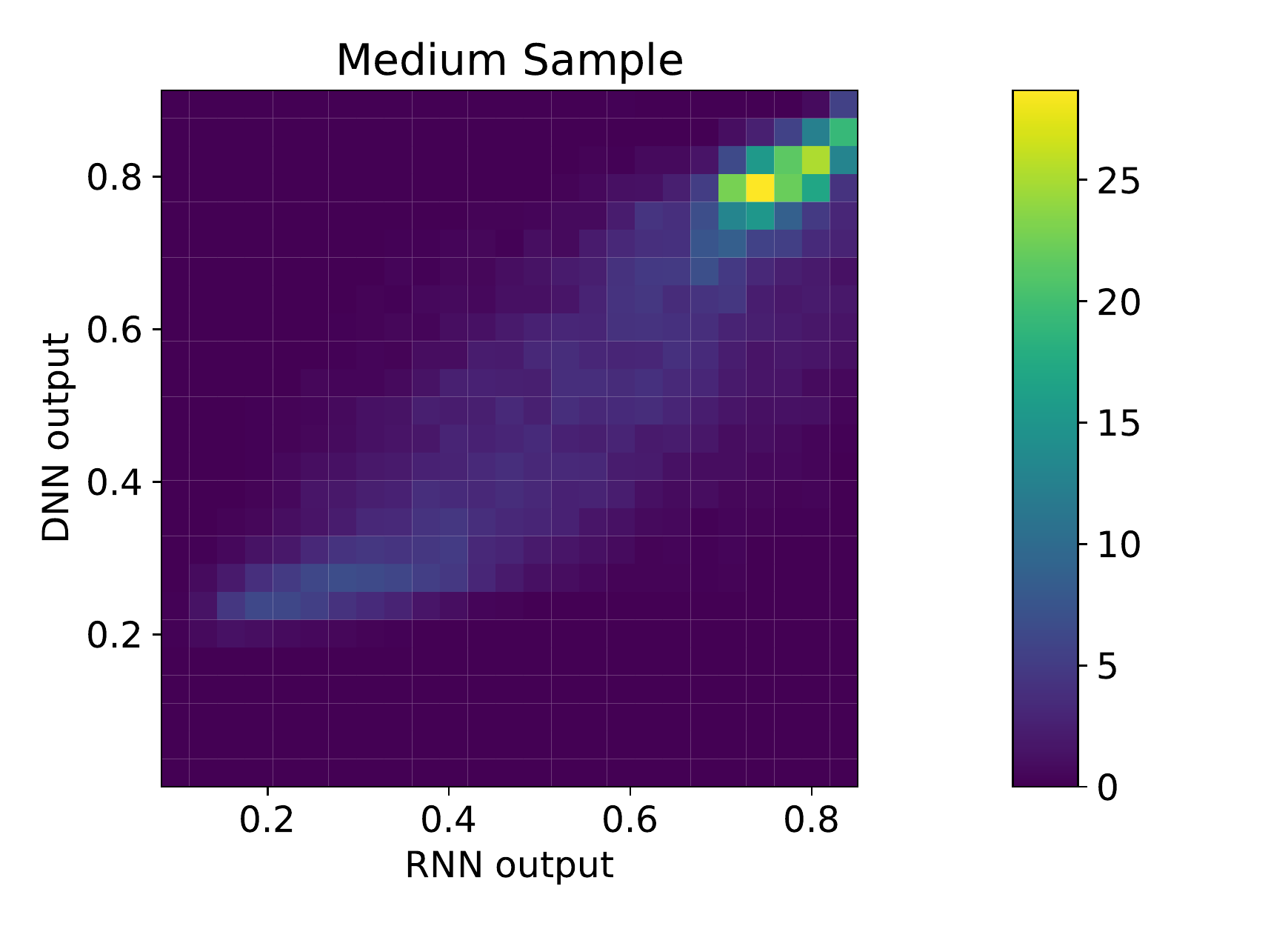}\\
    \includegraphics[width=0.32\textwidth]{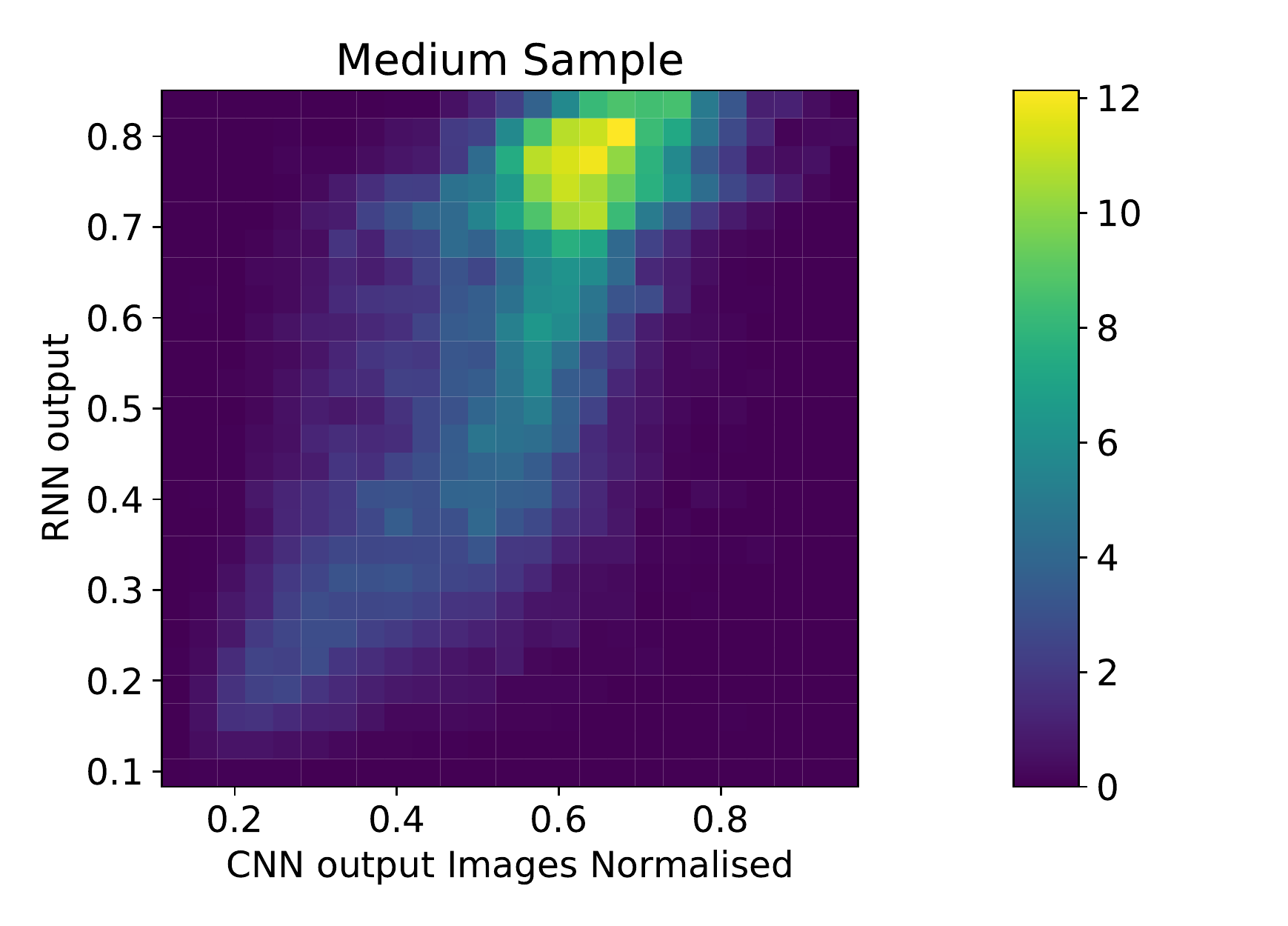}
    \includegraphics[width=0.32\textwidth]{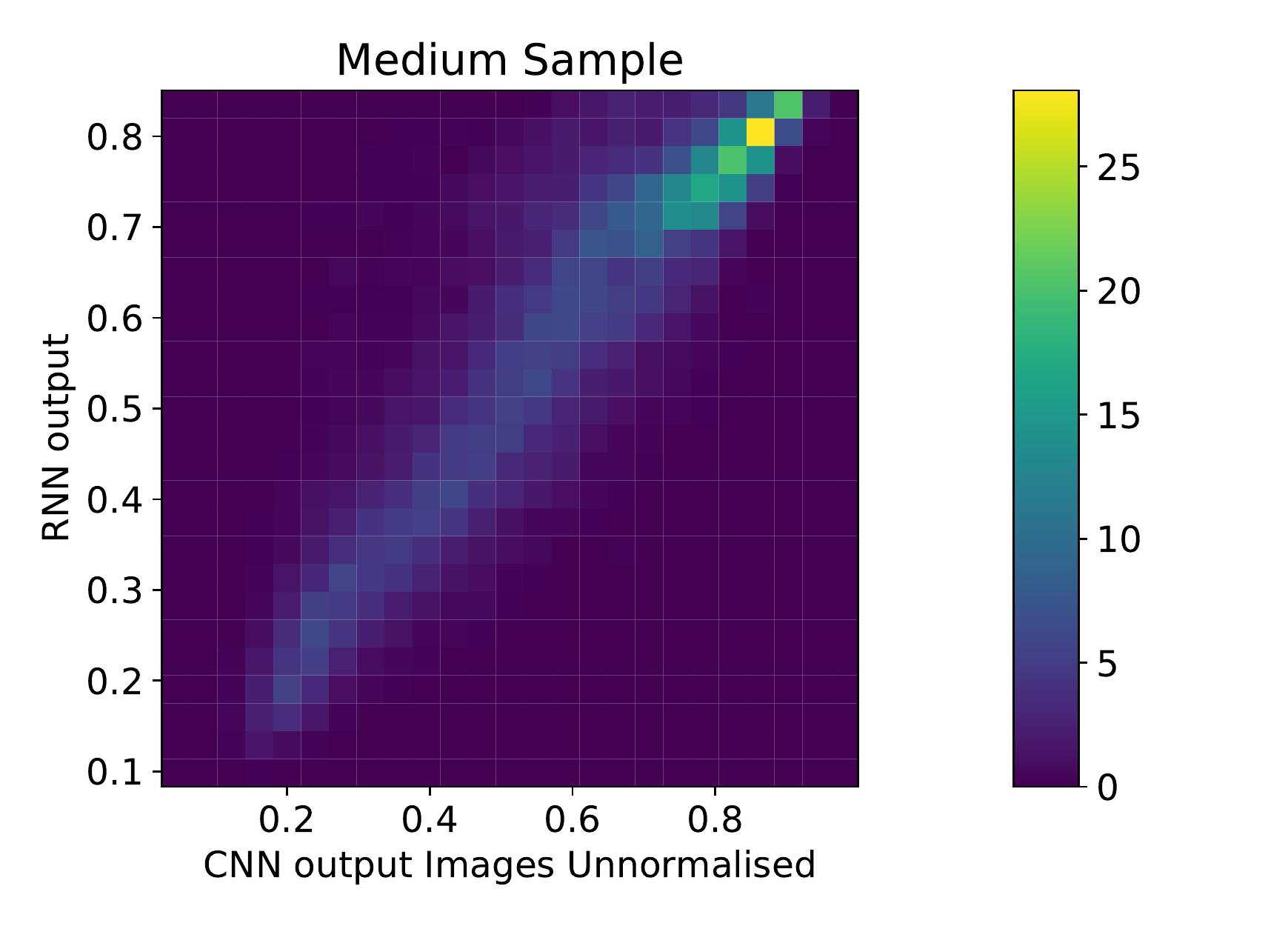}\\
    \includegraphics[width=0.32\textwidth]{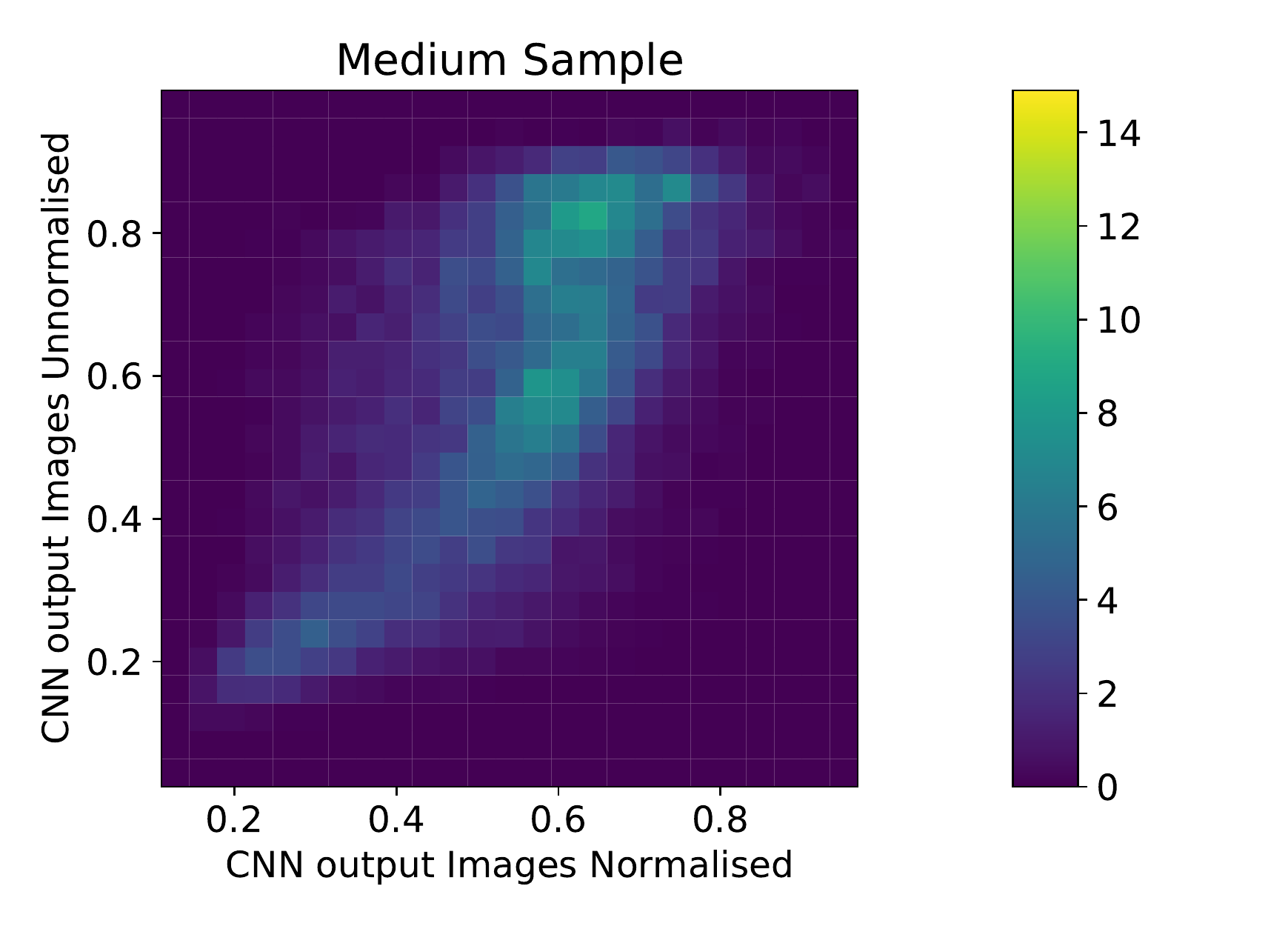}
    \caption{\label{fig:nnoutputscorr_medium} Bi-dimensional distributions of the Deep Neural Network outputs for the Medium sample.}
\end{figure}

We inspect the bi-dimensional distributions of the outputs of the DL models for all pair combinations of models in \cref{fig:nnoutputscorr_vac} and \cref{fig:nnoutputscorr_medium}, respectively for Vacuum and Medium. There is a strong linear correlation for the models which access the jet $p_T$ distribution, \emph{i.e.} the global DNN, the RNN and the CNN trained on unnormalised images, providing evidence that the underlying common features being learnt by the models are the distributions of jet $p_T$ and number of constituents. While still existent, the correlation between the output of these models and the output of the CNN on normalised images is more faint in the Medium sample, which corroborates the conclusion. Moreover, the output of the CNNs for Vacuum are significantly correlated.

\FloatBarrier
\section{Interpreting what the CNNs learnt}\label{app:CNNinterpretation}
\label{app:interpret}

Whilst most of the DNN architectures function as black-boxes once trained, methodologies have been developed to help us better understand what CNNs learn during training. After analysing some jet observables in~\cref{sec:Results}, we will now proceed to further identify which particular jet feature are CNNs using to classify the jet as experiencing jet quenching effects.

The first of such methods is the one of finding the images that produce maximal activations for the filters. As it was earlier argued in~\cref{sec:DL}, the CNN architecture used in this work is such that the last convolutional layer produces an array of dimensionality $N_{filter}$, \emph{i.e.} a dense vector of the high-level representation of the data, which is then passed on to a linear classifier in the last layer, being the output of the entire network
\begin{equation}
     \text{CNN} (X) = \sigma( \Vec{w} \cdot \text{CONV}(X) + b) \ ,\label{eq:CNNlinear}
\end{equation}
where $\sigma$ is the sigmoid function, $\Vec{w}$ the weights of the last layer, $b$ the bias of the last layer, and represents the whole convolution process:
\begin{equation}
    \text{CONV}(X) = \text{CONV}4 \circ \text{CONV}3 \circ \text{CONV}2 \circ \text{CONV}1 (X) \ ,
\end{equation}
where each CONV$_i$ represents a convolutional layer, c.f.~\cref{fig:cnn}.

Since the features produced by the last convolutional layer are afinely combined to produce the probability of belonging to the medium sample~\cref{eq:CNNlinear}, we know the contribution that each will impact the final score by inspecting the value of the weight in $\Vec{w}$ that multiplies it. Furthermore, we can find an image with the patterns that maximise these final features:
\begin{equation}
    \max_X \text{ CONV}(X)_i \ ,
\end{equation}
for each $i$ filter. This method is known as GradCAM \cite{selvaraju2017grad}, as it makes use of gradient descent to maximise an image, which is initialised at random, to produce an interpretable pattern of the learned features.

\begin{figure}[t]
    \centering 
    \includegraphics[width=0.32\textwidth]{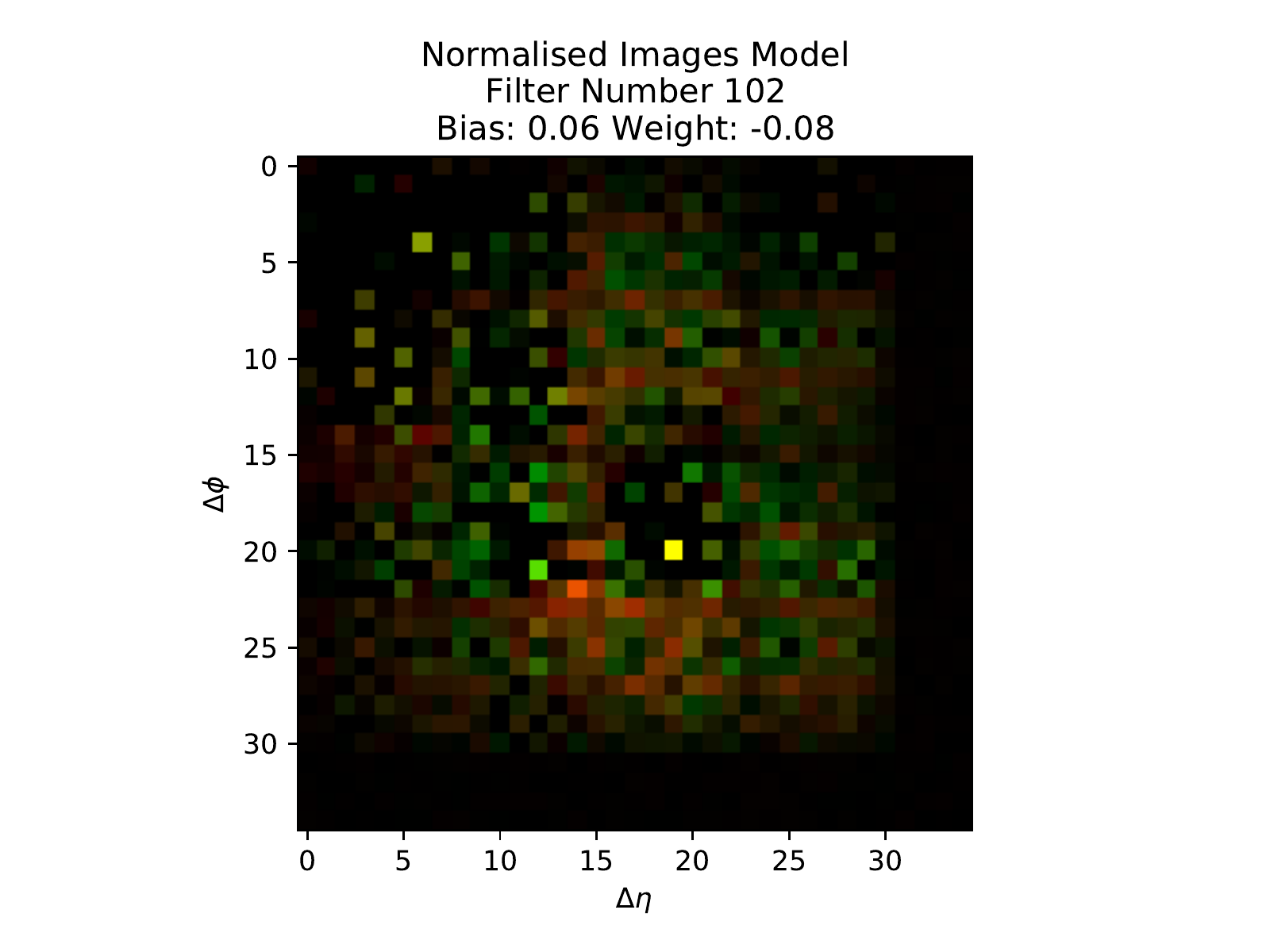}
    \includegraphics[width=0.32\textwidth]{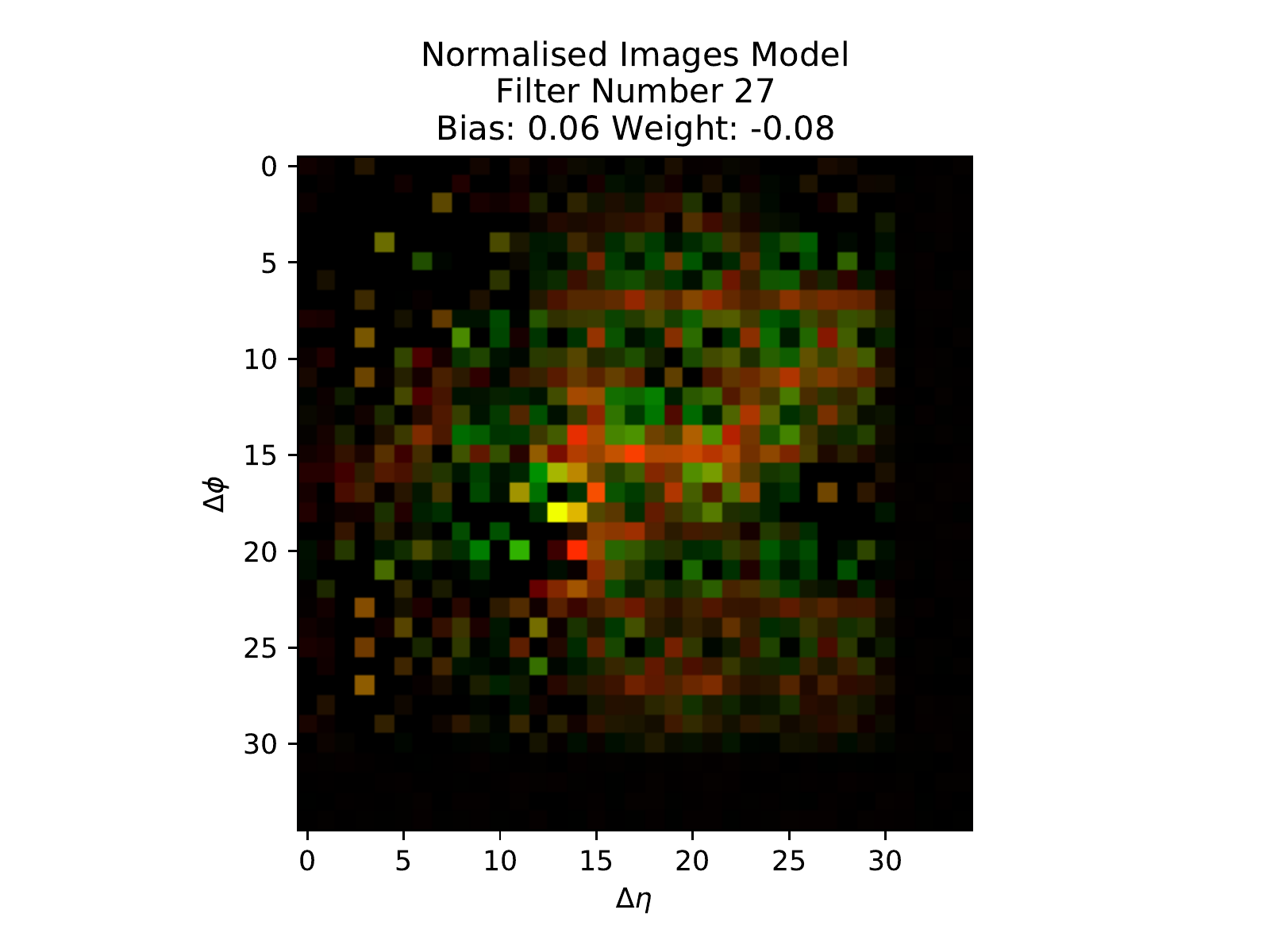}
    \includegraphics[width=0.32\textwidth]{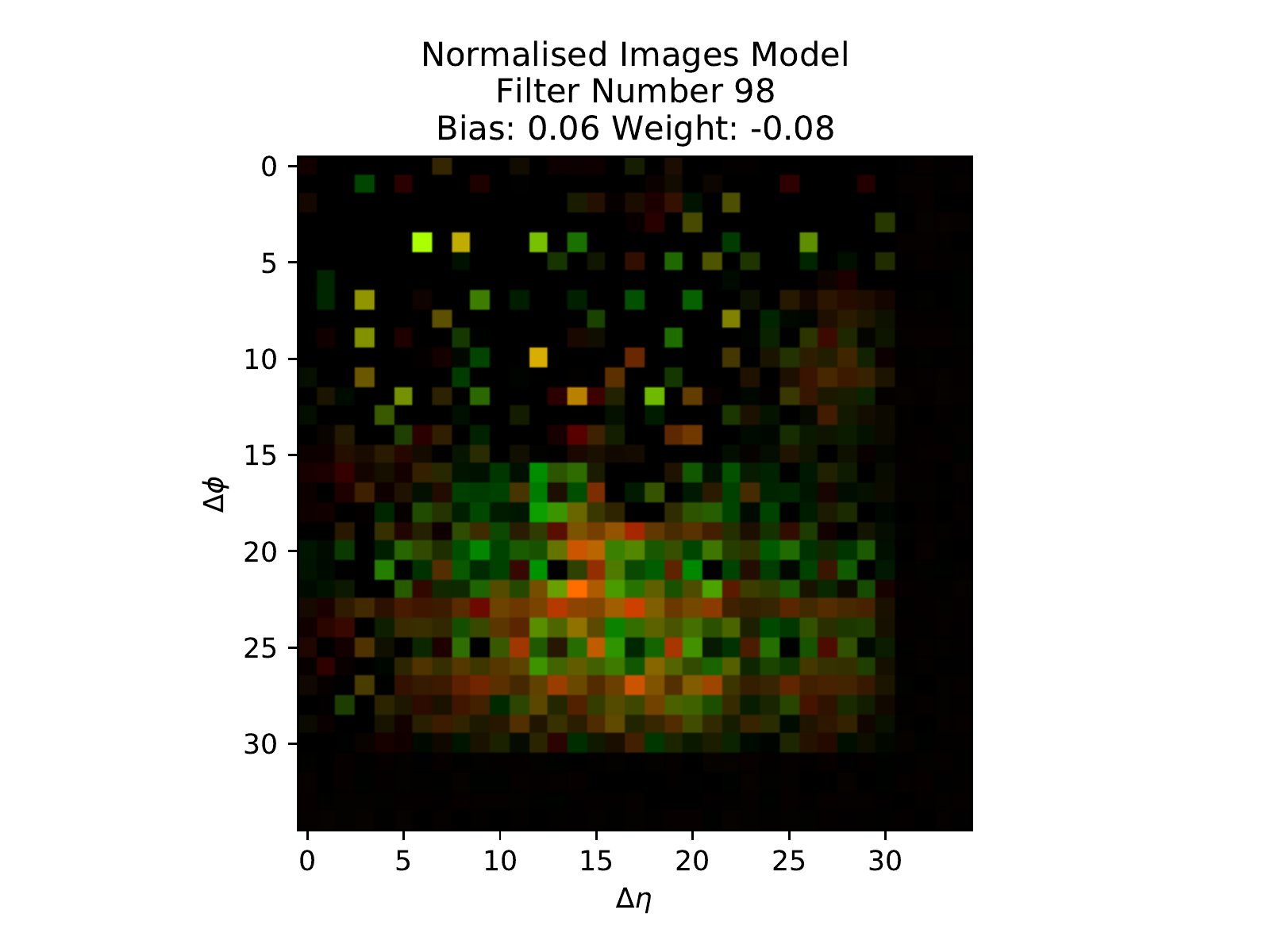} \\
    \includegraphics[width=0.32\textwidth]{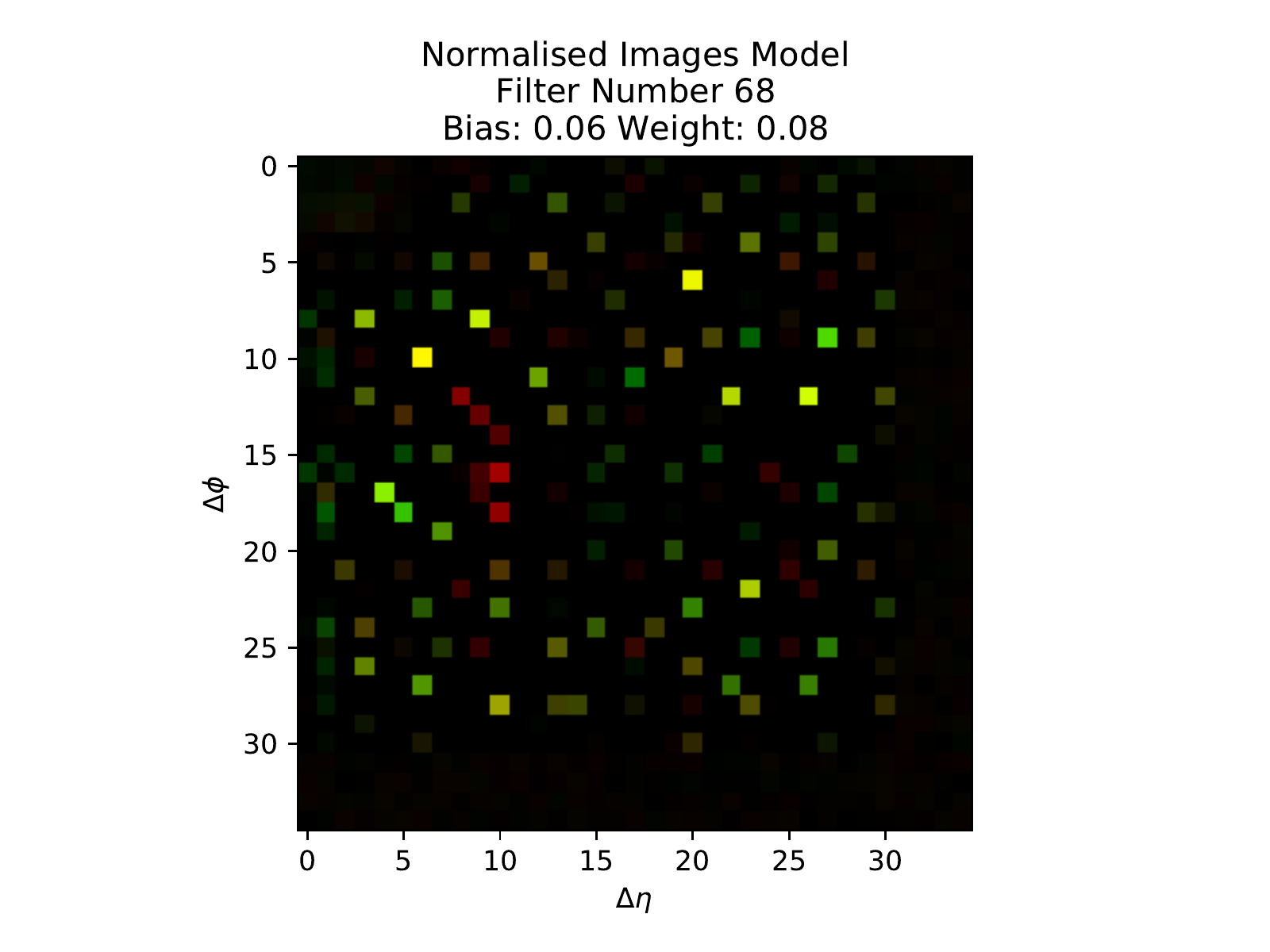}
    \includegraphics[width=0.32\textwidth]{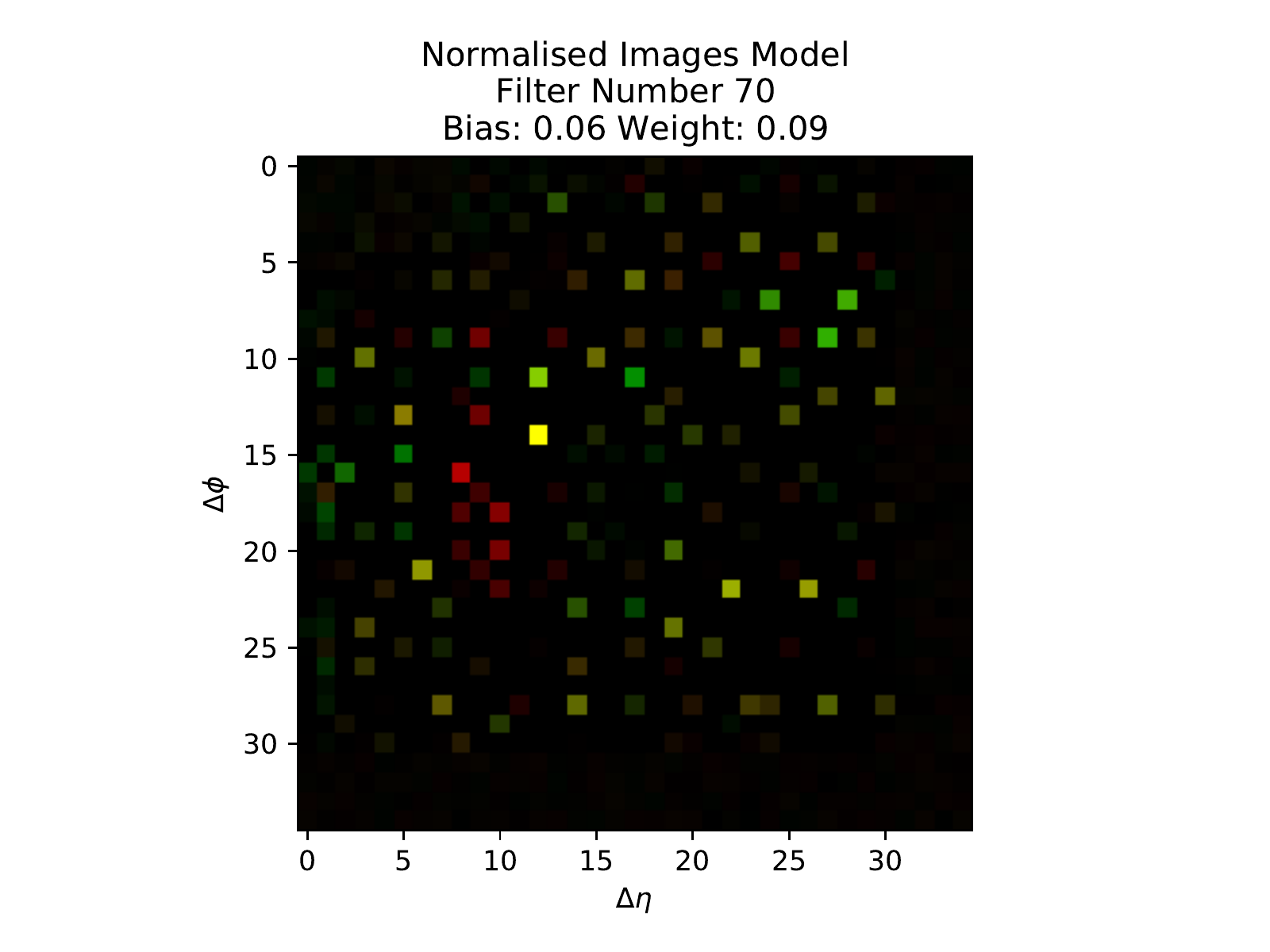}
    \includegraphics[width=0.32\textwidth]{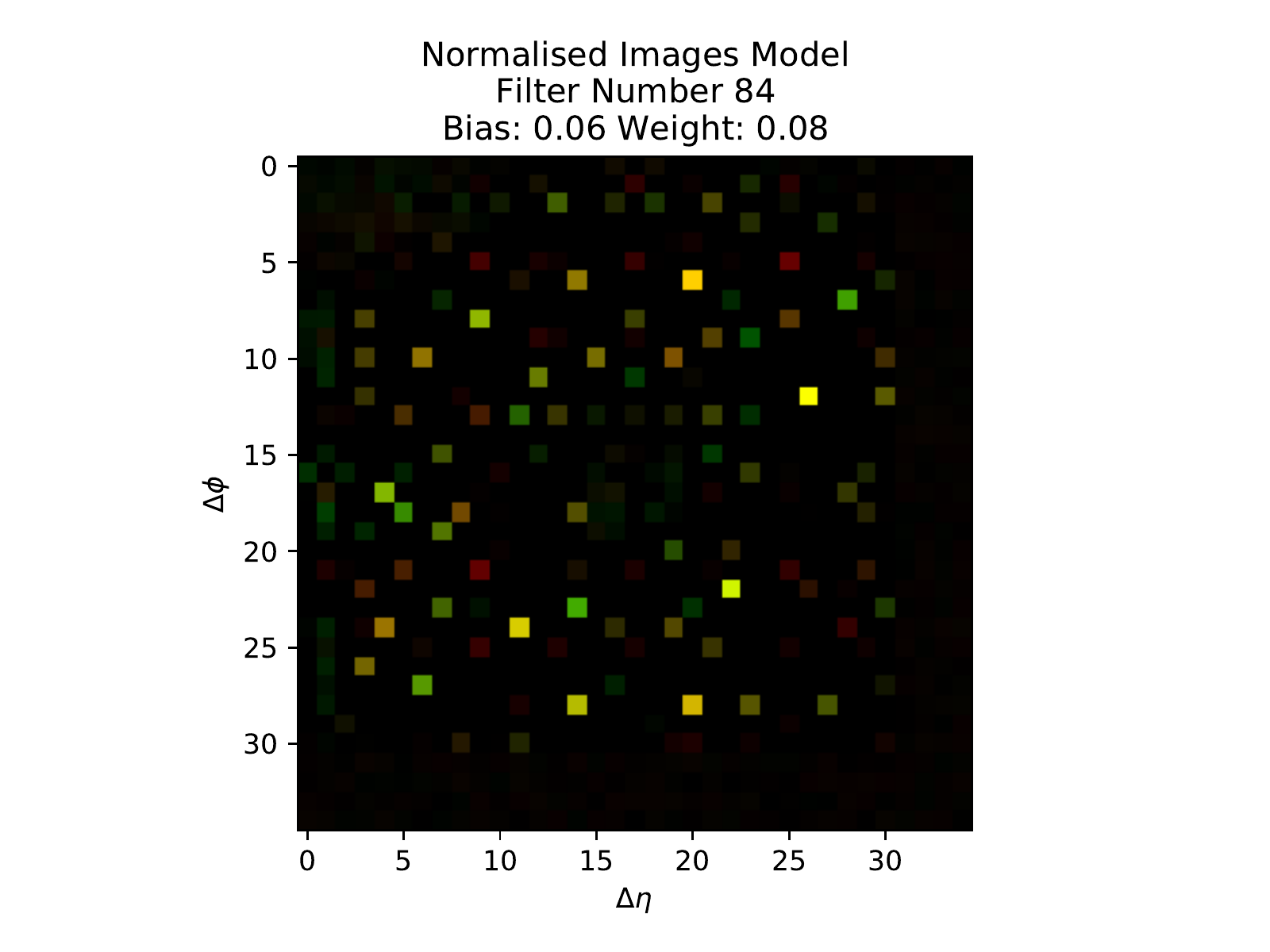}
    \caption{\label{fig:maxactiv-norm} The most discriminative patterns of the normalised images. Red represents the transverse momentum channel, green the multiplicity channel. Brighter pixels represent the pattern that maximised the activation of the respective filter. Since the top (bottom) filters have a negative (positive) weight, they are being triggered by vacuum (medium) sample jets.}
\end{figure}

In~\cref{fig:maxactiv-norm} we produce the patterns for the maximal activations for the three most discriminant high-level features, \emph{i.e.} the ones that have the largest and smallest associated weight $w_i$ in the final layer, for the CNN trained on normalised images. Negative (positive) values of the weights mean that the associated pattern is contributing to classify an image as vacuum (medium). We notice that for the patterns associated with vacuum discrimination, the CNN learnt to look for denser distributions of both transverse momentum and multiplicity across the grid cells, whereas for the medium sample it is looking for far more scattered patterns. In addition, we notice how many pixels have yellow hue, meaning that the network is looking at both channels jointly (otherwise the pixel would either be red or green, depending if the network focused solely on the $p_T$ or multiplicity channel respectively). The brighter the pixel (regardless if it is with respect to one or both channels) the more important it was to contribute to the activation of the filter.

\begin{figure}[h]
    \centering 
    \includegraphics[width=0.32\textwidth]{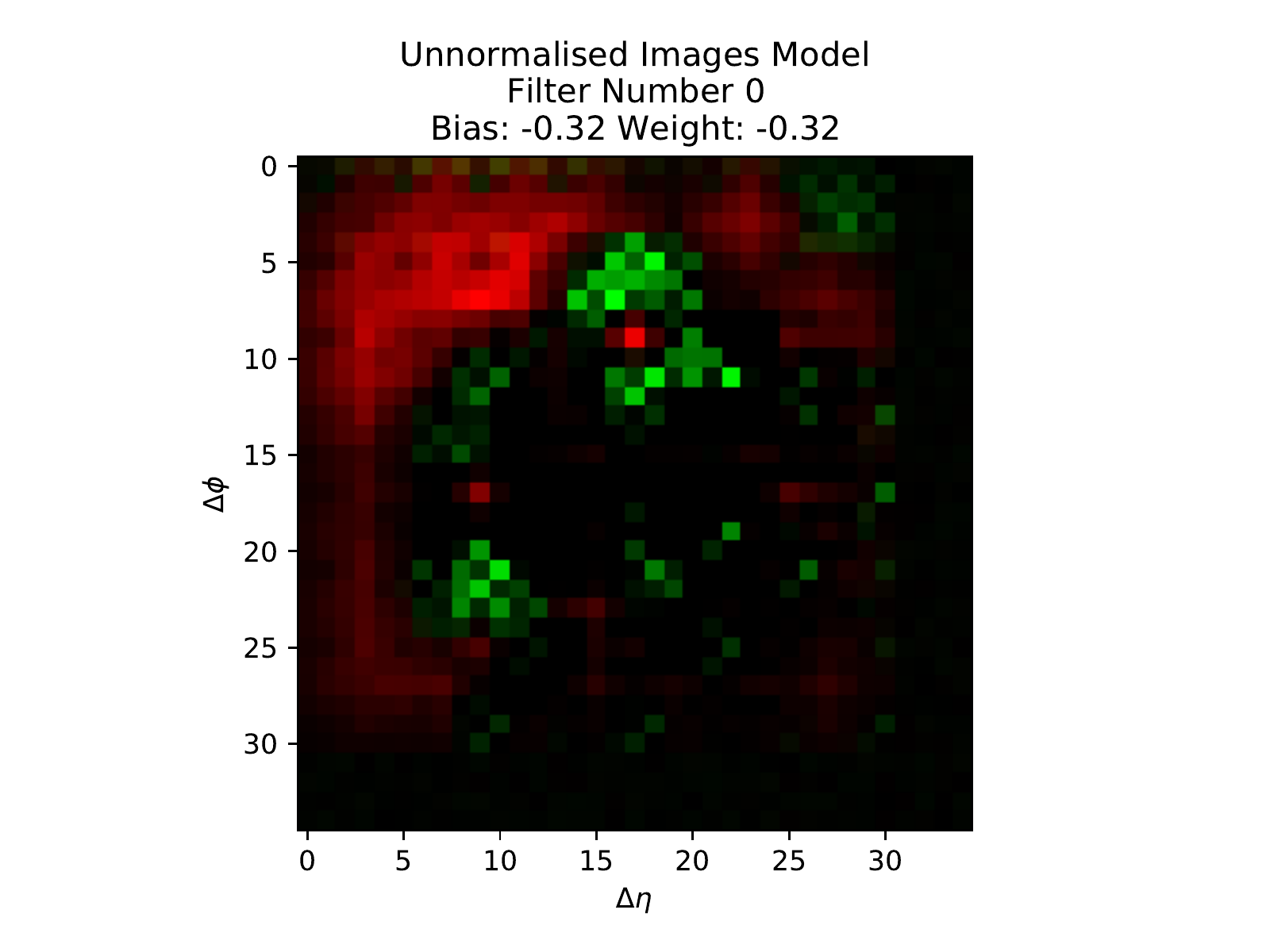}
    \includegraphics[width=0.32\textwidth]{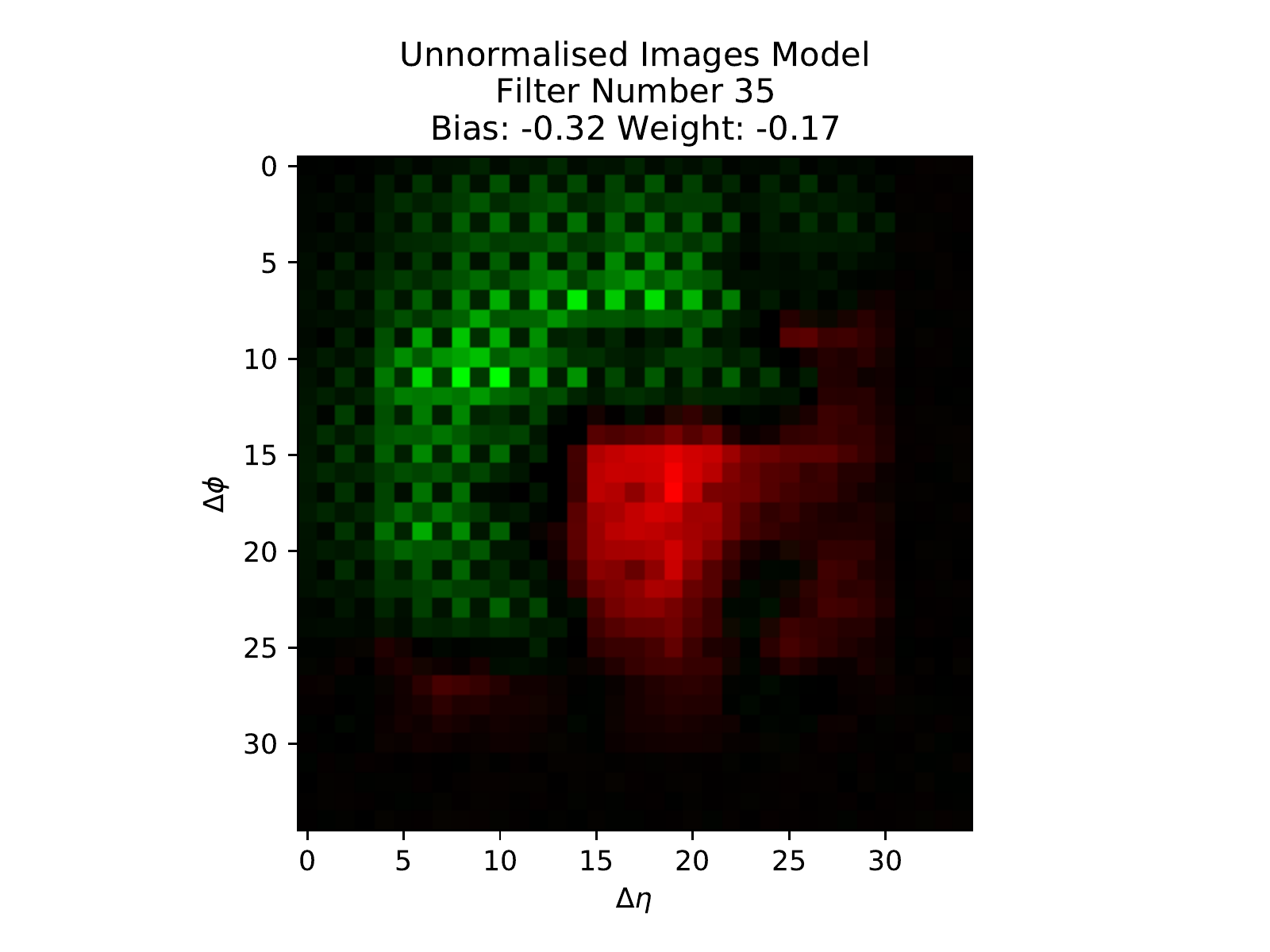}
    \includegraphics[width=0.32\textwidth]{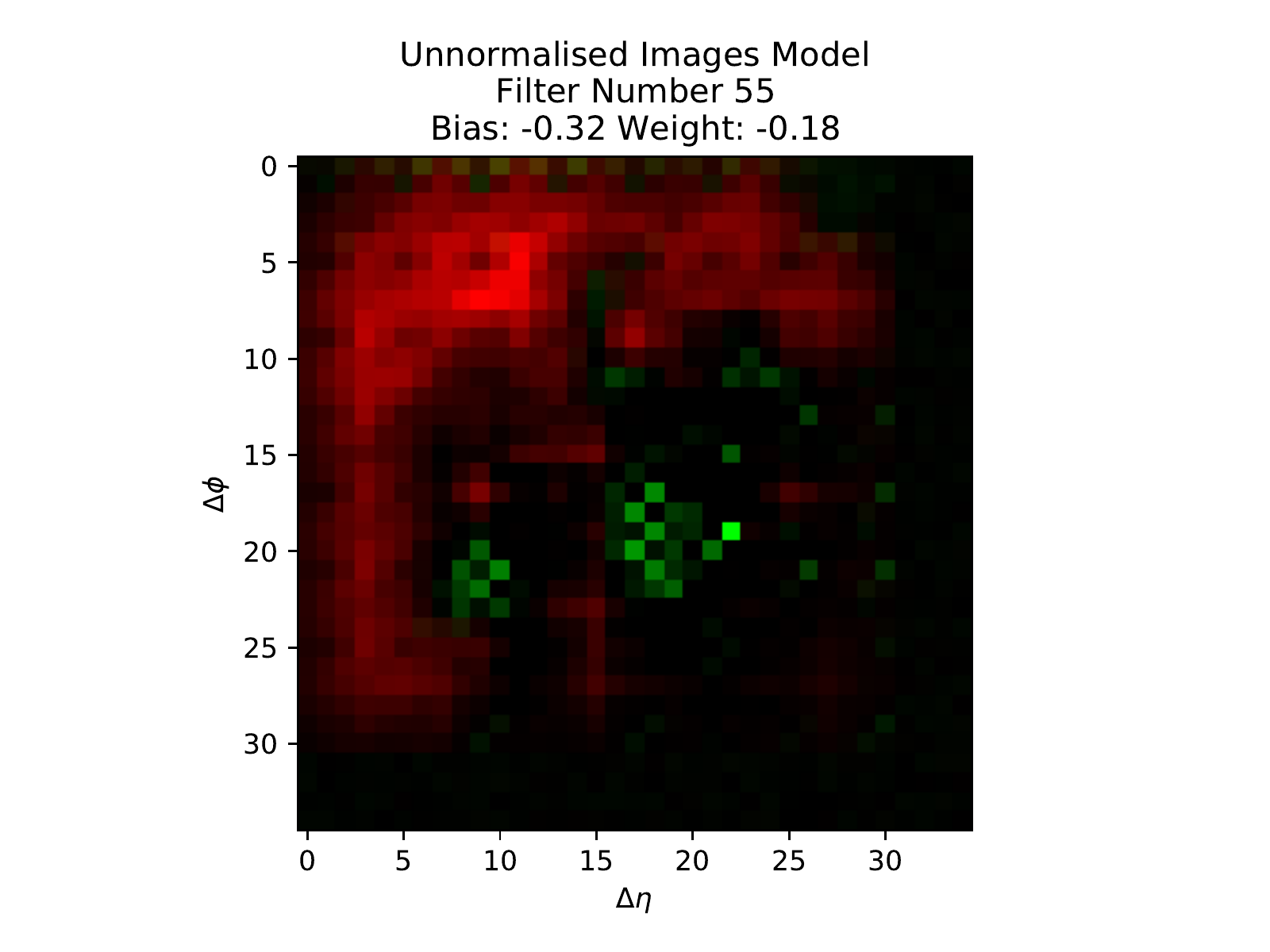}\\
    \includegraphics[width=0.32\textwidth]{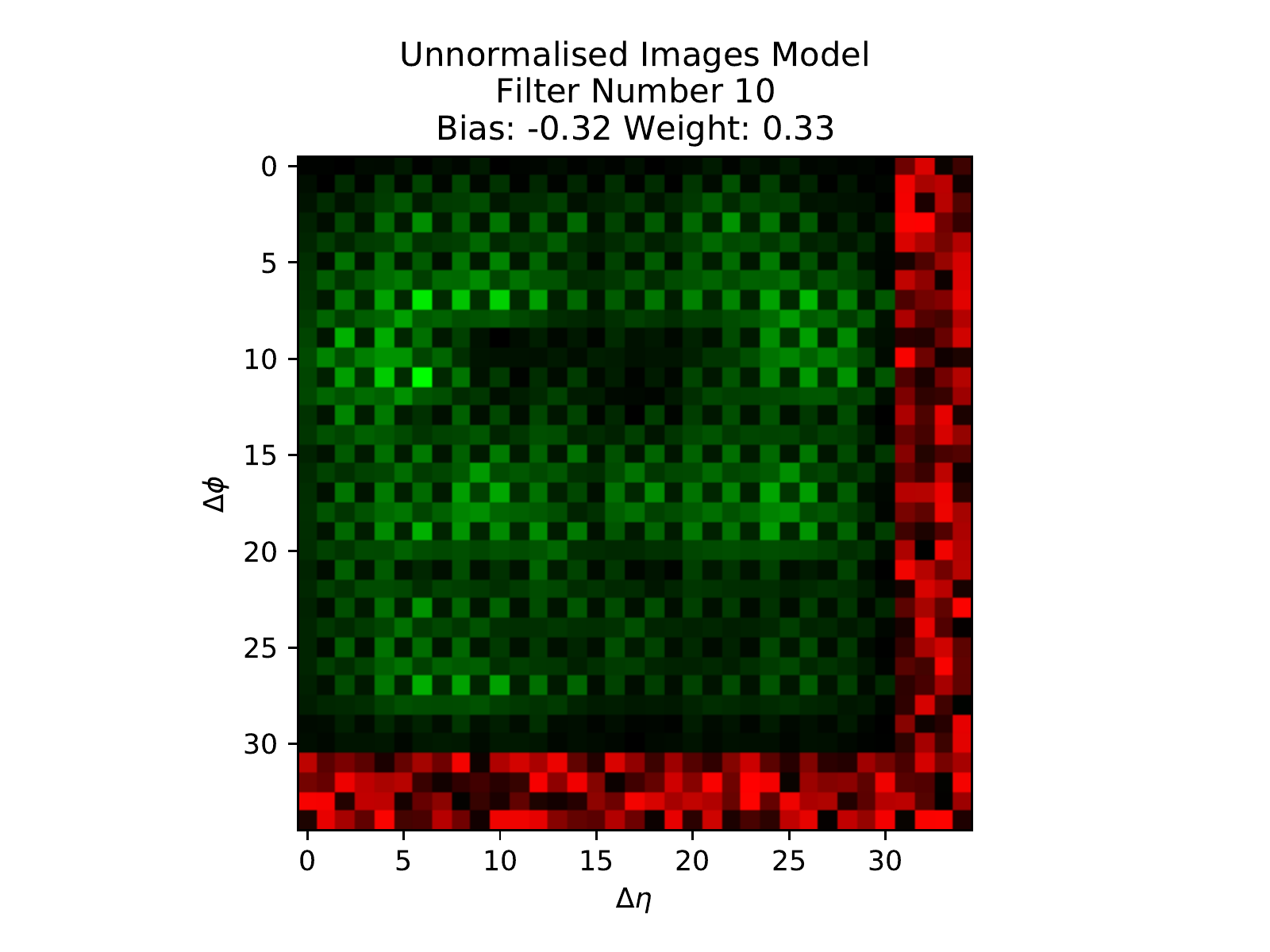}
    \includegraphics[width=0.32\textwidth]{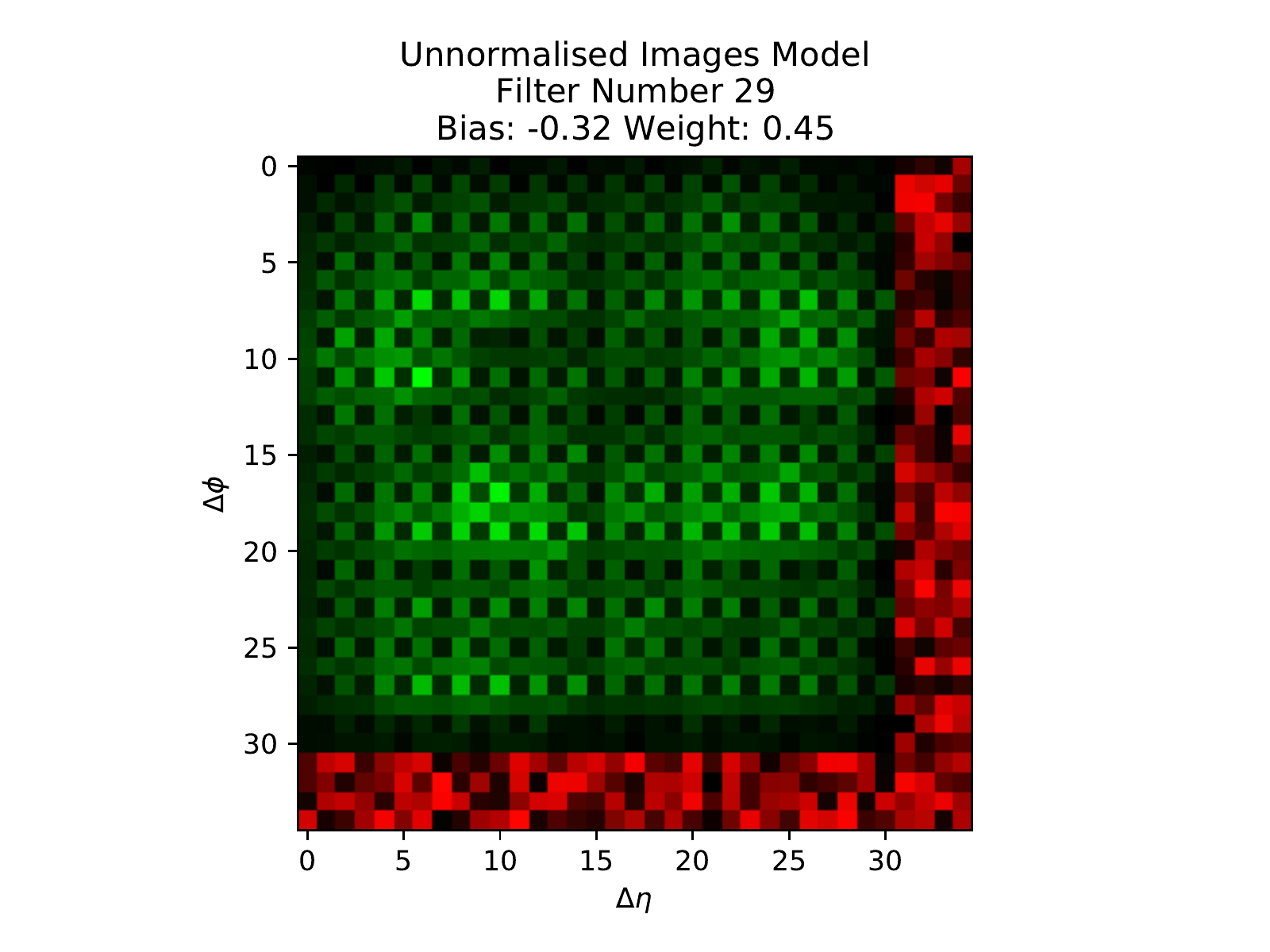}
    \includegraphics[width=0.32\textwidth]{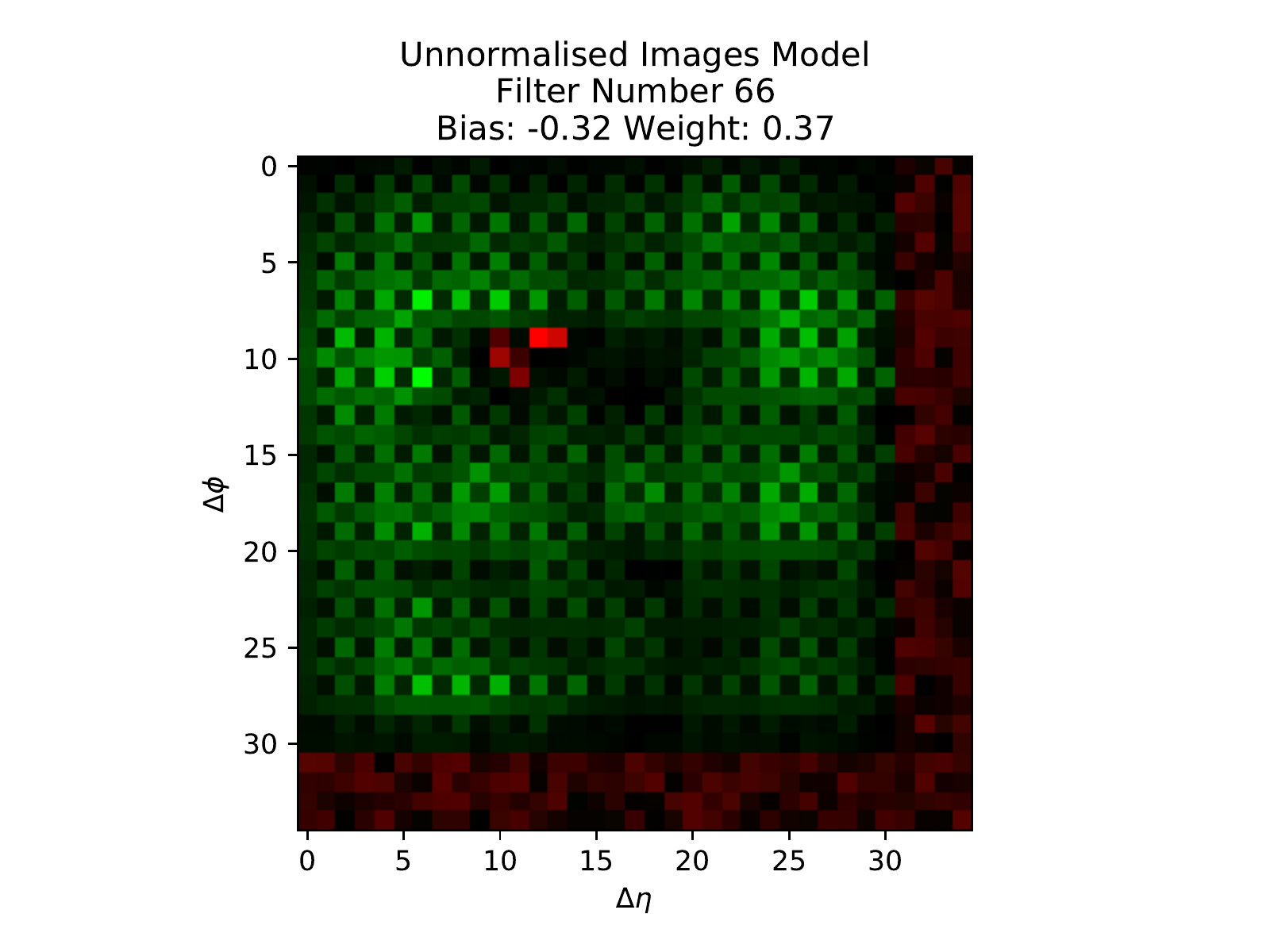} \\
    \caption{\label{fig:maxactiv-unnorm} The most discriminative patterns of the unnormalised images. Red represents the transverse momentum channel, green the multiplicity channel. Brighter pixels represent the pattern that maximised the activation of the respective filter. Since the top (bottom) filters have a negative (positive) weight, they are being triggered by vacuum (medium) sample jets.}
\end{figure}

In~\cref{fig:maxactiv-unnorm} we produce the patterns for the maximal activations for the three most discriminant high-level features, for the CNN trained on unnormalised images. We observe completely different patterns than those learnt by the CNN on normalised images. More concretely, there is no discerning trend to prefer denser patterns for vacuum and scattered ones for medium. Furthermore, at each pixel it is focusing either on $p_T$ or $n_{const}$, and the regions that it looks for the distribution of each are disjoint. This result is in agreement with~\cref{fig:ptZ}, where we observed that the output of the CNN on unnormalised images is sensitive to the scale of the $p_T$ of the $Z$ being emitted in the hard scattering.

While the previous study helps us understanding patterns which, once convoluted with the image, affect the final classification score, it does not provide any insight on specific images. Therefore, we need another method that produces an \emph{explanation} from the network on why it classified an example the way it did. This can be accomplished with a method called integrated gradients \cite{sundararajan2017axiomatic}, which schematically works as follows. First initialise a random image, \emph{i.e.} noise. Then, select a data image that has been properly classified, $X$, and produce $N+1$ linearly interpolated images between that image and the noise image:
\begin{equation}
    \{x_i : x_i = X_{noise} + \frac{i}{N} (X - X_{noise}), \ i \in [0, N] \} \ .
\end{equation}
This sequential interpolation approximates the morphing of the noise image into the data image. By computing the gradients of the prediction, \emph{i.e.} the output of the neural network for the correct class, with respect to each $x_i$ we will know what parts of the data image were relevant for the classification. By integrating them over $i$ using the trapezoidal rule, we will obtain an \emph{explanation} of the classification by isolating the pixels of the data image that contributed the most. We closely followed the implementation provided in the Keras website~\cite{intgrads}.

\begin{figure}[h]
    \centering 
    \includegraphics[width=0.9\textwidth]{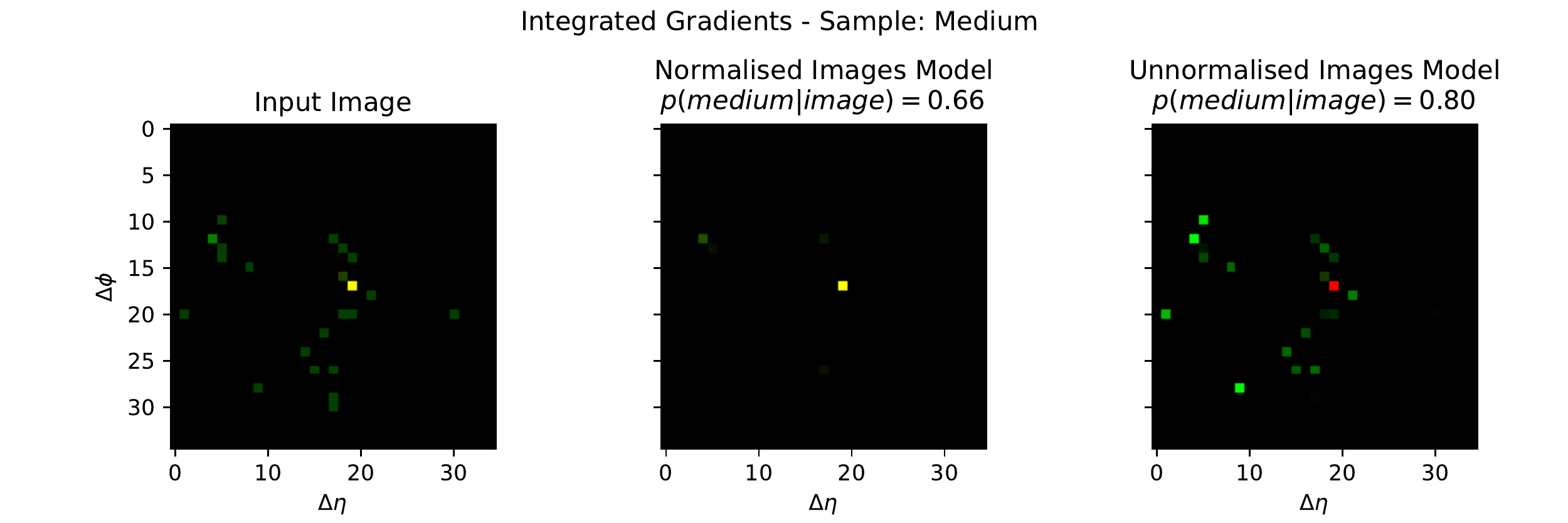}\\
    \includegraphics[width=0.9\textwidth]{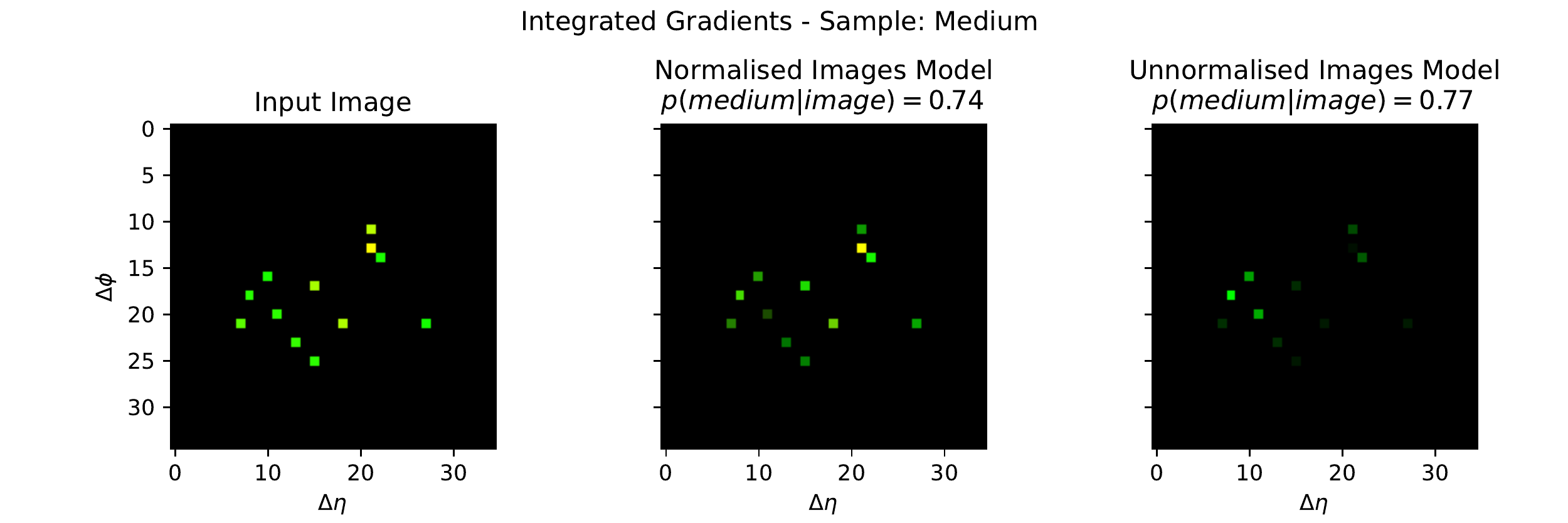}\\
    \includegraphics[width=0.9\textwidth]{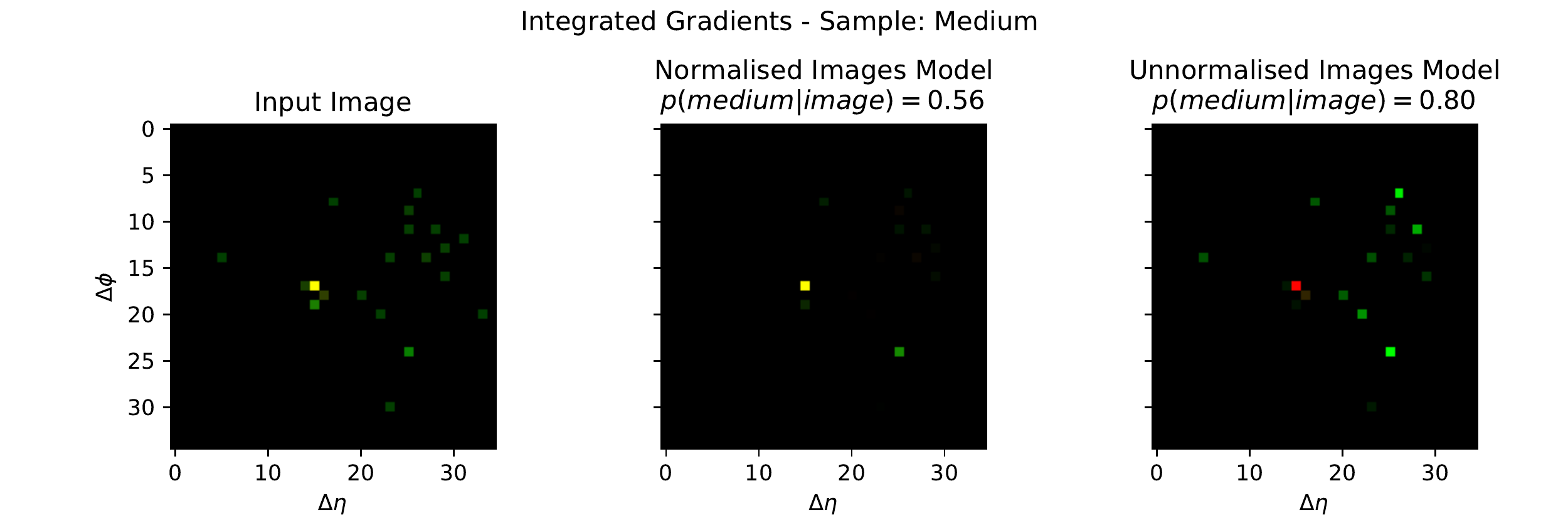}
    \caption{\label{fig:intgradmed} Integrated gradients for some correctly classified medium sample images.}
\end{figure}

In~\cref{fig:intgradmed} we can observe the integrated gradients for three examples of images from the medium sample from CNN models. For the normalised images, the model is looking for the pixel with the highest $p_T$ and $n_{const}$ fractions, \emph{i.e.} the bright yellow pixel on the input image, and it is looking at both channels of that pixel at the same point. On the otherhand, for the unnormalised images, the model is seemingly looking for the pixel with the most $p_T$ while jointly it learns the distribution of $n_{const}$ \emph{elsewhere} in the image. These are in agreement with the most discriminating patterns above, where we saw that the while the normalised images model is jointly looking at both channels, \emph{i.e.} the $p_T$ and $n_{const}$ spatial distributions, the unnormalised images model is focusing on these distributions at disjoint regions of the image.

\begin{figure}[h]
    \centering 
    \includegraphics[width=0.9\textwidth]{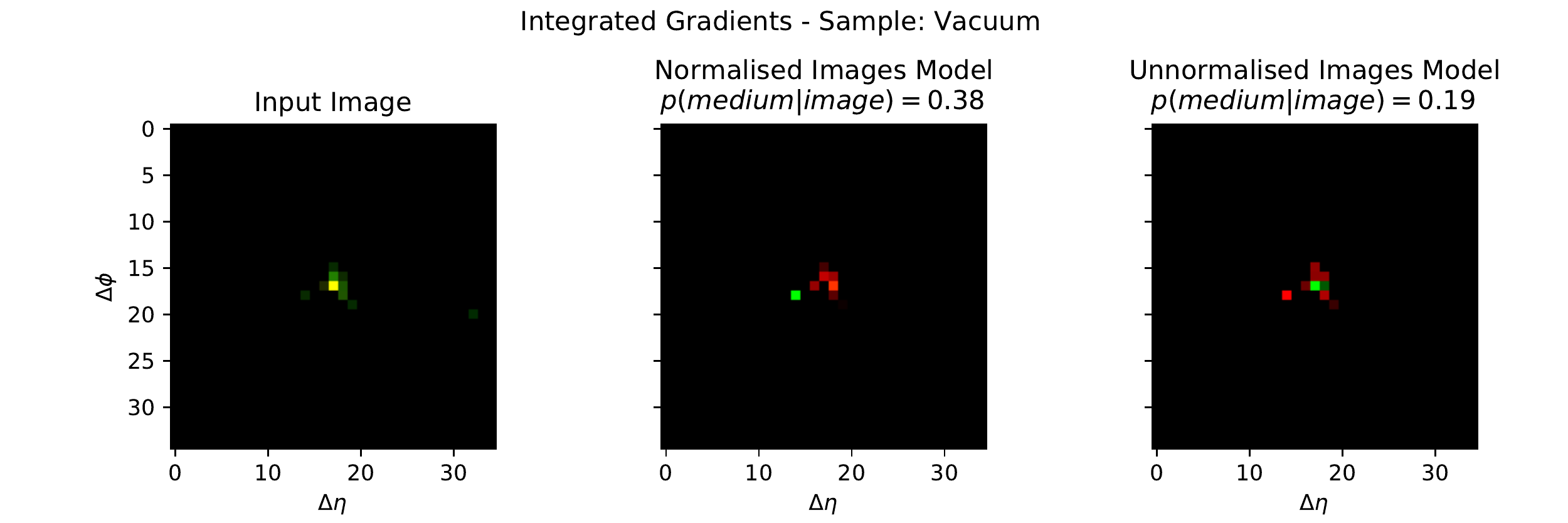}\\
    \includegraphics[width=0.9\textwidth]{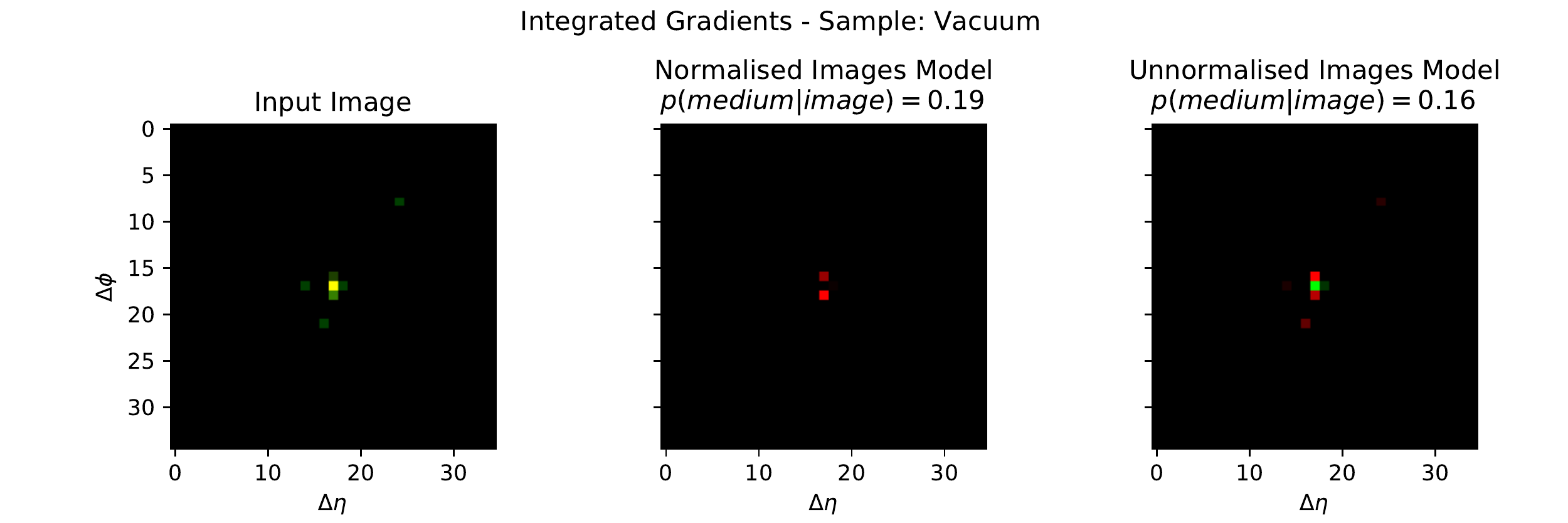}\\
    \includegraphics[width=0.9\textwidth]{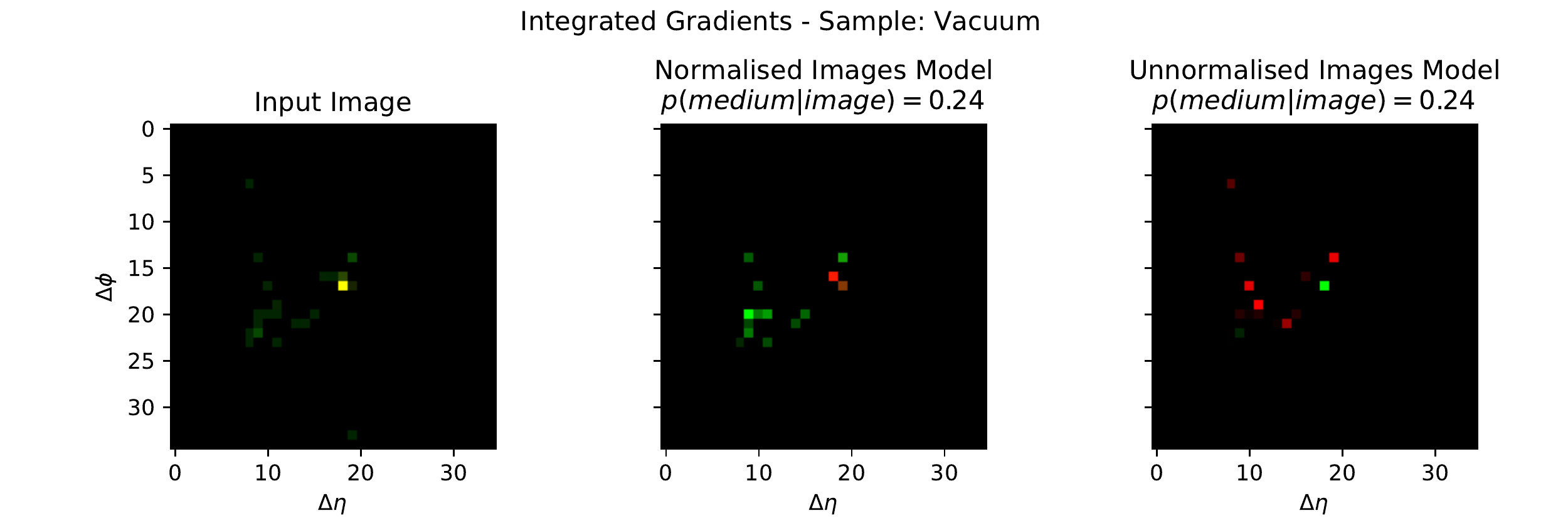}
    \caption{\label{fig:intgradvac} Integrated gradients for some correctly classified vacuum sample images.}
\end{figure}

Since we used a single sigmoid head for our models, in order to obtain the similar patterns for vacuum images we minimised, instead of maximising, the output of the CNN models for correctly classified vacuum images, instead of medium ones. The results are show in~\cref{fig:intgradvac}. The obtained integrated gradients are complementary to the ones we saw for the medium case. However, it seems that normalised images model is seemingly looking at $p_T$ and $n_{const}$ disjointly just like thr normalised images model. This might indicate that for vacuum images both networks are learning somehow similar features, while their learnt features for medium differ. This is corroborated by~\cref{fig:nnoutputscorr_vac} where we see that both CNN are highly correlated in vacuum, but the same is not true for the medium~\cref{fig:nnoutputscorr_medium}. 



\clearpage
\newpage

\bibliography{paper}{}
\bibliographystyle{JHEP}

\end{document}